\documentclass[twocolumn,traditabstract,longauth]{aa}

\usepackage{color}
\usepackage{graphicx}
\usepackage{epstopdf} 
\usepackage{txfonts}
\usepackage{booktabs} 
\usepackage{natbib}
\usepackage[switch, modulo]{lineno}
\usepackage{rotating}
\usepackage{xspace}
\usepackage{float}
\usepackage{url}
\usepackage[breaklinks=false]{hyperref} 
\newcommand{\g}{$\gamma$\xspace}
\newcommand{\wco}{$W_{\rm{CO}}$\xspace}
\newcommand{\hd}{H$_2$\xspace}
\newcommand{\hi}{\ion{H}{i}\xspace}
\newcommand{\hii}{\ion{H}{ii}\xspace}
\newcommand{\nh}{$N_{\rm{H}}$\xspace}
\newcommand{\nhi}{$N_{\ion{H}{i}}$\xspace}
\newcommand{\nhd}{$N_{\rm{H}_2}$\xspace}
\newcommand{\nhdnm}{$N_{\rm{H}_{\rm{DNM}}}$\xspace}
\newcommand{\nhgam}{$N_{\rm{H}\,\gamma}$\xspace}
\newcommand{\nhlam}{$N_{\rm{H}\,m\lambda}$\xspace}
\newcommand{\nhebv}{$N_{\rm{H}}/E(B-V)$\xspace}

\newcommand{\opa}{$\tau_{353}/N_{\rm{H}}$\xspace}
\newcommand{\opavgdnm}{$\overline{\tau_{353}/N}_{\rm{H}}^{\rm{DNM}}$\xspace}

\newcommand{\opaunit}{$10^{-27}$\,cm$^2$\xspace}
\newcommand{\avq}{$A_{V\rm{Q}}$\xspace}
\newcommand{\anh}{$A_{V\rm{Q}}/N_{\rm{H}}$\xspace}

\newcommand{\anhavgdnm}{$\overline{A_{V\rm{Q}}/N}_{\rm{H}}^{\rm{DNM}}$\xspace}

\newcommand{\anhunit}{$10^{-22}$\,mag\,cm$^2$\xspace}
\newcommand{\qhi}{$q_{\ion{H}{i}}$\xspace}
\newcommand{\qco}{$q_{\rm{CO}}$\xspace}
\newcommand{\qdnm}{$q_{\rm{DNM}}$\xspace}
\newcommand{\yhi}{$y_{\ion{H}{i}}$\xspace}
\newcommand{\yco}{$y_{\rm{CO}}$\xspace}
\newcommand{\ydnm}{$y_{\rm{DNM}}$\xspace}
\newcommand{\qcounit}{$10^{20}$\,cm$^{-2}$\,K$^{-1}$\,km$^{-1}$\,s\xspace}  
\newcommand{\qdnmQunit}{$10^{20}$\,cm$^{-2}$\,mag$^{-1}$\xspace} 
\newcommand{\qdnmTunit}{$10^{25}$\,cm$^{-2}$\xspace} 
\newcommand{\qdnmRunit}{$10^{31}$\,sr\,W$^{-1}$\xspace} 
\newcommand{\yhQunit}{$10^{-22}$\,mag\,cm$^2$\xspace} 
\newcommand{\yhTunit}{$10^{-26}$\,cm$^2$\xspace} 
\newcommand{\yhRunit}{$10^{-32}$\,W\,sr$^{-1}$\xspace} 
\newcommand{\ycoQunit}{$10^{-2}$\,mag\,K$^{-1}$\,km$^{-1}$\,s\xspace} 
\newcommand{\ycoTunit}{$10^{-6}$\,K$^{-1}$\,km$^{-1}$\,s\xspace} 
\newcommand{\ycoRunit}{$10^{-8}$\,W\,m$^{-2}$\,sr$^{-1}$\,K$^{-1}$\,km$^{-1}$\,s\xspace} 
\newcommand{\yisoQunit}{$10^{-2}$\,mag\xspace} 
\newcommand{\yisoTunit}{$10^{-6}$\xspace} 
\newcommand{\yisoRunit}{$10^{-8}$\,W\,m$^{-2}$\,sr$^{-1}$\xspace} 
\newcommand{\Igamunit}{$\gamma$\,cm$^{-2}$\,s$^{-1}$\,sr$^{-1}$\,MeV$^{-1}$\xspace}

\newcommand{\xco}{$X_{\rm{CO}}$\xspace}
\newcommand{\xcoG}{$X_{\rm{CO} \gamma}$\xspace}

\newcommand{\xcoGQ}{$X_{\rm{CO} \gamma}$\xspace}
\newcommand{\xcoT}{$X_{\rm{CO} \tau}$\xspace}
\newcommand{\xcoR}{$X_{\rm{CO} R}$\xspace}
\newcommand{\xcoQ}{$X_{\rm{CO} AvQ}$\xspace}
\newcommand{\xcounit}{$10^{20}$\,cm$^{-2}$\,K$^{-1}$\,km$^{-1}$\,s\xspace}
\newcommand{\taunu}{$\tau_{353}$\xspace}

\newcommand{\tautilde}{$\tilde{\tau}_{353}$\xspace}
\newcommand{\Rtilde}{$\tilde{R}$\xspace}
\newcommand{\avqtilde}{$\tilde{A}_{V\rm{Q}}$\xspace}
\newcommand{\Runit}{W\,m$^{-2}$\,sr$^{-1}$\xspace}
\newcommand{\spw}{$4\pi R/N_{\rm{H}}$\xspace}
\newcommand{\spwavgdnm}{$4\pi \overline{R/N}_{\rm{H}}^{\rm{DNM}}$\xspace}
\newcommand{\spwavgco}{$4\pi \overline{R/N}_{\rm{H}}^{{{\hbox{\hglue 0.7pt}{\rm CO}}}}$\xspace}
\newcommand{\spwunit}{$10^{-31}$\,W\xspace}
\newcommand{\wcounit}{K\,km\,s$^{-1}$\xspace}
\newcommand{\msq}{$^{-2} $\xspace}
\newcommand{\chisq}{$\chi^2 $\xspace}

\newcommand{\anaR}{$\gamma{+}R$\xspace}
\newcommand{\anaT}{$\gamma{+}\tau_{353}$\xspace}
\newcommand{\anaQ}{$\gamma{+}A_{V\rm{Q}}$\xspace}
\newcommand{\avavgco}{X}
\newcommand{\percc}{cm$^{-3}$\xspace}
\newcommand{\persqcm}{cm$^{-2}$\xspace}
\newcommand{\kmpers}{km\,s$^{-1}$\xspace}
\def\Fermi{\textit{Fermi}\xspace}
\def\Planck{\textit{Planck}\xspace}
\def\IRAS{\textit{IRAS}\xspace}
\def\WMAP{\textit{WMAP}\xspace}
\def\WISE{\textit{WISE}\xspace}
\def\FUSE{\textit{FUSE}\xspace}
\def\Herschel{\textit{Herschel}\xspace}

\begin{document} 

\def\setsymbol#1#2{\expandafter\def\csname #1\endcsname{#2}}
\def\getsymbol#1{\csname #1\endcsname}


\def\HeJT{$^4$He-JT}

\def\allearlypapers{\nocite{planck2011-1.1, planck2011-1.3, planck2011-1.4, planck2011-1.5, planck2011-1.6, planck2011-1.7, planck2011-1.10, planck2011-1.10sup, planck2011-5.1a, planck2011-5.1b, planck2011-5.2a, planck2011-5.2b, planck2011-5.2c, planck2011-6.1, planck2011-6.2, planck2011-6.3a, planck2011-6.4a, planck2011-6.4b, planck2011-6.6, planck2011-7.0, planck2011-7.2, planck2011-7.3, planck2011-7.7a, planck2011-7.7b, planck2011-7.12, planck2011-7.13}}

\def\all2013resultspapers{\nocite{planck2013-p01, planck2013-p02, planck2013-p02a, planck2013-p02d, planck2013-p02b, planck2013-p03, planck2013-p03c, planck2013-p03f, planck2013-p03d, planck2013-p03e, planck2013-p01a, planck2013-p06, planck2013-p03a, planck2013-pip88, planck2013-p08, planck2013-p11, planck2013-p12, planck2013-p13, planck2013-p14, planck2013-p15, planck2013-p05b, planck2013-p17, planck2013-p09, planck2013-p09a, planck2013-p20, planck2013-p19, planck2013-pipaberration, planck2013-p05, planck2013-p05a, planck2013-pip56, planck2013-p06b}}

\newbox\tablebox    \newdimen\tablewidth
\def\leaderfil{\leaders\hbox to 5pt{\hss.\hss}\hfil}
%
%
\def\endPlancktable{\tablewidth=\columnwidth 
    $$\hss\copy\tablebox\hss$$
    \vskip-\lastskip\vskip -2pt}
\def\endPlancktablewide{\tablewidth=\textwidth 
    $$\hss\copy\tablebox\hss$$
    \vskip-\lastskip\vskip -2pt}
\def\tablenote#1 #2\par{\begingroup \parindent=0.8em
    \abovedisplayshortskip=0pt\belowdisplayshortskip=0pt
    \noindent
    $$\hss\vbox{\hsize\tablewidth \hangindent=\parindent \hangafter=1 \noindent
    \hbox to \parindent{$^#1$\hss}\strut#2\strut\par}\hss$$
    \endgroup}
\def\doubleline{\vskip 3pt\hrule \vskip 1.5pt \hrule \vskip 5pt}

%
\def\L2{\ifmmode L_2\else $L_2$\fi}
\def\dtt{\Delta T/T}
\def\DeltaT{\ifmmode \Delta T\else $\Delta T$\fi}
\def\deltat{\ifmmode \Delta t\else $\Delta t$\fi}
\def\fknee{\ifmmode f_{\rm knee}\else $f_{\rm knee}$\fi}
\def\Fmax{\ifmmode F_{\rm max}\else $F_{\rm max}$\fi}
\def\solar{\ifmmode{\rm M}_{\mathord\odot}\else${\rm M}_{\mathord\odot}$\fi}
\def\Msolar{\ifmmode{\rm M}_{\mathord\odot}\else${\rm M}_{\mathord\odot}$\fi}
\def\Lsolar{\ifmmode{\rm L}_{\mathord\odot}\else${\rm L}_{\mathord\odot}$\fi}
\def\mag{\sup{m}}
\def\inv{\ifmmode^{-1}\else$^{-1}$\fi}
\def\mo{\ifmmode^{-1}\else$^{-1}$\fi}
\def\sup#1{\ifmmode ^{\rm #1}\else $^{\rm #1}$\fi}
\def\expo#1{\ifmmode \times 10^{#1}\else $\times 10^{#1}$\fi}
\def\,{\thinspace}
\def\lsim{\mathrel{\raise .4ex\hbox{\rlap{$<$}\lower 1.2ex\hbox{$\sim$}}}}
\def\gsim{\mathrel{\raise .4ex\hbox{\rlap{$>$}\lower 1.2ex\hbox{$\sim$}}}}
\let\lea=\lsim
\let\gea=\gsim
\def\simprop{\mathrel{\raise .4ex\hbox{\rlap{$\propto$}\lower 1.2ex\hbox{$\sim$}}}}
\def\deg{\ifmmode^\circ\else$^\circ$\fi}
\def\pdeg{\ifmmode $\setbox0=\hbox{$^{\circ}$}\rlap{\hskip.11\wd0 .}$^{\circ}
          \else \setbox0=\hbox{$^{\circ}$}\rlap{\hskip.11\wd0 .}$^{\circ}$\fi}
\def\arcs{\ifmmode {^{\scriptstyle\prime\prime}}
          \else $^{\scriptstyle\prime\prime}$\fi}
\def\arcm{\ifmmode {^{\scriptstyle\prime}}
          \else $^{\scriptstyle\prime}$\fi}
\newdimen\sa  \newdimen\sb
\def\parcs{\sa=.07em \sb=.03em
     \ifmmode \hbox{\rlap{.}}^{\scriptstyle\prime\kern -\sb\prime}\hbox{\kern -\sa}
     \else \rlap{.}$^{\scriptstyle\prime\kern -\sb\prime}$\kern -\sa\fi}
\def\parcm{\sa=.08em \sb=.03em
     \ifmmode \hbox{\rlap{.}\kern\sa}^{\scriptstyle\prime}\hbox{\kern-\sb}
     \else \rlap{.}\kern\sa$^{\scriptstyle\prime}$\kern-\sb\fi}
\def\ra[#1 #2 #3.#4]{#1\sup{h}#2\sup{m}#3\sup{s}\llap.#4}
\def\dec[#1 #2 #3.#4]{#1\deg#2\arcm#3\arcs\llap.#4}
\def\deco[#1 #2 #3]{#1\deg#2\arcm#3\arcs}
\def\rra[#1 #2]{#1\sup{h}#2\sup{m}}
\def\page{\vfill\eject}
\def\dots{\relax\ifmmode \ldots\else $\ldots$\fi}
%
%
\def\WHzsr{\ifmmode $W\,Hz\mo\,sr\mo$\else W\,Hz\mo\,sr\mo\fi}
\def\mHz{\ifmmode $\,mHz$\else \,mHz\fi}
\def\GHz{\ifmmode $\,GHz$\else \,GHz\fi}
\def\mKs{\ifmmode $\,mK\,s$^{1/2}\else \,mK\,s$^{1/2}$\fi}
\def\muKs{\ifmmode \,\mu$K\,s$^{1/2}\else \,$\mu$K\,s$^{1/2}$\fi}
\def\muKRJs{\ifmmode \,\mu$K$_{\rm RJ}$\,s$^{1/2}\else \,$\mu$K$_{\rm RJ}$\,s$^{1/2}$\fi}
\def\muKHz{\ifmmode \,\mu$K\,Hz$^{-1/2}\else \,$\mu$K\,Hz$^{-1/2}$\fi}
\def\MJysr{\ifmmode \,$MJy\,sr\mo$\else \,MJy\,sr\mo\fi}
\def\MJysrmK{\ifmmode \,$MJy\,sr\mo$\,mK$_{\rm CMB}\mo\else \,MJy\,sr\mo\,mK$_{\rm CMB}\mo$\fi}
\def\microns{\ifmmode \,\mu$m$\else \,$\mu$m\fi}
\def\micron{\microns}
\def\muK{\ifmmode \,\mu$K$\else \,$\mu$\hbox{K}\fi}
\def\microK{\ifmmode \,\mu$K$\else \,$\mu$\hbox{K}\fi}
\def\muW{\ifmmode \,\mu$W$\else \,$\mu$\hbox{W}\fi}
\def\kms{\ifmmode $\,km\,s$^{-1}\else \,km\,s$^{-1}$\fi}
\def\kmsMpc{\ifmmode $\,\kms\,Mpc\mo$\else \,\kms\,Mpc\mo\fi}
%
%


\setsymbol{LFI:center:frequency:70GHz:units}{70.3\,GHz}
\setsymbol{LFI:center:frequency:44GHz:units}{44.1\,GHz}
\setsymbol{LFI:center:frequency:30GHz:units}{28.5\,GHz}

\setsymbol{LFI:center:frequency:70GHz}{70.3}
\setsymbol{LFI:center:frequency:44GHz}{44.1}
\setsymbol{LFI:center:frequency:30GHz}{28.5}

\setsymbol{LFI:center:frequency:LFI18:Rad:M:units}{71.7\GHz}
\setsymbol{LFI:center:frequency:LFI19:Rad:M:units}{67.5\GHz}
\setsymbol{LFI:center:frequency:LFI20:Rad:M:units}{69.2\GHz}
\setsymbol{LFI:center:frequency:LFI21:Rad:M:units}{70.4\GHz}
\setsymbol{LFI:center:frequency:LFI22:Rad:M:units}{71.5\GHz}
\setsymbol{LFI:center:frequency:LFI23:Rad:M:units}{70.8\GHz}
\setsymbol{LFI:center:frequency:LFI24:Rad:M:units}{44.4\GHz}
\setsymbol{LFI:center:frequency:LFI25:Rad:M:units}{44.0\GHz}
\setsymbol{LFI:center:frequency:LFI26:Rad:M:units}{43.9\GHz}
\setsymbol{LFI:center:frequency:LFI27:Rad:M:units}{28.3\GHz}
\setsymbol{LFI:center:frequency:LFI28:Rad:M:units}{28.8\GHz}
\setsymbol{LFI:center:frequency:LFI18:Rad:S:units}{70.1\GHz}
\setsymbol{LFI:center:frequency:LFI19:Rad:S:units}{69.6\GHz}
\setsymbol{LFI:center:frequency:LFI20:Rad:S:units}{69.5\GHz}
\setsymbol{LFI:center:frequency:LFI21:Rad:S:units}{69.5\GHz}
\setsymbol{LFI:center:frequency:LFI22:Rad:S:units}{72.8\GHz}
\setsymbol{LFI:center:frequency:LFI23:Rad:S:units}{71.3\GHz}
\setsymbol{LFI:center:frequency:LFI24:Rad:S:units}{44.1\GHz}
\setsymbol{LFI:center:frequency:LFI25:Rad:S:units}{44.1\GHz}
\setsymbol{LFI:center:frequency:LFI26:Rad:S:units}{44.1\GHz}
\setsymbol{LFI:center:frequency:LFI27:Rad:S:units}{28.5\GHz}
\setsymbol{LFI:center:frequency:LFI28:Rad:S:units}{28.2\GHz}

\setsymbol{LFI:center:frequency:LFI18:Rad:M}{71.7}
\setsymbol{LFI:center:frequency:LFI19:Rad:M}{67.5}
\setsymbol{LFI:center:frequency:LFI20:Rad:M}{69.2}
\setsymbol{LFI:center:frequency:LFI21:Rad:M}{70.4}
\setsymbol{LFI:center:frequency:LFI22:Rad:M}{71.5}
\setsymbol{LFI:center:frequency:LFI23:Rad:M}{70.8}
\setsymbol{LFI:center:frequency:LFI24:Rad:M}{44.4}
\setsymbol{LFI:center:frequency:LFI25:Rad:M}{44.0}
\setsymbol{LFI:center:frequency:LFI26:Rad:M}{43.9}
\setsymbol{LFI:center:frequency:LFI27:Rad:M}{28.3}
\setsymbol{LFI:center:frequency:LFI28:Rad:M}{28.8}
\setsymbol{LFI:center:frequency:LFI18:Rad:S}{70.1}
\setsymbol{LFI:center:frequency:LFI19:Rad:S}{69.6}
\setsymbol{LFI:center:frequency:LFI20:Rad:S}{69.5}
\setsymbol{LFI:center:frequency:LFI21:Rad:S}{69.5}
\setsymbol{LFI:center:frequency:LFI22:Rad:S}{72.8}
\setsymbol{LFI:center:frequency:LFI23:Rad:S}{71.3}
\setsymbol{LFI:center:frequency:LFI24:Rad:S}{44.1}
\setsymbol{LFI:center:frequency:LFI25:Rad:S}{44.1}
\setsymbol{LFI:center:frequency:LFI26:Rad:S}{44.1}
\setsymbol{LFI:center:frequency:LFI27:Rad:S}{28.5}
\setsymbol{LFI:center:frequency:LFI28:Rad:S}{28.2}


\setsymbol{LFI:white:noise:sensitivity:70GHz:units}{134.7\muKs}
\setsymbol{LFI:white:noise:sensitivity:44GHz:units}{164.7\muKs}
\setsymbol{LFI:white:noise:sensitivity:30GHz:units}{143.4\muKs}

\setsymbol{LFI:white:noise:sensitivity:70GHz}{134.7}
\setsymbol{LFI:white:noise:sensitivity:44GHz}{164.7}
\setsymbol{LFI:white:noise:sensitivity:30GHz}{143.4}


\setsymbol{LFI:white:noise:sensitivity:LFI18:Rad:M:units}{512.0\muKs}
\setsymbol{LFI:white:noise:sensitivity:LFI19:Rad:M:units}{581.4\muKs}
\setsymbol{LFI:white:noise:sensitivity:LFI20:Rad:M:units}{590.8\muKs}
\setsymbol{LFI:white:noise:sensitivity:LFI21:Rad:M:units}{455.2\muKs}
\setsymbol{LFI:white:noise:sensitivity:LFI22:Rad:M:units}{492.0\muKs}
\setsymbol{LFI:white:noise:sensitivity:LFI23:Rad:M:units}{507.7\muKs}
\setsymbol{LFI:white:noise:sensitivity:LFI24:Rad:M:units}{462.2\muKs}
\setsymbol{LFI:white:noise:sensitivity:LFI25:Rad:M:units}{413.6\muKs}
\setsymbol{LFI:white:noise:sensitivity:LFI26:Rad:M:units}{478.6\muKs}
\setsymbol{LFI:white:noise:sensitivity:LFI27:Rad:M:units}{277.7\muKs}
\setsymbol{LFI:white:noise:sensitivity:LFI28:Rad:M:units}{312.3\muKs}
\setsymbol{LFI:white:noise:sensitivity:LFI18:Rad:S:units}{465.7\muKs}
\setsymbol{LFI:white:noise:sensitivity:LFI19:Rad:S:units}{555.6\muKs}
\setsymbol{LFI:white:noise:sensitivity:LFI20:Rad:S:units}{623.2\muKs}
\setsymbol{LFI:white:noise:sensitivity:LFI21:Rad:S:units}{564.1\muKs}
\setsymbol{LFI:white:noise:sensitivity:LFI22:Rad:S:units}{534.4\muKs}
\setsymbol{LFI:white:noise:sensitivity:LFI23:Rad:S:units}{542.4\muKs}
\setsymbol{LFI:white:noise:sensitivity:LFI24:Rad:S:units}{399.2\muKs}
\setsymbol{LFI:white:noise:sensitivity:LFI25:Rad:S:units}{392.6\muKs}
\setsymbol{LFI:white:noise:sensitivity:LFI26:Rad:S:units}{418.6\muKs}
\setsymbol{LFI:white:noise:sensitivity:LFI27:Rad:S:units}{302.9\muKs}
\setsymbol{LFI:white:noise:sensitivity:LFI28:Rad:S:units}{285.3\muKs}

\setsymbol{LFI:white:noise:sensitivity:LFI18:Rad:M}{512.0}
\setsymbol{LFI:white:noise:sensitivity:LFI19:Rad:M}{581.4}
\setsymbol{LFI:white:noise:sensitivity:LFI20:Rad:M}{590.8}
\setsymbol{LFI:white:noise:sensitivity:LFI21:Rad:M}{455.2}
\setsymbol{LFI:white:noise:sensitivity:LFI22:Rad:M}{492.0}
\setsymbol{LFI:white:noise:sensitivity:LFI23:Rad:M}{507.7}
\setsymbol{LFI:white:noise:sensitivity:LFI24:Rad:M}{462.2}
\setsymbol{LFI:white:noise:sensitivity:LFI25:Rad:M}{413.6}
\setsymbol{LFI:white:noise:sensitivity:LFI26:Rad:M}{478.6}
\setsymbol{LFI:white:noise:sensitivity:LFI27:Rad:M}{277.7}
\setsymbol{LFI:white:noise:sensitivity:LFI28:Rad:M}{312.3}
\setsymbol{LFI:white:noise:sensitivity:LFI18:Rad:S}{465.7}
\setsymbol{LFI:white:noise:sensitivity:LFI19:Rad:S}{555.6}
\setsymbol{LFI:white:noise:sensitivity:LFI20:Rad:S}{623.2}
\setsymbol{LFI:white:noise:sensitivity:LFI21:Rad:S}{564.1}
\setsymbol{LFI:white:noise:sensitivity:LFI22:Rad:S}{534.4}
\setsymbol{LFI:white:noise:sensitivity:LFI23:Rad:S}{542.4}
\setsymbol{LFI:white:noise:sensitivity:LFI24:Rad:S}{399.2}
\setsymbol{LFI:white:noise:sensitivity:LFI25:Rad:S}{392.6}
\setsymbol{LFI:white:noise:sensitivity:LFI26:Rad:S}{418.6}
\setsymbol{LFI:white:noise:sensitivity:LFI27:Rad:S}{302.9}
\setsymbol{LFI:white:noise:sensitivity:LFI28:Rad:S}{285.3}


\setsymbol{LFI:knee:frequency:70GHz:units}{29.5\mHz}
\setsymbol{LFI:knee:frequency:44GHz:units}{56.2\mHz}
\setsymbol{LFI:knee:frequency:30GHz:units}{113.7\mHz}

\setsymbol{LFI:knee:frequency:70GHz}{29.5}
\setsymbol{LFI:knee:frequency:44GHz}{56.2}
\setsymbol{LFI:knee:frequency:30GHz}{113.7}

\setsymbol{LFI:knee:frequency:LFI18:Rad:M:units}{16.3\mHz}
\setsymbol{LFI:knee:frequency:LFI19:Rad:M:units}{15.1\mHz}
\setsymbol{LFI:knee:frequency:LFI20:Rad:M:units}{18.7\mHz}
\setsymbol{LFI:knee:frequency:LFI21:Rad:M:units}{37.2\mHz}
\setsymbol{LFI:knee:frequency:LFI22:Rad:M:units}{12.7\mHz}
\setsymbol{LFI:knee:frequency:LFI23:Rad:M:units}{34.6\mHz}
\setsymbol{LFI:knee:frequency:LFI24:Rad:M:units}{46.2\mHz}
\setsymbol{LFI:knee:frequency:LFI25:Rad:M:units}{24.9\mHz}
\setsymbol{LFI:knee:frequency:LFI26:Rad:M:units}{67.6\mHz}
\setsymbol{LFI:knee:frequency:LFI27:Rad:M:units}{187.4\mHz}
\setsymbol{LFI:knee:frequency:LFI28:Rad:M:units}{122.2\mHz}
\setsymbol{LFI:knee:frequency:LFI18:Rad:S:units}{17.7\mHz}
\setsymbol{LFI:knee:frequency:LFI19:Rad:S:units}{22.0\mHz}
\setsymbol{LFI:knee:frequency:LFI20:Rad:S:units}{8.7\mHz}
\setsymbol{LFI:knee:frequency:LFI21:Rad:S:units}{25.9\mHz}
\setsymbol{LFI:knee:frequency:LFI22:Rad:S:units}{15.8\mHz}
\setsymbol{LFI:knee:frequency:LFI23:Rad:S:units}{129.8\mHz}
\setsymbol{LFI:knee:frequency:LFI24:Rad:S:units}{100.9\mHz}
\setsymbol{LFI:knee:frequency:LFI25:Rad:S:units}{38.9\mHz}
\setsymbol{LFI:knee:frequency:LFI26:Rad:S:units}{58.9\mHz}
\setsymbol{LFI:knee:frequency:LFI27:Rad:S:units}{104.4\mHz}
\setsymbol{LFI:knee:frequency:LFI28:Rad:S:units}{40.7\mHz}

\setsymbol{LFI:knee:frequency:LFI18:Rad:M}{16.3}
\setsymbol{LFI:knee:frequency:LFI19:Rad:M}{15.1}
\setsymbol{LFI:knee:frequency:LFI20:Rad:M}{18.7}
\setsymbol{LFI:knee:frequency:LFI21:Rad:M}{37.2}
\setsymbol{LFI:knee:frequency:LFI22:Rad:M}{12.7}
\setsymbol{LFI:knee:frequency:LFI23:Rad:M}{34.6}
\setsymbol{LFI:knee:frequency:LFI24:Rad:M}{46.2}
\setsymbol{LFI:knee:frequency:LFI25:Rad:M}{24.9}
\setsymbol{LFI:knee:frequency:LFI26:Rad:M}{67.6}
\setsymbol{LFI:knee:frequency:LFI27:Rad:M}{187.4}
\setsymbol{LFI:knee:frequency:LFI28:Rad:M}{122.2}
\setsymbol{LFI:knee:frequency:LFI18:Rad:S}{17.7}
\setsymbol{LFI:knee:frequency:LFI19:Rad:S}{22.0}
\setsymbol{LFI:knee:frequency:LFI20:Rad:S}{8.7}
\setsymbol{LFI:knee:frequency:LFI21:Rad:S}{25.9}
\setsymbol{LFI:knee:frequency:LFI22:Rad:S}{15.8}
\setsymbol{LFI:knee:frequency:LFI23:Rad:S}{129.8}
\setsymbol{LFI:knee:frequency:LFI24:Rad:S}{100.9}
\setsymbol{LFI:knee:frequency:LFI25:Rad:S}{38.9}
\setsymbol{LFI:knee:frequency:LFI26:Rad:S}{58.9}
\setsymbol{LFI:knee:frequency:LFI27:Rad:S}{104.4}
\setsymbol{LFI:knee:frequency:LFI28:Rad:S}{40.7}


\setsymbol{LFI:slope:70GHz:units}{$-1.03$\mHz}
\setsymbol{LFI:slope:44GHz:units}{$-0.89$\mHz}
\setsymbol{LFI:slope:30GHz:units}{$-0.87$\mHz}

\setsymbol{LFI:slope:70GHz}{$-1.03$}
\setsymbol{LFI:slope:44GHz}{$-0.89$}
\setsymbol{LFI:slope:30GHz}{$-0.87$}

\setsymbol{LFI:slope:LFI18:Rad:M:units}{$-1.04$\mHz}
\setsymbol{LFI:slope:LFI19:Rad:M:units}{$-1.09$\mHz}
\setsymbol{LFI:slope:LFI20:Rad:M:units}{$-0.69$\mHz}
\setsymbol{LFI:slope:LFI21:Rad:M:units}{$-1.56$\mHz}
\setsymbol{LFI:slope:LFI22:Rad:M:units}{$-1.01$\mHz}
\setsymbol{LFI:slope:LFI23:Rad:M:units}{$-0.96$\mHz}
\setsymbol{LFI:slope:LFI24:Rad:M:units}{$-0.83$\mHz}
\setsymbol{LFI:slope:LFI25:Rad:M:units}{$-0.91$\mHz}
\setsymbol{LFI:slope:LFI26:Rad:M:units}{$-0.95$\mHz}
\setsymbol{LFI:slope:LFI27:Rad:M:units}{$-0.87$\mHz}
\setsymbol{LFI:slope:LFI28:Rad:M:units}{$-0.88$\mHz}
\setsymbol{LFI:slope:LFI18:Rad:S:units}{$-1.15$\mHz}
\setsymbol{LFI:slope:LFI19:Rad:S:units}{$-1.00$\mHz}
\setsymbol{LFI:slope:LFI20:Rad:S:units}{$-0.95$\mHz}
\setsymbol{LFI:slope:LFI21:Rad:S:units}{$-0.92$\mHz}
\setsymbol{LFI:slope:LFI22:Rad:S:units}{$-1.01$\mHz}
\setsymbol{LFI:slope:LFI23:Rad:S:units}{$-0.95$\mHz}
\setsymbol{LFI:slope:LFI24:Rad:S:units}{$-0.73$\mHz}
\setsymbol{LFI:slope:LFI25:Rad:S:units}{$-1.16$\mHz}
\setsymbol{LFI:slope:LFI26:Rad:S:units}{$-0.79$\mHz}
\setsymbol{LFI:slope:LFI27:Rad:S:units}{$-0.82$\mHz}
\setsymbol{LFI:slope:LFI28:Rad:S:units}{$-0.91$\mHz}

\setsymbol{LFI:slope:LFI18:Rad:M}{$-1.04$}
\setsymbol{LFI:slope:LFI19:Rad:M}{$-1.09$}
\setsymbol{LFI:slope:LFI20:Rad:M}{$-0.69$}
\setsymbol{LFI:slope:LFI21:Rad:M}{$-1.56$}
\setsymbol{LFI:slope:LFI22:Rad:M}{$-1.01$}
\setsymbol{LFI:slope:LFI23:Rad:M}{$-0.96$}
\setsymbol{LFI:slope:LFI24:Rad:M}{$-0.83$}
\setsymbol{LFI:slope:LFI25:Rad:M}{$-0.91$}
\setsymbol{LFI:slope:LFI26:Rad:M}{$-0.95$}
\setsymbol{LFI:slope:LFI27:Rad:M}{$-0.87$}
\setsymbol{LFI:slope:LFI28:Rad:M}{$-0.88$}
\setsymbol{LFI:slope:LFI18:Rad:S}{$-1.15$}
\setsymbol{LFI:slope:LFI19:Rad:S}{$-1.00$}
\setsymbol{LFI:slope:LFI20:Rad:S}{$-0.95$}
\setsymbol{LFI:slope:LFI21:Rad:S}{$-0.92$}
\setsymbol{LFI:slope:LFI22:Rad:S}{$-1.01$}
\setsymbol{LFI:slope:LFI23:Rad:S}{$-0.95$}
\setsymbol{LFI:slope:LFI24:Rad:S}{$-0.73$}
\setsymbol{LFI:slope:LFI25:Rad:S}{$-1.16$}
\setsymbol{LFI:slope:LFI26:Rad:S}{$-0.79$}
\setsymbol{LFI:slope:LFI27:Rad:S}{$-0.82$}
\setsymbol{LFI:slope:LFI28:Rad:S}{$-0.91$}


\setsymbol{LFI:FWHM:70GHz:units}{13\parcm01}
\setsymbol{LFI:FWHM:44GHz:units}{27\parcm92}
\setsymbol{LFI:FWHM:30GHz:units}{32\parcm65}

\setsymbol{LFI:FWHM:70GHz}{13.01}
\setsymbol{LFI:FWHM:44GHz}{27.92}
\setsymbol{LFI:FWHM:30GHz}{32.65}

\setsymbol{LFI:FWHM:LFI18:units}{13\parcm39}
\setsymbol{LFI:FWHM:LFI19:units}{13\parcm01}
\setsymbol{LFI:FWHM:LFI20:units}{12\parcm75}
\setsymbol{LFI:FWHM:LFI21:units}{12\parcm74}
\setsymbol{LFI:FWHM:LFI22:units}{12\parcm87}
\setsymbol{LFI:FWHM:LFI23:units}{13\parcm27}
\setsymbol{LFI:FWHM:LFI24:units}{22\parcm98}
\setsymbol{LFI:FWHM:LFI25:units}{30\parcm46}
\setsymbol{LFI:FWHM:LFI26:units}{30\parcm31}
\setsymbol{LFI:FWHM:LFI27:units}{32\parcm65}
\setsymbol{LFI:FWHM:LFI28:units}{32\parcm66}

\setsymbol{LFI:FWHM:LFI18}{13.39}
\setsymbol{LFI:FWHM:LFI19}{13.01}
\setsymbol{LFI:FWHM:LFI20}{12.75}
\setsymbol{LFI:FWHM:LFI21}{12.74}
\setsymbol{LFI:FWHM:LFI22}{12.87}
\setsymbol{LFI:FWHM:LFI23}{13.27}
\setsymbol{LFI:FWHM:LFI24}{22.98}
\setsymbol{LFI:FWHM:LFI25}{30.46}
\setsymbol{LFI:FWHM:LFI26}{30.31}
\setsymbol{LFI:FWHM:LFI27}{32.65}
\setsymbol{LFI:FWHM:LFI28}{32.66}



\setsymbol{LFI:FWHM:uncertainty:LFI18:units}{0.170\arcm}
\setsymbol{LFI:FWHM:uncertainty:LFI19:units}{0.174\arcm}
\setsymbol{LFI:FWHM:uncertainty:LFI20:units}{0.170\arcm}
\setsymbol{LFI:FWHM:uncertainty:LFI21:units}{0.156\arcm}
\setsymbol{LFI:FWHM:uncertainty:LFI22:units}{0.164\arcm}
\setsymbol{LFI:FWHM:uncertainty:LFI23:units}{0.171\arcm}
\setsymbol{LFI:FWHM:uncertainty:LFI24:units}{0.652\arcm}
\setsymbol{LFI:FWHM:uncertainty:LFI25:units}{1.075\arcm}
\setsymbol{LFI:FWHM:uncertainty:LFI26:units}{1.131\arcm}
\setsymbol{LFI:FWHM:uncertainty:LFI27:units}{1.266\arcm}
\setsymbol{LFI:FWHM:uncertainty:LFI28:units}{1.287\arcm}

\setsymbol{LFI:FWHM:uncertainty:LFI18}{0.170}
\setsymbol{LFI:FWHM:uncertainty:LFI19}{0.174}
\setsymbol{LFI:FWHM:uncertainty:LFI20}{0.170}
\setsymbol{LFI:FWHM:uncertainty:LFI21}{0.156}
\setsymbol{LFI:FWHM:uncertainty:LFI22}{0.164}
\setsymbol{LFI:FWHM:uncertainty:LFI23}{0.171}
\setsymbol{LFI:FWHM:uncertainty:LFI24}{0.652}
\setsymbol{LFI:FWHM:uncertainty:LFI25}{1.075}
\setsymbol{LFI:FWHM:uncertainty:LFI26}{1.131}
\setsymbol{LFI:FWHM:uncertainty:LFI27}{1.266}
\setsymbol{LFI:FWHM:uncertainty:LFI28}{1.287}


\setsymbol{HFI:center:frequency:100GHz:units}{100\,GHz}
\setsymbol{HFI:center:frequency:143GHz:units}{143\,GHz}
\setsymbol{HFI:center:frequency:217GHz:units}{217\,GHz}
\setsymbol{HFI:center:frequency:353GHz:units}{353\,GHz}
\setsymbol{HFI:center:frequency:545GHz:units}{545\,GHz}
\setsymbol{HFI:center:frequency:857GHz:units}{857\,GHz}

\setsymbol{HFI:center:frequency:100GHz}{100}
\setsymbol{HFI:center:frequency:143GHz}{143}
\setsymbol{HFI:center:frequency:217GHz}{217}
\setsymbol{HFI:center:frequency:353GHz}{353}
\setsymbol{HFI:center:frequency:545GHz}{545}
\setsymbol{HFI:center:frequency:857GHz}{857}


\setsymbol{HFI:Ndetectors:100GHz}{8}
\setsymbol{HFI:Ndetectors:143GHz}{11}
\setsymbol{HFI:Ndetectors:217GHz}{12}
\setsymbol{HFI:Ndetectors:353GHz}{12}
\setsymbol{HFI:Ndetectors:545GHz}{3}
\setsymbol{HFI:Ndetectors:857GHz}{4}


\setsymbol{HFI:FWHM:Maps:100GHz:units}{9\parcm88}
\setsymbol{HFI:FWHM:Maps:143GHz:units}{7\parcm18}
\setsymbol{HFI:FWHM:Maps:217GHz:units}{4\parcm87}
\setsymbol{HFI:FWHM:Maps:353GHz:units}{4\parcm65}
\setsymbol{HFI:FWHM:Maps:545GHz:units}{4\parcm72}
\setsymbol{HFI:FWHM:Maps:857GHz:units}{4\parcm39}
\setsymbol{HFI:FWHM:Maps:100GHz}{9.88}
\setsymbol{HFI:FWHM:Maps:143GHz}{7.18}
\setsymbol{HFI:FWHM:Maps:217GHz}{4.87}
\setsymbol{HFI:FWHM:Maps:353GHz}{4.65}
\setsymbol{HFI:FWHM:Maps:545GHz}{4.72}
\setsymbol{HFI:FWHM:Maps:857GHz}{4.39}


\setsymbol{HFI:beam:ellipticity:Maps:100GHz}{1.15}
\setsymbol{HFI:beam:ellipticity:Maps:143GHz}{1.01}
\setsymbol{HFI:beam:ellipticity:Maps:217GHz}{1.06}
\setsymbol{HFI:beam:ellipticity:Maps:353GHz}{1.05}
\setsymbol{HFI:beam:ellipticity:Maps:545GHz}{1.14}
\setsymbol{HFI:beam:ellipticity:Maps:857GHz}{1.19}


\setsymbol{HFI:FWHM:Mars:100GHz:units}{9\parcm37}
\setsymbol{HFI:FWHM:Mars:143GHz:units}{7\parcm04}
\setsymbol{HFI:FWHM:Mars:217GHz:units}{4\parcm68}
\setsymbol{HFI:FWHM:Mars:353GHz:units}{4\parcm43}
\setsymbol{HFI:FWHM:Mars:545GHz:units}{3\parcm80}
\setsymbol{HFI:FWHM:Mars:857GHz:units}{3\parcm67}

\setsymbol{HFI:FWHM:Mars:100GHz}{9.37}
\setsymbol{HFI:FWHM:Mars:143GHz}{7.04}
\setsymbol{HFI:FWHM:Mars:217GHz}{4.68}
\setsymbol{HFI:FWHM:Mars:353GHz}{4.43}
\setsymbol{HFI:FWHM:Mars:545GHz}{3.80}
\setsymbol{HFI:FWHM:Mars:857GHz}{3.67}


\setsymbol{HFI:beam:ellipticity:Mars:100GHz}{1.18}
\setsymbol{HFI:beam:ellipticity:Mars:143GHz}{1.03}
\setsymbol{HFI:beam:ellipticity:Mars:217GHz}{1.14}
\setsymbol{HFI:beam:ellipticity:Mars:353GHz}{1.09}
\setsymbol{HFI:beam:ellipticity:Mars:545GHz}{1.25}
\setsymbol{HFI:beam:ellipticity:Mars:857GHz}{1.03}


\setsymbol{HFI:CMB:relative:calibration:100GHz}{$\lsim 1\%$}
\setsymbol{HFI:CMB:relative:calibration:143GHz}{$\lsim 1\%$}
\setsymbol{HFI:CMB:relative:calibration:217GHz}{$\lsim 1\%$}
\setsymbol{HFI:CMB:relative:calibration:353GHz}{$\lsim 1\%$}
\setsymbol{HFI:CMB:relative:calibration:545GHz}{}
\setsymbol{HFI:CMB:relative:calibration:857GHz}{}


\setsymbol{HFI:CMB:absolute:calibration:100GHz}{$\lsim 2\%$}
\setsymbol{HFI:CMB:absolute:calibration:143GHz}{$\lsim 2\%$}
\setsymbol{HFI:CMB:absolute:calibration:217GHz}{$\lsim 2\%$}
\setsymbol{HFI:CMB:absolute:calibration:353GHz}{$\lsim 2\%$}
\setsymbol{HFI:CMB:absolute:calibration:545GHz}{}
\setsymbol{HFI:CMB:absolute:calibration:857GHz}{}


\setsymbol{HFI:FIRAS:gain:calibration:accuracy:statistical:100GHz}{}
\setsymbol{HFI:FIRAS:gain:calibration:accuracy:statistical:143GHz}{}
\setsymbol{HFI:FIRAS:gain:calibration:accuracy:statistical:217GHz}{}
\setsymbol{HFI:FIRAS:gain:calibration:accuracy:statistical:353GHz}{2.5\%}
\setsymbol{HFI:FIRAS:gain:calibration:accuracy:statistical:545GHz}{1\%}
\setsymbol{HFI:FIRAS:gain:calibration:accuracy:statistical:857GHz}{0.5\%}


\setsymbol{HFI:FIRAS:gain:calibration:accuracy:systematic:100GHz}{}
\setsymbol{HFI:FIRAS:gain:calibration:accuracy:systematic:143GHz}{}
\setsymbol{HFI:FIRAS:gain:calibration:accuracy:systematic:217GHz}{}
\setsymbol{HFI:FIRAS:gain:calibration:accuracy:systematic:353GHz}{}
\setsymbol{HFI:FIRAS:gain:calibration:accuracy:systematic:545GHz}{7\%}
\setsymbol{HFI:FIRAS:gain:calibration:accuracy:systematic:857GHz}{7\%}


\setsymbol{HFI:FIRAS:zero:point:accuracy:100GHz:units}{0.8\MJysr}
\setsymbol{HFI:FIRAS:zero:point:accuracy:143GHz:units}{}
\setsymbol{HFI:FIRAS:zero:point:accuracy:217GHz:units}{}
\setsymbol{HFI:FIRAS:zero:point:accuracy:353GHz:units}{1.4\MJysr}
\setsymbol{HFI:FIRAS:zero:point:accuracy:545GHz:units}{2.2\MJysr}
\setsymbol{HFI:FIRAS:zero:point:accuracy:857GHz:units}{1.7\MJysr}

\setsymbol{HFI:FIRAS:zero:point:accuracy:100GHz}{0.8}
\setsymbol{HFI:FIRAS:zero:point:accuracy:143GHz}{}
\setsymbol{HFI:FIRAS:zero:point:accuracy:217GHz}{}
\setsymbol{HFI:FIRAS:zero:point:accuracy:353GHz}{1.4}
\setsymbol{HFI:FIRAS:zero:point:accuracy:545GHz}{2.2}
\setsymbol{HFI:FIRAS:zero:point:accuracy:857GHz}{1.7}


\setsymbol{HFI:unit:conversion:100GHz:units}{0.2415\MJysrmK}
\setsymbol{HFI:unit:conversion:143GHz:units}{0.3694\MJysrmK}
\setsymbol{HFI:unit:conversion:217GHz:units}{0.4811\MJysrmK}
\setsymbol{HFI:unit:conversion:353GHz:units}{0.2883\MJysrmK}
\setsymbol{HFI:unit:conversion:545GHz:units}{0.05826\MJysrmK}
\setsymbol{HFI:unit:conversion:857GHz:units}{0.002238\MJysrmK}

\setsymbol{HFI:unit:conversion:100GHz}{0.2415}
\setsymbol{HFI:unit:conversion:143GHz}{0.3694}
\setsymbol{HFI:unit:conversion:217GHz}{0.4811}
\setsymbol{HFI:unit:conversion:353GHz}{0.2883}
\setsymbol{HFI:unit:conversion:545GHz}{0.05826}
\setsymbol{HFI:unit:conversion:857GHz}{0.002238}


\setsymbol{HFI:colour:correction:alpha=-2:V1.01:100GHz}{0.9893}
\setsymbol{HFI:colour:correction:alpha=-2:V1.01:143GHz}{0.9759}
\setsymbol{HFI:colour:correction:alpha=-2:V1.01:217GHz}{1.0007}
\setsymbol{HFI:colour:correction:alpha=-2:V1.01:353GHz}{1.0028}
\setsymbol{HFI:colour:correction:alpha=-2:V1.01:545GHz}{1.0019}
\setsymbol{HFI:colour:correction:alpha=-2:V1.01:857GHz}{0.9889}


\setsymbol{HFI:colour:correction:alpha=0:V1.01:100GHz}{1.0008}
\setsymbol{HFI:colour:correction:alpha=0:V1.01:143GHz}{1.0148}
\setsymbol{HFI:colour:correction:alpha=0:V1.01:217GHz}{0.9909}
\setsymbol{HFI:colour:correction:alpha=0:V1.01:353GHz}{0.9888}
\setsymbol{HFI:colour:correction:alpha=0:V1.01:545GHz}{0.9878}
\setsymbol{HFI:colour:correction:alpha=0:V1.01:857GHz}{1.0014}

   \title{\textit{Planck} intermediate results. XXVIII. Interstellar gas and dust in the Chamaeleon clouds as seen by \textit{Fermi} LAT and \textit{Planck} }

\author{\small
Planck and Fermi Collaborations:
P.~A.~R.~Ade\inst{79}
\and
N.~Aghanim\inst{55}
\and
G.~Aniano\inst{55}
\and
M.~Arnaud\inst{67}
\and
M.~Ashdown\inst{64, 5}
\and
J.~Aumont\inst{55}
\and
C.~Baccigalupi\inst{78}
\and
A.~J.~Banday\inst{85, 9}
\and
R.~B.~Barreiro\inst{61}
\and
N.~Bartolo\inst{28}
% ...
}
\institute{\small
APC, AstroParticule et Cosmologie, Universit\'{e} Paris Diderot, CNRS/IN2P3, CEA/lrfu, Observatoire de Paris, Sorbonne Paris Cit\'{e}, 10, rue Alice Domon et L\'{e}onie Duquet, 75205 Paris Cedex 13, France\goodbreak
\and
African Institute for Mathematical Sciences, 6-8 Melrose Road, Muizenberg, Cape Town, South Africa\goodbreak
\and
Agenzia Spaziale Italiana Science Data Center, Via del Politecnico snc, 00133, Roma, Italy\goodbreak
\and
Agenzia Spaziale Italiana, Viale Liegi 26, Roma, Italy\goodbreak
\and
Astrophysics Group, Cavendish Laboratory, University of Cambridge, J J Thomson Avenue, Cambridge CB3 0HE, U.K.\goodbreak
\and
Astrophysics \& Cosmology Research Unit, School of Mathematics, Statistics \& Computer Science, University of KwaZulu-Natal, Westville Campus, Private Bag X54001, Durban 4000, South Africa\goodbreak
\and
Atacama Large Millimeter/submillimeter Array, ALMA Santiago Central Offices, Alonso de Cordova 3107, Vitacura, Casilla 763 0355, Santiago, Chile\goodbreak
\and
CITA, University of Toronto, 60 St. George St., Toronto, ON M5S 3H8, Canada\goodbreak
\and
CNRS, IRAP, 9 Av. colonel Roche, BP 44346, F-31028 Toulouse cedex 4, France\goodbreak
\and
California Institute of Technology, Pasadena, California, U.S.A.\goodbreak
\and
Centro de Estudios de F\'{i}sica del Cosmos de Arag\'{o}n (CEFCA), Plaza San Juan, 1, planta 2, E-44001, Teruel, Spain\goodbreak
\and
Computational Cosmology Center, Lawrence Berkeley National Laboratory, Berkeley, California, U.S.A.\goodbreak
\and
Consejo Superior de Investigaciones Cient\'{\i}ficas (CSIC), Madrid, Spain\goodbreak
\and
DSM/Irfu/SPP, CEA-Saclay, F-91191 Gif-sur-Yvette Cedex, France\goodbreak
\and
DTU Space, National Space Institute, Technical University of Denmark, Elektrovej 327, DK-2800 Kgs. Lyngby, Denmark\goodbreak
\and
D\'{e}partement de Physique Th\'{e}orique, Universit\'{e} de Gen\`{e}ve, 24, Quai E. Ansermet,1211 Gen\`{e}ve 4, Switzerland\goodbreak
\and
Departamento de F\'{\i}sica, Universidad de Oviedo, Avda. Calvo Sotelo s/n, Oviedo, Spain\goodbreak
\and
Department of Astrophysics/IMAPP, Radboud University Nijmegen, P.O. Box 9010, 6500 GL Nijmegen, The Netherlands\goodbreak
\and
Department of Physics \& Astronomy, University of British Columbia, 6224 Agricultural Road, Vancouver, British Columbia, Canada\goodbreak
\and
Department of Physics and Astronomy, Dana and David Dornsife College of Letter, Arts and Sciences, University of Southern California, Los Angeles, CA 90089, U.S.A.\goodbreak
\and
Department of Physics and Astronomy, University College London, London WC1E 6BT, U.K.\goodbreak
\and
Department of Physics, Florida State University, Keen Physics Building, 77 Chieftan Way, Tallahassee, Florida, U.S.A.\goodbreak
\and
Department of Physics, Gustaf H\"{a}llstr\"{o}min katu 2a, University of Helsinki, Helsinki, Finland\goodbreak
\and
Department of Physics, Nagoya University, Chikusa-ku, Nagoya, 464-8602, Japan\goodbreak
\and
Department of Physics, Princeton University, Princeton, New Jersey, U.S.A.\goodbreak
\and
Department of Physics, University of California, Santa Barbara, California, U.S.A.\goodbreak
\and
Department of Physics, University of Illinois at Urbana-Champaign, 1110 West Green Street, Urbana, Illinois, U.S.A.\goodbreak
\and
Dipartimento di Fisica e Astronomia G. Galilei, Universit\`{a} degli Studi di Padova, via Marzolo 8, 35131 Padova, Italy\goodbreak
\and
Dipartimento di Fisica e Scienze della Terra, Universit\`{a} di Ferrara, Via Saragat 1, 44122 Ferrara, Italy\goodbreak
\and
Dipartimento di Fisica, Universit\`{a} La Sapienza, P. le A. Moro 2, Roma, Italy\goodbreak
\and
Dipartimento di Fisica, Universit\`{a} degli Studi di Milano, Via Celoria, 16, Milano, Italy\goodbreak
\and
Dipartimento di Fisica, Universit\`{a} degli Studi di Trieste, via A. Valerio 2, Trieste, Italy\goodbreak
\and
Dipartimento di Fisica, Universit\`{a} di Roma Tor Vergata, Via della Ricerca Scientifica, 1, Roma, Italy\goodbreak
\and
Discovery Center, Niels Bohr Institute, Blegdamsvej 17, Copenhagen, Denmark\goodbreak
\and
Dpto. Astrof\'{i}sica, Universidad de La Laguna (ULL), E-38206 La Laguna, Tenerife, Spain\goodbreak
\and
European Southern Observatory, ESO Vitacura, Alonso de Cordova 3107, Vitacura, Casilla 19001, Santiago, Chile\goodbreak
\and
European Space Agency, ESAC, Planck Science Office, Camino bajo del Castillo, s/n, Urbanizaci\'{o}n Villafranca del Castillo, Villanueva de la Ca\~{n}ada, Madrid, Spain\goodbreak
\and
European Space Agency, ESTEC, Keplerlaan 1, 2201 AZ Noordwijk, The Netherlands\goodbreak
\and
Facolt\`{a} di Ingegneria, Universit\`{a} degli Studi e-Campus, Via Isimbardi 10, Novedrate (CO), 22060, Italy\goodbreak
\and
HGSFP and University of Heidelberg, Theoretical Physics Department, Philosophenweg 16, 69120, Heidelberg, Germany\goodbreak
\and
Helsinki Institute of Physics, Gustaf H\"{a}llstr\"{o}min katu 2, University of Helsinki, Helsinki, Finland\goodbreak
\and
INAF - Osservatorio Astrofisico di Catania, Via S. Sofia 78, Catania, Italy\goodbreak
\and
INAF - Osservatorio Astronomico di Padova, Vicolo dell'Osservatorio 5, Padova, Italy\goodbreak
\and
INAF - Osservatorio Astronomico di Roma, via di Frascati 33, Monte Porzio Catone, Italy\goodbreak
\and
INAF - Osservatorio Astronomico di Trieste, Via G.B. Tiepolo 11, Trieste, Italy\goodbreak
\and
INAF/IASF Bologna, Via Gobetti 101, Bologna, Italy\goodbreak
\and
INAF/IASF Milano, Via E. Bassini 15, Milano, Italy\goodbreak
\and
INFN, Sezione di Bologna, Via Irnerio 46, I-40126, Bologna, Italy\goodbreak
\and
INFN, Sezione di Roma 1, Universit\`{a} di Roma Sapienza, Piazzale Aldo Moro 2, 00185, Roma, Italy\goodbreak
\and
INFN/National Institute for Nuclear Physics, Via Valerio 2, I-34127 Trieste, Italy\goodbreak
\and
IPAG: Institut de Plan\'{e}tologie et d'Astrophysique de Grenoble, Universit\'{e} Grenoble Alpes, IPAG, F-38000 Grenoble, France, CNRS, IPAG, F-38000 Grenoble, France\goodbreak
\and
Imperial College London, Astrophysics group, Blackett Laboratory, Prince Consort Road, London, SW7 2AZ, U.K.\goodbreak
\and
Infrared Processing and Analysis Center, California Institute of Technology, Pasadena, CA 91125, U.S.A.\goodbreak
\and
Institut Universitaire de France, 103, bd Saint-Michel, 75005, Paris, France\goodbreak
\and
Institut d'Astrophysique Spatiale, CNRS (UMR8617) Universit\'{e} Paris-Sud 11, B\^{a}timent 121, Orsay, France\goodbreak
\and
Institut d'Astrophysique de Paris, CNRS (UMR7095), 98 bis Boulevard Arago, F-75014, Paris, France\goodbreak
\and
Institute for Space Sciences, Bucharest-Magurale, Romania\goodbreak
\and
Institute of Astronomy, University of Cambridge, Madingley Road, Cambridge CB3 0HA, U.K.\goodbreak
\and
Institute of Theoretical Astrophysics, University of Oslo, Blindern, Oslo, Norway\goodbreak
\and
Instituto de Astrof\'{\i}sica de Canarias, C/V\'{\i}a L\'{a}ctea s/n, La Laguna, Tenerife, Spain\goodbreak
\and
Instituto de F\'{\i}sica de Cantabria (CSIC-Universidad de Cantabria), Avda. de los Castros s/n, Santander, Spain\goodbreak
\and
Jet Propulsion Laboratory, California Institute of Technology, 4800 Oak Grove Drive, Pasadena, California, U.S.A.\goodbreak
\and
Jodrell Bank Centre for Astrophysics, Alan Turing Building, School of Physics and Astronomy, The University of Manchester, Oxford Road, Manchester, M13 9PL, U.K.\goodbreak
\and
Kavli Institute for Cosmology Cambridge, Madingley Road, Cambridge, CB3 0HA, U.K.\goodbreak
\and
LAL, Universit\'{e} Paris-Sud, CNRS/IN2P3, Orsay, France\goodbreak
\and
LERMA, CNRS, Observatoire de Paris, 61 Avenue de l'Observatoire, Paris, France\goodbreak
\and
Laboratoire AIM, IRFU/Service d'Astrophysique - CEA/DSM - CNRS - Universit\'{e} Paris Diderot, B\^{a}t. 709, CEA-Saclay, F-91191 Gif-sur-Yvette Cedex, France\goodbreak
\and
Laboratoire Traitement et Communication de l'Information, CNRS (UMR 5141) and T\'{e}l\'{e}com ParisTech, 46 rue Barrault F-75634 Paris Cedex 13, France\goodbreak
\and
Laboratoire de Physique Subatomique et de Cosmologie, Universit\'{e} Joseph Fourier Grenoble I, CNRS/IN2P3, Institut National Polytechnique de Grenoble, 53 rue des Martyrs, 38026 Grenoble cedex, France\goodbreak
\and
Laboratoire de Physique Th\'{e}orique, Universit\'{e} Paris-Sud 11 \& CNRS, B\^{a}timent 210, 91405 Orsay, France\goodbreak
\and
Lawrence Berkeley National Laboratory, Berkeley, California, U.S.A.\goodbreak
\and
Max-Planck-Institut f\"{u}r Astrophysik, Karl-Schwarzschild-Str. 1, 85741 Garching, Germany\goodbreak
\and
Max-Planck-Institut f\"{u}r Extraterrestrische Physik, Giessenbachstra{\ss}e, 85748 Garching, Germany\goodbreak
\and
McGill Physics, Ernest Rutherford Physics Building, McGill University, 3600 rue University, Montr\'{e}al, QC, H3A 2T8, Canada\goodbreak
\and
National University of Ireland, Department of Experimental Physics, Maynooth, Co. Kildare, Ireland\goodbreak
\and
Niels Bohr Institute, Blegdamsvej 17, Copenhagen, Denmark\goodbreak
\and
Observational Cosmology, Mail Stop 367-17, California Institute of Technology, Pasadena, CA, 91125, U.S.A.\goodbreak
\and
SISSA, Astrophysics Sector, via Bonomea 265, 34136, Trieste, Italy\goodbreak
\and
School of Physics and Astronomy, Cardiff University, Queens Buildings, The Parade, Cardiff, CF24 3AA, U.K.\goodbreak
\and
Space Research Institute (IKI), Russian Academy of Sciences, Profsoyuznaya Str, 84/32, Moscow, 117997, Russia\goodbreak
\and
Space Sciences Laboratory, University of California, Berkeley, California, U.S.A.\goodbreak
\and
Special Astrophysical Observatory, Russian Academy of Sciences, Nizhnij Arkhyz, Zelenchukskiy region, Karachai-Cherkessian Republic, 369167, Russia\goodbreak
\and
Sub-Department of Astrophysics, University of Oxford, Keble Road, Oxford OX1 3RH, U.K.\goodbreak
\and
UPMC Univ Paris 06, UMR7095, 98 bis Boulevard Arago, F-75014, Paris, France\goodbreak
\and
Universit\'{e} de Toulouse, UPS-OMP, IRAP, F-31028 Toulouse cedex 4, France\goodbreak
% ...
}


\date{Received September 10, 2014; accepted February 25, 2015}

\abstract{
   The nearby Chamaeleon clouds have been observed in \g rays by the \Fermi Large Area Telescope (LAT) and in thermal dust emission by \Planck
    and \IRAS. Cosmic rays and large dust grains, if smoothly mixed with gas, can jointly serve with the \hi and $^{12}$CO radio data to (i) map the hydrogen column densities, \nh, in the different gas phases, in particular at the dark neutral medium (DNM) transition between the \hi-bright and CO-bright media; (ii) constrain the CO-to-\hd conversion factor, \xco; and (iii) probe the dust properties per gas nucleon in each phase and map their spatial variations across the clouds.
   We have separated clouds at local, intermediate, and Galactic velocities in \hi and $^{12}$CO line emission to model in parallel the \g-ray intensity recorded between 0.4 and 100 GeV; the dust optical depth at 353\,GHz, \taunu; the thermal radiance of the large grains; and an estimate of the dust extinction, \avq, 
   empirically corrected for the starlight intensity. 
   The dust and $\gamma$-ray models have been coupled to account for the DNM gas.
   The consistent \g-ray emissivity spectra recorded in the different phases confirm that the GeV--TeV cosmic rays probed by the LAT uniformly permeate all gas phases up to the $^{12}$CO cores. 
      The dust and cosmic rays both reveal large amounts of DNM gas, with comparable spatial distributions and twice as much mass as in the CO-bright clouds.
      We give constraints on the \ion{H}{i}-DNM-CO transitions for five separate clouds. CO-dark \hd dominates the molecular columns up to $A_V\simeq0.9$ and its mass often exceeds the one-third of the molecular mass expected by theory.
      The corrected \avq extinction largely provides the best fit to the total gas traced by the \g rays. Nevertheless, we find evidence for a marked rise in \anh with increasing \nh and molecular fraction, and with decreasing dust temperature. The rise in \opa is even steeper. We observe variations of lesser amplitude and orderliness for the specific power of the grains, except for a coherent decline by half in the CO cores. This combined information suggests grain evolution. We provide average values for the dust properties per gas nucleon in the different phases. 
      The \g rays and dust radiance yield consistent \xco estimates near $0.7\,{\times}\,$\xcounit. The \avq and \taunu tracers yield biased values because of the large rise in grain opacity in the CO clouds. These results clarify a recurrent disparity in the \g-ray versus dust calibration of \xco, but they confirm the factor of 2 difference found between the \xco estimates in nearby clouds and in the neighbouring spiral arms.
  }

\titlerunning{Gas \& dust in the Chamaeleon complex}
\authorrunning{Fermi LAT and Planck Collaborations}
\maketitle

\section{Introduction}
\label{intro}

The interstellar gas reserves of the Milky Way are commonly evaluated by means of a large set of multiwavelength tracers. Frequently used are the ubiquitous 21 cm line emission from atomic hydrogen \citep[\hi, see][]{kalberla10}, the widespread 2.6 mm line emission from $^{12}$CO as a proxy for \hd molecules \citep{dame01,planck14_co}, submillimetre to infrared thermal emission from dust grains mixed with the gas \citep{planck11_3D}, and \g rays with energies above a few hundred MeV spawned by cosmic rays (CRs) permeating the gas and interacting with its nucleons \citep{strong88}. Knowledge of the mass, physical state, volume distribution, and dynamics of the different gas phases is the key to understanding the life cycle of the interstellar medium (ISM) in our Galaxy. To this end we need to carefully investigate the validity domain of the total-gas tracers and to quantify their departure from a linear behaviour due to radiation transfer and/or environmental evolution. In this context, the synergy between the \Planck
\footnote{\Planck\ (\href{http://www.esa.int/Planck}{http://www.esa.int/Planck}) is a project of the European Space Agency (ESA) with instruments provided by two scientific consortia funded by ESA member states (in particular the lead countries France and Italy), with contributions from NASA (USA) and telescope reflectors provided by a collaboration between ESA and a scientific consortium led and funded by Denmark.}  and \Fermi LAT all-sky surveys offers new perspectives to study the properties and limitations of these tracers in the multi-phase complexity of clouds down to parsec scales in the solar neighbourhood. 

\subsection{Specific goals}

The ISM is optically thin to thermal dust emission at far infrared to millimetre wavelengths. The emission arises from large grains in thermal equilibrium with the ambient interstellar radiation field (ISRF). 
Several studies have reported an apparent increase in dust emissivity (intensity radiated per gas nucleon) and opacity (optical depth per gas nucleon) with increasing gas column density in both the atomic and molecular gas \citep{stepnik03,planck11_cirrus,martin12,roy13,ysard13,planck14_tau,planck13_pip82}. Interestingly, this might be a hint of dust evolution across the gas phases. Alternatively, dust opacities can be underestimated because of irradiation and temperature changes along the lines of sight, and overestimated by underrating the total gas for reasons that include significant \hi opacity, insufficient sensitivity to CO emission, significant amounts of CO-dark \hd, and opaque CO in dense regions. In this context, the joint analysis of the interstellar \g radiation and thermal dust emission can help constrain the total gas column density, \nh, in order to follow variations of the dust properties.

For a uniform CR irradiation through a cloud, the \g rays provide a measure of the total gas, regardless of its thermodynamic and chemical state, and without absorption limitations across the whole Galaxy. They thereby give valuable insight into (i) saturation corrections to \nhi column densities in the cloud; (ii) the in-situ CO-to-\hd conversion for the derivation of \hd column densities; and (iii) the mass content of the dark neutral medium (DNM) that escapes radio and millimetre surveys in the form of optically thick \hi and/or CO-dark \hd. Irregular CR depletion or concentration inside a cloud can be tested using spectral variations because of the energy dependent propagation of the particles through the magnetic field as they resonantly diffuse on small-scale magnetic turbulence or by focusing or mirroring on the larger-scale structure of the magnetic field. The current \g-ray observations span two to three decades in particle energy and can be used to test these effects. 

The integrated $J\,{=}\,1\,{\rightarrow}\,0$ CO line intensity, \wco, is often assumed to scale linearly with the \nhd column density \citep{dame87}, but the value of the conversion factor, $X_{\rm{CO}}\equiv N_{\rm{H}_2}/W_{\rm{CO}}$, remains uncertain, both in the solar neighbourhood \citep{LAT10_Cep,pineda10,planck11_dark,LAT12_Cham} and at large scales in the Galaxy along the metallicity and UV-flux gradients \citep{strong04,LAT10_Cep,LAT11_3rdQ,pineda13,bolatto13}. Cloud-to-cloud variations in  average \xco can reflect dynamical differences in the relative mass contained in the molecular envelopes (more exposed to CO photodissociation, thus with a higher \xco) and in well-shielded cores \citep[with lower \xco,][]{sheffer08}. Dust and \g-ray proxies for the total gas have been used separately to measure \xco in different locations, at different angular resolutions, and with different methods,  leading to discrepant values (see \citealt{bolatto13} for a review of past references). We aim to compare the calibration of \xco with dust and \g rays in the same cloud and with the same method for the first time.

At the atomic-molecular interface of the ISM, a combination of \hi and \hd gas with little or no CO can escape the \hi and CO surveys because of high levels of \hi self-absorption and low levels of CO excitation. Such a mix of dark neutral medium (DNM) has been theoretically predicted in translucent clouds ($1 \leq A_V \leq 5$ mag) or translucent envelopes of giant molecular clouds \citep{dishoeckblack88}. In this zone, a large fraction of \hd is associated with C$^0$ and C$^+$ instead of with CO because \hd is more efficient at self-shielding against UV dissociation than CO.
The lack of correlation between the OH column-density and \wco suggests large quantities of \hd that are either unseen in CO surveys \citep{barriault10,allen12} or detectable only by summing lines over wide regions without any mapping \citep{pineda10}.

The \g-ray studies have revealed the ubiquity of the DNM, both in mass fraction and spatial extent \citep{grenier05}. In the solar neighbourhood, it appears to be as extended as the dense \hi and as massive as the CO-bright \hd. 
Recent analyses of \Fermi data have confirmed its ubiquity in nearby clouds \citep{LAT10_Cep,LAT12_Cham}. It contributes almost one million solar masses in the star-forming complex of Cygnus X \citep{LAT12_Cyg}. 
The DNM presence has been repeatedly suggested in dust studies as emission excesses over the \nhi and \wco expectations \citep{blitz90,reach94,reach98,magnani03,lee12,planck11_3D}. According to the \Planck data, little CO emission has been missed outside the boundaries of the present 2.6 mm surveys, down to a sensitivity of 1 or 2\,\wcounit \citep{planck14_co}. Fainter CO cannot account for the brightness of the excesses seen off the Galactic plane \citep{planck11_dark}. DNM mass fractions, however, remain uncertain for various causes: from dust emission because of the potential emissivity variations mentioned above \citep{planck11_3D,planck11_dark}; from dust stellar reddening because of the uncertain colour distribution of the background star population, the contamination of unreddened foreground stars, and some incompleteness along the lines of sight \citep{paradis12,LAT12_Cyg}; and from C$^+$ line emission at 158\,$\mu$m because of the difficult separation of the contributions from the DNM, the atomic cold neutral medium (CNM), and photon-dominated regions \citep[PDR;][]{pineda13,langer14}.

In this context, we aim to couple the total gas tracing capability of the CRs and of dust emission to extract reliable column densities in the DNM and to characterize the transition between the \hi-bright, DNM, and CO-bright media in a nearby cloud complex.

\subsection{Choice of cloud}
\label{sect_choice}

\begin{figure*}[!ht]
\includegraphics[width=\hsize]{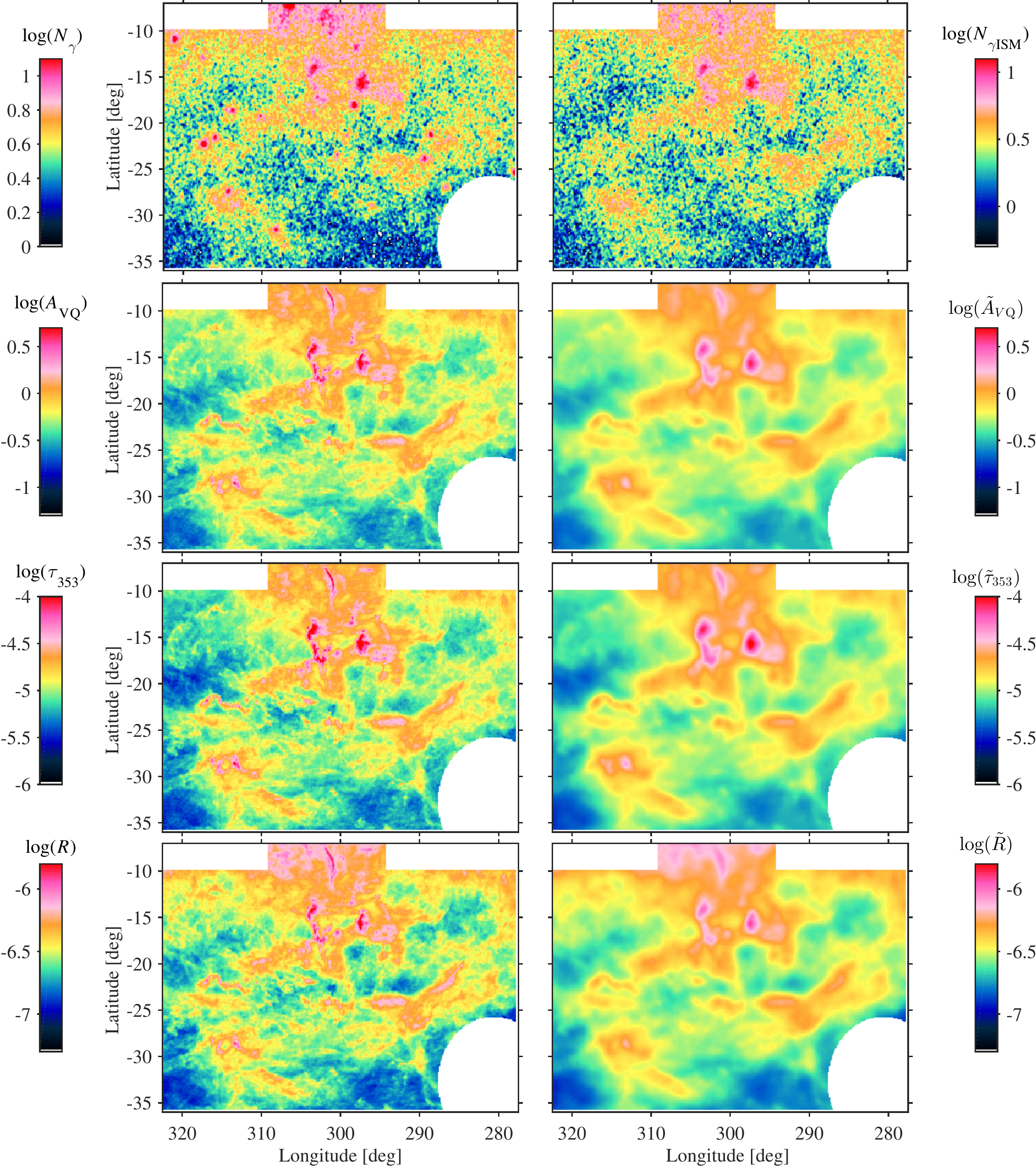} 
\caption{Maps toward the Chamaeleon region of the \g-ray counts recorded in the $0.4-100$ GeV band and of the dust quantities (modified extinction \avq in magnitudes, optical depth \taunu, and radiance $R$ in \Runit). The total \g-ray photon counts are shown on the left and those spawned by cosmic-ray interactions with gas (after subtraction of other ancillary components) on the right. The \g-ray maps have been constructed on a 7\farcm5 pixel grid and smoothed with a Gaussian kernel of 0\fdg1 for display. The dust quantities are shown at 5$\arcmin$ resolution in the left panels, and at the \Fermi LAT resolution on the right (after convolution with the energy-dependent response function of the LAT, assuming the local interstellar \g-ray spectrum over the 0.4--100 GeV band, tilded variables). Regions excluded from the analysis have been masked out.}
\label{dustgam_maps}
\end{figure*}

With its proximity, its moderate molecular mass of the order of $10^4$~M$_{\sun}$ \citep{mizuno01}, and its moderate star-formation activity, the Chamaeleon-Musca complex provides a useful target to probe gas tracers in the $10^{20-22}$\,\persqcm range in \nh. The clouds lie at distances of 140--180\,pc \citep{mizuno01} or 120--150\,pc \citep{corradi04}. We adopt a distance of 150\,pc for mass derivations, but we note that the \nh measurements do not depend on this choice. 

Because of its location at relatively high Galactic latitudes and with typical linear sizes of 10--20\,pc, variations in column density are more likely to reflect changes in volume density than pile-up along the line of sight or confusion with background structures. The available observations have angular resolutions ranging from 5$\arcmin$ to 15$\arcmin$ that limit the cross-talk between the structures of the different gas phases.

The Chamaeleon clouds should be bathed in a relatively uniform ISRF. The lack of OB stellar clusters ensures a relatively quiet environment in terms of: (i) UV irradiation for dust heating; (ii) photo-ionization with little \hii mass; (iii) stellar-wind turbulence for standard CR diffusion (unlike in the turbulent Cygnus X, \citealt{LAT11_cocoon}); and (iv) lack of internal CR sources in the form of supernova remnants. Early \Fermi LAT analyses have shown that the clouds are pervaded by a CR flux close to the average in the local ISM and with an energy spectrum, the so-called Local Interstellar Spectrum (LIS), that is consistent with particle measurements in the solar system \citep{LAT12_Cham}.

The derivation of the dust spectral energy distribution (SED) so far from the ecliptic plane is minimally affected by uncertainties in the zodiacal light removal from the \IRAS and \Planck data \citep{planck14_tau}. The subtraction of the cosmological microwave background and fluctuations in the cosmic infrared background do not significantly affect the bright SEDs \citep{planck14_tau}. The clouds also lie conveniently away from the \Fermi bubbles that dominate the \g-ray sky at energies above a few GeV \citep{su10,LAT14_bubbles}.  

\subsection{Analysis rationale}
\label{sect_rationale}

We can take advantage of the  sensitivity, angular resolution, and broad frequency coverage of \Planck and \Fermi LAT to reassess the relationship between GeV \g rays, dust emission, and \hi and CO line intensities. 
We defer the joint analysis of \g rays and dust extinction or reddening to later work. We use instead two spectral characterizations of the dust thermal emission recently proposed to match the \Planck, \IRAS, and Wide-field Infrared Survey Explorer (\WISE) data. The first is based on modified blackbody spectra parametrized by the optical depth at 353\,GHz, \taunu, the temperature $T$, and spectral index $\beta$ \citep{planck14_tau}. The second uses the physical model of \citet{draineli07} to estimate the dust optical extinction and to renormalize it according to the starlight intensity ($U_{\rm min}$, defined in Sect.~\ref{sect_dustdata}) to better match reddening measurements from quasars \citep{planck14_pip103}. We denote this corrected extinction \avq hereinafter. To follow spatial variations in the dust heating rate, we have also considered a third dust tracer, the radiance $R$, which is the bolometric integral of the thermal intensity \citep{planck14_tau}.

The atomic gas largely dominates the mass budget. Consequently it is the largest contributor to the \g-ray and dust signals. Atomic clouds in different locations and states may have different CR or dust content, so we have developed a careful kinematical separation of the different \hi structures present in the region under study. We have distinguished the \hi gas associated with the star-forming CO clouds, an intermediate-velocity \hi arc crossing the field, and the Galactic \hi background. 

The \g-ray emission detected toward the Chamaeleon region is shown in Fig.~\ref{dustgam_maps}. It is dominated by hadronic interactions between CR and gas nuclei. The ISM itself is transparent to \g rays at these energies. Earlier studies have indicated that the bulk of the Galactic CRs radiating in the energy bands selected for this work have diffusion lengths far exceeding typical cloud dimensions \citep{hunter97,LAT10_Cep,LAT11_3rdQ}. They also indicate an efficient CR penetration in all the gas phases studied here (\hi, DNM, and CO-bright). \textit{The interstellar part of the \g-ray emission can therefore be modelled, to first order, as a linear combination of the gas column densities summed for the various gas phases and  different clouds present along the lines of sight.} The \g-ray intensity $I(l,b,E)$ in the $(l,b)$ Galactic direction and at energy $E$ can be expressed as 
$I(l,b,E) = \sum_{i\in\{{\rm HI1, HI2, \dots, CO, DNM}\}} q_{i}(E)\, N_i(l,b) + \dots$ 
The $q_i(E)$ parameters are to be determined by fits to the \Fermi data. They bear information on the CR flux and gas mass in the different interstellar structures. The model includes other sources of non-gaseous origin (e.g. point sources) that are detailed in Sect.~\ref{sect_gamodel}.

The ISM is also optically thin to the thermal emission of large dust grains. \textit{For a uniform dust-to-gas mass ratio, $R_{\rm DG}$, and uniform mass emission coefficient, $\kappa_{\nu}$, of the grains in a cloud, the dust column density can be modelled to first order as a linear combination of the gas column densities in the different phases and clouds:} if we denote with $D\in\{A_{VQ},\tau_{353},R\}$ any of the three dust tracers, we can express it as 
 $D(l,b)=\sum_{i\in\{{\rm HI1, HI2, \dots, CO, DNM}\}} y_{i}\,N_{i}(l,b) +
\dots$
The $y_i$ coefficients are to be fitted to the data of the $D(l,b)$ tracer. They give measures of the average dust properties per gas nucleon in the different interstellar structures, namely the \anh ratio, the opacity \opa at 353\,GHz, and the specific power \spw of the grains. The models are detailed in Sect.~\ref{sect_dustmodel}.

The interstellar \g-ray emission and the dust tracers shown in Fig.~\ref{dustgam_maps} exhibit very strong structural similarities. They reflect the common presence of CRs and dust in the \hi and CO bright media, but also in the DNM, for which we have no independent template. This inconspicuous phase, however, shows up jointly \textit{as \g-ray and dust emission excesses over \nhi and \wco expectations, with comparable spatial distributions}. We have therefore iteratively coupled the \g-ray and dust models to account for the DNM contribution to the total gas. The method is described in Sect.~\ref{sect_DNMiteration}. The use of the \g rays and of three different dust tracers enables tests of the robustness of the DNM reconstruction.

In order to show the spatial distributions of the dust and \g rays at the angular resolution sampled by the LAT, we have convolved the dust maps with the energy-dependent response of the LAT. To do so, we have assumed the \g-ray emissivity spectrum $q_{\rm LIS}$ of the local interstellar matter. The maps of the LAT-averaged quantities, \avqtilde, \tautilde, and \Rtilde, are shown in Fig.~\ref{dustgam_maps} for the overall energy band. They illustrate the close resemblance in spatial distribution between the dust and \g-ray photon counts of interstellar origin. Figure~\ref{dustgam_maps} also shows that the three dust maps largely agree on the overall distribution of the grains at the original 5$\arcmin$ resolution, but that they significantly differ in contrast (see e.g. at latitudes $b > -15^{\circ}$). The radiance has 3 times less dynamical range than the optical depth, in particular toward the densest molecular zones. The dynamical range of \avq is intermediate between that in $R$ and \taunu. These differences are still present when seen at the LAT resolution. They signal potential variations of the dust properties per gas nucleon that can be tested against the independent \g rays. 

\subsection{Contents}

The paper is organized as follows. Section~\ref{sect_data} presents the \g-ray, dust, \hi, and CO data; Sect.~\ref{sect_models} summarizes the models developed to study  the \hi, CO, and DNM contributions to the dust and \g-ray data, and how the DNM templates are built. In Sect.~\ref{sect_results}, we describe the results of the model fits, their errors, and the impact of the \hi optical depth correction. Sections~\ref{sect_spec} and \ref{sect_DNM} focus on the CR spectrum pervading the different gas phases and on the column-density maps inferred for the DNM. In Sects.~\ref{sect_xcopaspw} and \ref{sect_anhopaspw} we discuss the results on the \xco factors and the average dust properties in each phase. In Sect.~\ref{sect_dustevol}, we present evidence for a marked evolution in dust opacity and a milder evolution in \anh ratio and specific power as the gas becomes denser. In Sect.\ref{sect_phase}, we discuss the transitions between the different gas phases in five separate clouds within the local complex. We summarize the main conclusions and discuss follow-on studies in the last section. Appendices~\ref{sect_HIcomp} to \ref{sect_coeff} present additional information on the kinematical separation of the \hi structures, checks on the \wco calibration, fits without a DNM contribution, and the table of $q_i$ and $y_i$ coefficients.

\section{Data}
\label{sect_data}

We have selected a region around the Chamaeleon complex at Galactic longitudes $277.5^{\circ} \leq l \leq 322.5^{\circ}$ and latitudes $-36^{\circ} \leq b \leq -7^{\circ}$, and we have masked a disc around the Large Magellanic Cloud and toward regions with large contamination from gas in the Galactic disc, at $-10^{\circ} \leq b \leq -7^{\circ}$ and $l < 294\degr$ or $l > 309\degr$. The analysis region is shown in Fig.~\ref{dustgam_maps}. We have selected a broad region to provide enough contrast in the diffuse \hi gas. All maps are projected on the same $0.125^{\circ}$-spaced Cartesian grid as that of the CO survey from the Harvard-Smithsonian Center for Astrophysics (CfA). A finer grid would oversample the Parkes Galactic All Sky Survey \citep[GASS,][]{mcclure09} \hi data and yield too many empty bins, with zero photons, at high energy in \g rays.

\subsection{\g-ray data}
\label{sect_gamma}

We have analysed five years of LAT survey data, starting on 5 August 2008. We have applied tight rejection criteria (\texttt{CLEAN} class selection, photon arrival directions within $100\degr$ of the Earth zenith and time intervals when the LAT rocking angle was below $52\degr$ ) in order to reduce the contamination by residual CRs and \g rays from the Earth atmospheric limb (see \citealt{LAT12_2FGL} for details). To improve the LAT angular resolution below 1.6 GeV, we have kept only the photons that converted to pairs in the front section of the tracker \citep{LAT09_instrument}. At higher energy, we took all photons that produced pairs in the front and back sections of the tracker. We have used the reprocessed Pass 7 photon data, its associated instrument response functions (IRFs, version P7REP-V15) and the related isotropic spectrum\footnote{\href{http://fermi.gsfc.nasa.gov/ssc/data/access/lat/BackgroundModels.html}{http://fermi.gsfc.nasa.gov/ssc/data/access/lat/BackgroundModels.html}}. The LAT exposure was calculated for the adopted photon selection with the RELEASE-09-32-05 of the LAT Science Tools\footnote{The Science Tools are available from the {\it Fermi} Science Support Center, \href{http://fermi.gsfc.nasa.gov/ssc/}{http://fermi.gsfc.nasa.gov/ssc/}}. Systematic errors on the LAT sensitive area increase from 5\,\% to 10\,\% with energy over the 0.4--100 GeV range selected for the analysis \citep{LAT12_perf}.

We have calculated the effective point-spread-function (PSF), the exposure map, the \g-ray emissivity spectrum $q_{\rm LIS}$ of the local interstellar gas \citep{casandjian12} and the spectrum of the isotropic background in 12 energy bins, with a bin width of 0.2 dex, and centred from $10^{2.7}$ to $10^{4.9}$ MeV. To ensure statistics robust enough to follow details in the spatial distributions of the different interstellar components, we have analysed the data in 4 broader and independent energy bands, bounded by $10^{2.6}$, $10^{2.8}$, $10^{3.2}$, $10^{3.6}$, and $10^5$ MeV. We have also analysed the entire $10^{2.6}$--$10^5$ MeV interval as a single band. The LAT energy resolution decreases from 15\,\% to 8\,\% across these energies. Given the large width of the analysis bands, we have not corrected the fluxes for the energy resolution.

The LAT PSF is strongly energy-dependent \citep{LAT09_instrument}. For the local interstellar spectrum (LIS), the half widths at half maximum of the effective PSF are respectively 0\fdg44, 0\fdg27, 0\fdg16, 0\fdg07 with increasing energy in the four bands, and 0\fdg14 in the overall band. To account for the spillover of \g-ray emission produced outside the analysis region, we have modelled both point sources and interstellar contributions in a region 4$\degr$ wider than the region of analysis. This choice corresponds to the 99.5\,\% containment radius of the PSF in the lowest energy band.

The observed \g-ray emission also includes a contribution from the large-scale Galactic inverse Compton (IC) emission emanating from the interactions of CR electrons with the ISRF. It can be modelled with \texttt{GALPROP}\footnote{\href{http://galprop.stanford.edu/}{http://galprop.stanford.edu/}}, version 5.4. The run 54-LRYusifovXCO4z6R30-Ts150-mag2 has been tested against the LAT data \citep{LAT12_diffpaper2}. It assumes a 30\,kpc radius for the Galaxy and a radial distribution of CR sources such as pulsars in the Galactic plane. The particles are allowed to diffuse in the plane and into a halo that is 4 kpc high. We have used this run to generate an energy-dependent template of the Galactic inverse Compton emission across the field of view.

\subsection{Dust data}
\label{sect_dustdata}

We have used the all-sky maps of the dust optical depth \taunu, temperature $T$, and spectral index $\beta$, which were constructed at an angular resolution of 5$\arcmin$ from the combined analysis of the \Planck 857, 545, and 353\,GHz data, and of the \IRAS 100\,$\mu$m data (product release 5, \citealt{planck14_tau}). Compared to previous works (e.g. \citealt{schlegel98}), the use of the \Planck data has greatly improved in precision and in angular resolution the spectral characterization of the dust emission, in particular in regions of large temperature contrast inside molecular clouds and near stellar clusters or IR sources. We summarize here important aspects of this characterization. 

Modified blackbody intensity spectra, $I_{\nu}\,{=}\,\tau_{\nu_0} B_{\nu}(T) (\nu /\nu_0)^{\hbox{\hglue 1pt}\beta}$, where
$B_{\nu}(T )$ is the Planck function for dust at temperature $T$, were fitted to the observed SED in each direction. The fits were performed at 30$\arcmin$ resolution with $\tau_{\nu_0}$, $T$, and $\beta$ as free parameters. The fits were then repeated at 5$\arcmin$ resolution by fixing $\beta$ as obtained in the first step. This procedure limited the noise impact on the $T$--$\beta$ degeneracy. SEDs were checked to be 
consistent with the data at all frequencies (see Fig. 11 in \citealt{planck14_tau}), in particular in bright interstellar areas such as the Chamaeleon region . We note that the contamination from CO line emission in the 353\,GHz filter band, amounting to a few per cent of the signal, was not removed, so as to avoid adding large noise in all directions away from CO clouds.

The derivation of the optical depth, $\tau_{\nu}$, and opacity, $\sigma_{\nu}$, at frequency $\nu$ follows the relations
   \begin{equation}
      \tau_{\nu} = \frac{I_{\nu}}{B_{\nu}(T)} = \sigma_{\nu} \, N_{\rm{H}} = \kappa_0 \left( \frac{\nu}{\nu_0} \right)^{\beta} R_{\rm DG} \, \mu_{\rm{H}} N_{\rm{H}}
   \end{equation}
for the observed specific intensity $I_{\nu}$ of the dust emission, the Planck function $B_{\nu}(T)$ at temperature $T$, the hydrogen column density \nh, the mean gas mass per hydrogen atom $\mu_{\rm{H}}\,{=}\,2.27\,{\times}\,10^{-27}$ kg, the dust-to-gas mass ratio $R_{\rm DG}$, and the mass emission or absorption coefficient $\kappa_0$ at reference frequency $\nu_0$. We have used the map of optical depth, \taunu, estimated at 353\,GHz, and its associated uncertainty. 

The radiance, in \Runit, gives the integral in frequency of the thermal spectrum and it relates to the specific power, $\Pi$, radiated per gas nucleon as
   \begin{equation}
      R = \tau_{353} \int_0^{\infty} \left( \frac{\nu}{\nu_{353}} \right)^{\beta} B_{\nu}(T) d\nu = \frac{\Pi \, N_{\rm{H}}}{4 \pi}.
   \end{equation}
We have propagated the errors on $\tau_{353}$, $T$, and $\beta$ to calculate the uncertainties on the radiance. These uncertainties are upper limits to the real values, since we could not include the negative covariance terms between the anti-correlated $T$ and $\beta$ \citep{planck14_tau}. Within the region of analysis, the optical depth uncertainties range from 2\,\% to 4\,\% and the radiance uncertainties range from 10\,\% to 20\,\%, with a strong peak in frequency around 14\,\%.

The dust model of \citet{draineli07} has also been fitted to the SEDs recorded by \Planck, \IRAS, and \WISE from 12 to 850\,$\mu$m \citep{planck14_pip103}. All-sky maps were thereby constructed for the mass surface density of the dust, the optical extinction, the mass fraction locked up in PAH grains, and the lower $U_{\rm min}$ cutoff in the $U^{-2}$ distribution of starlight intensities heating the bulk of the grains. The comparison between the resulting extinction values and independent estimates based on quasar colours has revealed deviations that significantly correlate with $U_{\rm min}$. The modelled extinction has thus been renormalized according to $U_{\rm min}$ to compensate for this bias \citep{planck14_pip103}. For our work, we have used the renormalized \avq extinction map at 5$\arcmin$ resolution (denoted QDL07 by \citealt{planck14_pip103}). We stress that \avq is a quantity drawn from the thermal emission of the grains, in spite of its absorption-related name. We also note that the physical parameters of the \citet{draineli07} model yield poorer fits to the observed SEDs than modified blackbody spectra. Nonetheless, we show below that, after renormalization, the \avq map is better correlated with the interstellar \g rays than the optical depth deduced from the modified blackbody characterization (see Fig.~\ref{dustgam_maps} and the results in Sect.~\ref{sect_models}). 

The \taunu, radiance, and \avq maps have been derived with the Planck data from the first release.  We have checked that the results of the present work are not significantly changed when we use the most recent version of the Planck data available within the Planck consortium.


\subsection{\hi data and kinematical component separation}


\begin{figure*}
\sidecaption
\includegraphics[width=12cm]{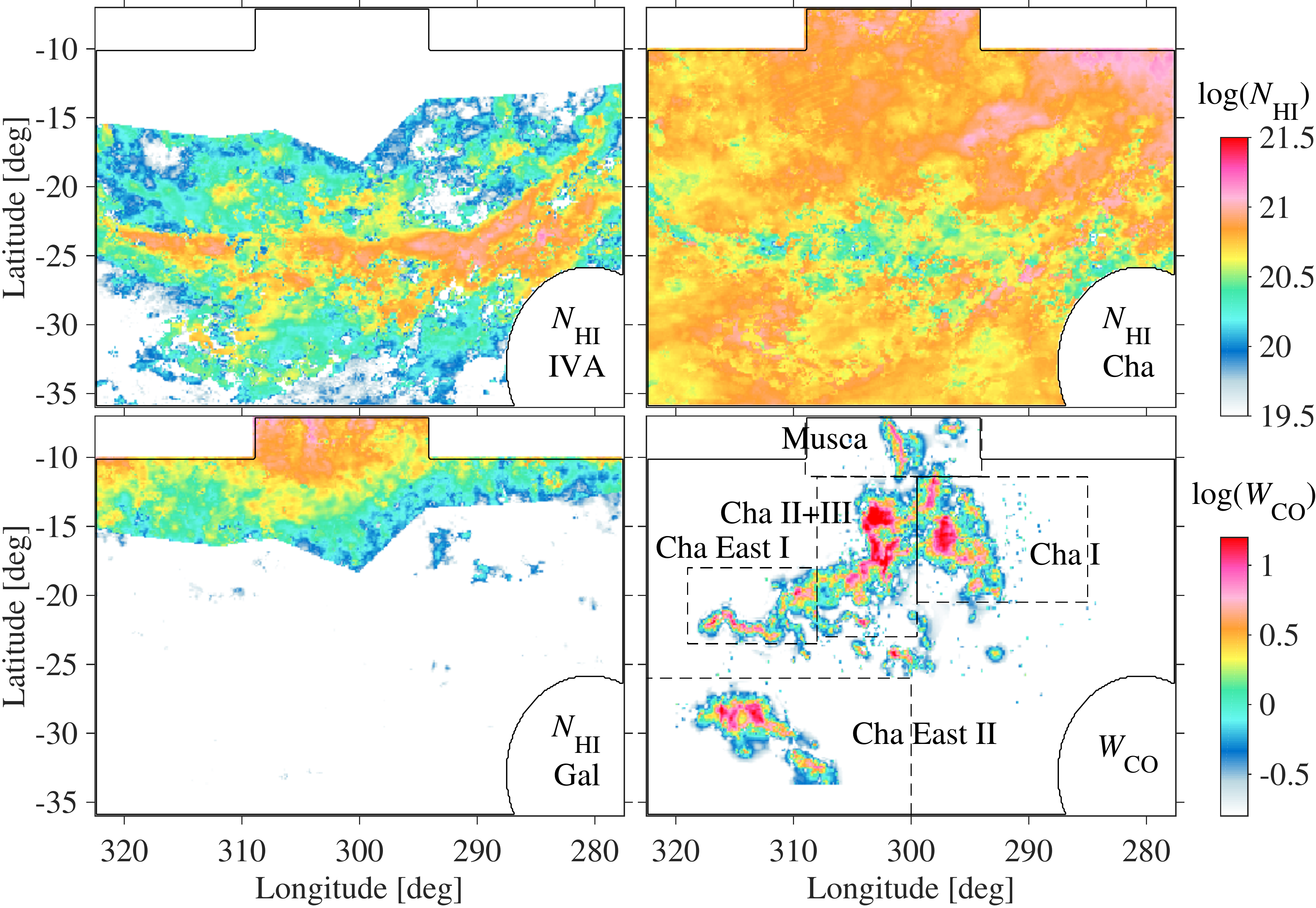}
\caption{Maps 
 of the \nhi column densities (in cm$^{-2}$) and \wco line intensities (in \wcounit) obtained for different velocity components in the analysis region: \nhi for the intermediate velocity arc; \nhi and \wco for the local Chamaeleon complex; and \nhi for the Galactic disc background. The \nhi column densities have been obtained for optically thin emission. The solid black contour marks the analysis region. The dashed rectangles outline the boundaries of the five substructures described in Sect.~\ref{sec_clouds} and analysed in Sect.~\ref{sect_phase}.}
\label{NHI_WCO}
\end{figure*}

The \hi Galactic All Sky Survey (GASS) is the most sensitive and highest resolution survey of 21 cm line emission of the southern sky \citep{mcclure09}. We have used the GASS data corrected for stray radiation, instrumental baselines and radio interference contamination, and with both IFs to remove the negative ghosts occasionally caused in frequency-switching mode by the presence of high-velocity-cloud lines in one of the bands \citep{kalberla10}. We have used the GASS data server\footnote{\href{http://www.astro.uni-bonn.de/hisurvey/gass/index.php}{http://www.astro.uni-bonn.de/hisurvey/gass/index.ph}} 
to resample the original data cubes onto our spatial grid. Our choice of 0\fdg1 for the Gaussian interpolation kernel gives an effective FWHM resolution of 14\farcm5 and an rms noise of 0.07\,K per channel. We have kept the original velocity resolution of 0.82\,\kmpers in the 3D (longitude, latitude, velocity) cube. All velocities mentioned hereinafter are given with respect to the local standard of rest.

Line profiles in the 3D cube have been used to kinematically separate the four main structures that can be distinguished in velocity (see Fig.~\ref{NHI_slices}), namely:
   \begin{itemize}
      \item the local atomic gas in the Chamaeleon complex; 
      \item the gas in an intermediate velocity arc (IVA), crossing the whole region around $-25\degr$ in latitude; 
      \item the more distant gas at large height above the Galactic plane;
      \item gas from the Large Magellanic Cloud (LMC) and its tidal tails. 
   \end{itemize}
The wide velocity range of the IVA component, spanning the $-40 \leq v \leq -4$\,\kmpers interval, is due to very broad line wings in addition to a small velocity gradient along the structure. The origin and distance of this dynamically unusual cloud are unknown;  it is half as massive as the nearby Chamaeleon region  if it is at the same distance.
 
The four features are well defined in the longitude, latitude, velocity $(l,b,v)$ cube, but they occasionally merge because of the gas dynamics and large line widths. In order to  separate them, we have developed a specific separation scheme which is described in Appendix~\ref{sect_HIcomp}. It is based on fitting each \hi spectrum as a sum of lines with pseudo-Voigt profiles. The prior detection of line peaks and shoulders in each spectrum limits the number of lines to be fitted and it provides objective initial values for their velocity centroids. All fits match the data to better than 80 or 90\,\% of the total intensity. In order to preserve the total intensity exactly, the small residuals between the modelled and observed spectra have been distributed among the fitted lines according to their relative strength in each channel. 

We have defined 3D boundaries in longitude, latitude, and velocity for each of the four components. The spatial separations between the IVA and Galactic disc components on the one hand, and between the Galactic disc and LMC components on the other, run along curves of minimum intensity at medium latitudes. The details are given in Appendix~\ref{sect_HIcomp}. 

We have constructed the \nhi column-density map of each component by selecting the lines with \textit{centroids} falling within the appropriate velocity interval, depending on the $(l,b)$ direction, and by integrating their individual profiles in velocity. This procedure gives more reliable column-density estimates than a direct integration of the \hi spectra over the chosen velocity interval. The difference is exemplified in the case of two partially overlapping lines with different peak temperatures. Integrating the observed spectrum in velocity on both sides of a boundary set between the lines would incorrectly attribute the intensity of the wings spilling over the boundary. The large over- (under-)estimation of \nhi from the weak (bright) line would affect the derivation of average cloud properties per gas nucleon in both components. The method used here corrects for the line spillover across velocity boundaries. It also avoids sharp spatial jumps across the resulting maps. This approach thereby enables the exploration of differences in CR and dust volume densities in different structures along the line of sight. 

We have checked that changes in velocity cuts of a few \kmpers have little impact on the resulting \nhi maps. The lines of the local and IVA components strongly overlap in velocity around $l\,{=}\,283\degr$ and $b\,{=}\,{-}25\degr$. Changing the velocity cut by 1 or 2\,\kmpers results in a 3 to 6\,\% change in the total mass in the corresponding velocity range. The difference arises mainly from the region of strong overlap. 

We have integrated the line profiles for a given choice of spin temperature ($T_s$) to correct for the \hi optical depth. The same temperature correction has been applied to all \hi components. In addition to the optically thin case, maps have been produced for uniform spin temperatures of 125, 200, 300, 400, 500, 600, 700, and 800\,K. 

The maps obtained for the optically thin case are shown in Fig.~\ref{NHI_WCO}. Within the region of analysis, the local, IVA, and Galactic disc components exhibit a comparable range of column-densities, with peak values in slight excess of $10^{21}$\,\persqcm. With comparable intensities, but distinct spatial distributions, they can be treated as independent components contributing to the overall dust or \g-ray emission. 

We have checked that the anti-correlation that can be seen in Fig.~\ref{NHI_WCO} between the local and IVA components  corresponds to the presence of two lines of different brightness along those directions. Examples are given in Appendix~\ref{sect_HIcomp}. The trough that crosses the local Chamaeleon map is visible in Fig.~\ref{NHI_slices} at positive velocities prior to any component separation. One may speculate that a large-scale shock has expelled gas from the low velocity Chamaeleon region  and caused both the anti-correlation and the unsually broad wings of the IVA lines. 

\subsection{$^{12}$CO data}


To trace the distribution of the $^{12}$CO $(J\,{=}\,1\,{\rightarrow}\,0)$ emission at 115 GHz, we have used the NANTEN observations of the Chamaeleon clouds with a 2\farcm6 beam, 8$\arcmin$ spacing grid, 0.1\,\kmpers velocity resolution, and a typical noise below 0.4\,K per channel \citep{mizuno01}. Because of the undersampling, we have checked the NANTEN \wco intensities against the fully sampled CfA survey data \citep[8\farcm8 FWHM beam with 7\farcm5 spacing, from][]{boulanger98} across the subset of clouds that have been observed by both instruments (Cham I, II, and III). After removing negative ghosts and flattening baselines in the NANTEN data cube (see Appendix~\ref{sect_COcalib} for details), we have obtained  consistent intensities between it and data from the CfA survey. Unlike what was found in other high-latitude regions \citep{planck11_dark}, Fig.~\ref{WCO_NANTEN_CfA} shows that the NANTEN and CfA photometries fully agree in this region. The derivation of the \xco factor from the present analyses therefore can be directly compared to previous estimations based on CfA data in the solar neigbourhood \citep{LAT10_Cep,LAT12_Cham,LAT12_Orion,bolatto13}.

The ground-based data were preferred over the \Planck CO products for the present work because of the high noise level in the \Planck \texttt{TYPE\,1} CO map and because the dust optical depth was used in the component separation to extract the \Planck \texttt{TYPE\,3} CO map  \citep{planck14_co}. Figure~\ref{WCO_NANTEN_CfA} also shows a systematic photometric difference between the measurements by \Planck and the two radio telescopes. Re-analysing the \Planck data for CO in this specific region is beyond the scope of this paper, however. 

The most sensitive (\texttt{TYPE\,3}) \Planck CO map shows only three tiny clumps beyond the boundary of the NANTEN survey. They lie at low latitude to the west of the Cha East II cloud in Fig.~\ref{NHI_WCO}. Because of their small intensity, $<$ 5\,\wcounit, and small extent, $< 0.25$\,deg$^2$, and because of the photometry mismatch between the \Planck and radio line data, they were not added to the \wco map. Their absence does not affect the \xco results or the CO-cloud masses presented below. 

The CO line velocities span $-12$ to $+8$\,\kmpers (see Fig. 3 of \citealt{mizuno01}). The cloudlet detected at $-12 < v < -4$\,\kmpers appears to be an extension of the local complex rather than a molecular counterpart to the intermediate velocity arc. We did not attempt to separate its small contribution as an independent component. We have thus integrated the CO lines over the whole $-12 \leq v \leq +8$\,\kmpers interval to produce the \wco intensity map shown in Fig.~\ref{NHI_WCO}. 

We have also used the moment-masked CfA CO survey of the Galactic plane \citep{dame01,dame11} to complement the NANTEN data at low latitudes. We have checked that, when convolved with the LAT PSF, the contribution of the Galactic disc emission inside the analysis region is too faint to be detected as an additional component in the \g-ray analyses presented below. This is even more true for the dust analyses because of their better angular precision, so we have dropped the Galactic disc contribution from these analyses. 

\subsection{Individual substructures}
\label{sec_clouds}

In order to study the relative contributions of the different gas phases to the total column density, we have considered five separate substructures in the complex, away from the zone where \hi lines may overlap between the local and IVA components: 
\begin{itemize}
\item Musca at $294\degr \leq l \leq 309\degr$, $b > -11\fdg4$; 
\item Cha I at $285\degr \leq l < 299\fdg5$, $-20\fdg5 \leq b \leq -11\fdg4$; 
\item Cha II+III at $299\fdg5 \leq l < 308\degr$, $-23\degr \leq b \leq -11\fdg4$; 
\item Cha East I at $308\degr \leq l \leq 319\degr$, $-23\fdg5 \leq b \leq -18\degr$; 
\item Cha East II at $l \ge 300\degr$, $b \leq -26\degr$.
\end{itemize}
These limits and names, which are shown in Fig.~\ref{NHI_WCO}, approximately follow  \citet{mizuno01}.

\subsection{Ionized gas}

In view of the very faint diffuse free-free emission detected at 40\,GHz across this field in the nine years of observations of \WMAP, we have ignored the contribution from the warm ionized gas in this study. To verify this assumption, we have taken the 9-year free-free map, based on the maximum-entropy separation and the extinction-corrected H$\alpha$ map as a prior \citep{gold11}. We have translated the intensities into \hii column densities for a gas temperature of $10^4$\,K and effective electron densities of 2 or 10\,\percc \citep{sodroski97}. The resulting column densities, in the $10^{14-15}$\,\persqcm range, show little spatial contrast. Such a quantitatively small and spatially smooth contribution to the total gas column density would not be detected against the other more massive and more structured gaseous components.

\section{Models and analyses}
\label{sect_models}

\subsection{Gas components on test}
\label{sect_gascomp}

All the analyses use only four \hi and CO maps:
\begin{itemize}
\item the \nhi map from the local Chamaeleon clouds;
\item the \nhi map from the intermediate velocity arc;
\item the \nhi map from the Galactic disc;
\item the \wco map from the local Chamaeleon clouds.
\end{itemize}
Faint \hi emission from the LMC outskirts and its streams is present in the analysis region. This emission has not been detected in the \g-ray and dust fits presented in Sects.~\ref{sect_gamodel} and \ref{sect_dustmodel}. There is no detection either of the faint CO emission from the Galactic disc background near the low latitude edge of the region. Both these components have thus been dropped from the analyses.
In addition to the four \hi and CO components listed above, we have constructed DNM templates from the \g-ray data and dust tracers, so that any analysis uses a total of five gaseous components. 

We have performed multivariate fits to separate and study the individual contribution of each component to the \g-ray and dust data shown in Fig.~\ref{dustgam_maps}. We have performed three studies in parallel, jointly analysing either the \g rays and \avq maps (\anaQ ), the \g rays and dust optical depth (\anaT), or the \g rays and dust radiance (\anaR). 

\subsection{\g-ray model}
\label{sect_gamodel}

\begin{figure}
\includegraphics[width=\hsize]{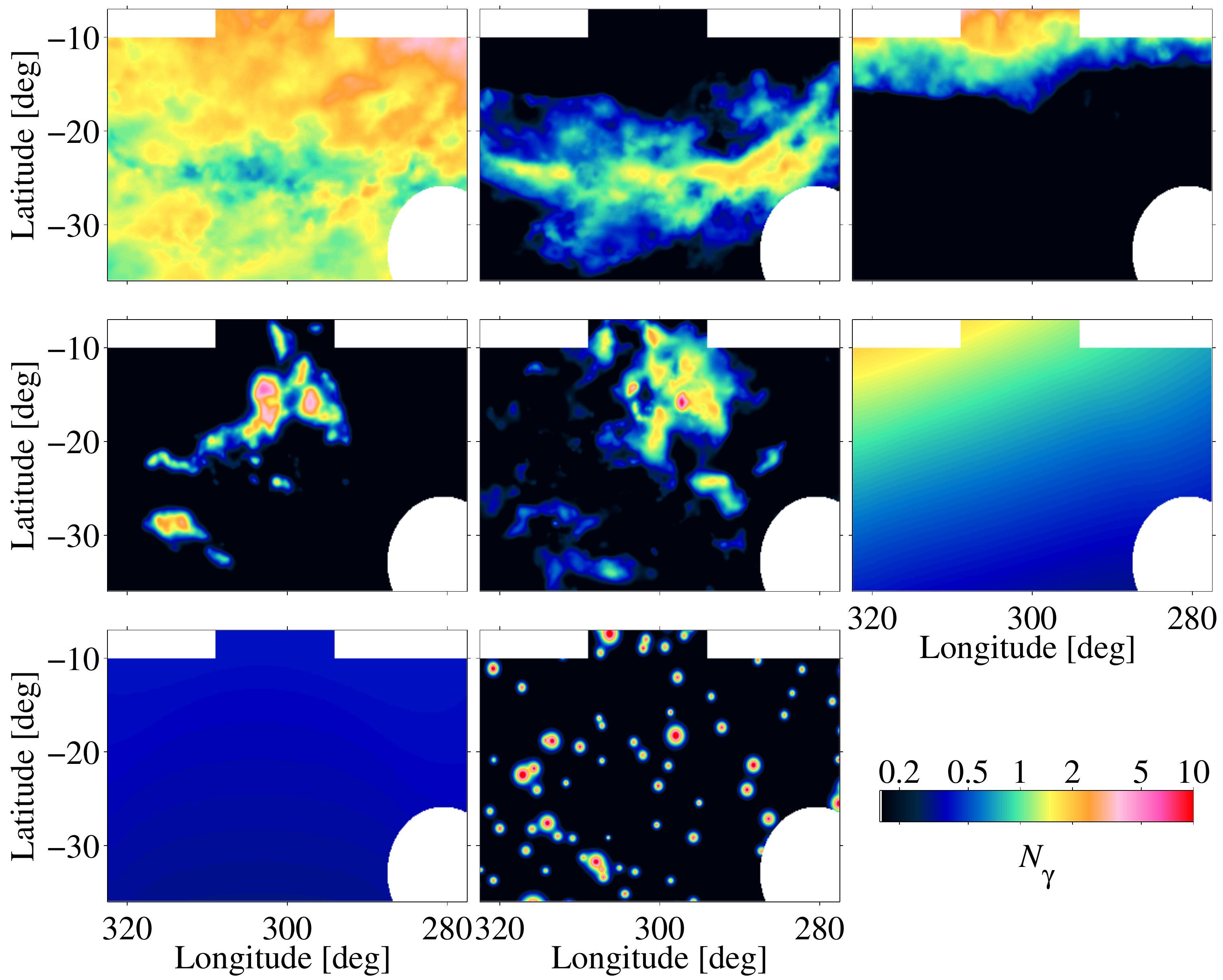}
\caption{Photon yields, on a 
0\fdg125-pixel grid, from the various components of the \g-ray model in the 0.4--100 GeV band. From left to right and from top to bottom, the yields come from the \nhi column densities in the local, IVA, and Galactic disc clouds, the \wco intensity in the local clouds, the \nhdnm column density derived from \avq, the IC emission, the isotropic background, and point sources.}
\label{ISMCMP}
\end{figure}

Because of the arguments presented in Sect.~\ref{sect_rationale} on the ISM transparency to \g rays and on the smooth penetration of cosmic rays through the different forms of gas probed by the \hi and CO lines or in the intermediate DNM phase, we have modelled the \g-ray emission 
as a linear combination of template maps representing the different ISM components. The model also includes a contribution from the Galactic IC emission, point sources of non-interstellar origin, and an isotropic flux to account for the extragalactic \g-ray background and for any residual cosmic rays misclassified as \g rays. 

The \g-ray intensity in each $(l,b)$ direction, $I(l,b,E)$ in \Igamunit, is modelled at each energy $E$ as
\begin{eqnarray}
      I(l,b,E) &=& q_{\rm LIS}(E) \times \bigr[\, \sum_{i=1}^3 q_{\ion{H}{i},i}(E) \, N_{\ion{H}{i},i}(l,b) \nonumber\\
      &+&\, q_{\rm{CO}}(E) \, W_{\rm{CO}}(l,b) + \, q_{\rm{DNM}}(E) \, D^{\rm{DNM}}(l,b)\,\bigr] \nonumber \\
      &+&\, q_{\rm{IC}}(E) \, I_{\rm IC}(l,b,E) + q_{\rm iso}(E) \, I_{\rm iso}(E) \nonumber\\
       &+&\, \sum_j q_{S_j}(E) \, S_j(E) \, \delta(l_j,b_j) + q_{S{\rm ext}} \, S_{\rm ext}(l,b,E),
   \label{equa_gam}
\end{eqnarray} 
where $N_{\ion{H}{i}\,i}$ denotes the three \hi maps listed in Sect.~\ref{sect_gascomp} and $D^{\rm{DNM}}$ stands for the DNM map derived from the dust data. The derivation of the DNM templates is described in Sect.~\ref{sect_DNMiteration}. The $q$ coefficients of the model are to be determined from fits to the \Fermi LAT data.

The $q_{\ion{H}{i},i}$, $q_{\rm{IC}}$, and $q_{\rm iso}$ parameters are simple normalization factors to account for possible deviations from the input spectra taken for the CR-gas interactions ($q_{\rm{LIS}}(E)$), for the isotropic intensity ($I_{\rm iso}(E)$), and for the IC intensity ($I_{\rm IC}(l,b,E)$). We check that there are no spectral deviations form the LIS, as they may signal a CR penetration or exclusion problem between the different gas phases or clouds.

Together with the LIS, the $q_{\ion{H}{i},i}$ parameters give estimates of the average \g-ray emissivity per nucleon in the different atomic clouds. With the further assumption of a uniform CR flux, they serve to scale the mass probed by the \g rays in the other phases. 

As a reliable input for the gas emissivity spectrum, we have used the $q_{\rm{LIS}}$ emission rate, in photons s$^{-1}$ sr$^{-1}$ MeV$^{-1}$ per nucleon, measured with five years of LAT data with the same IRFs and the same selection criteria in instrumental and Earth-limb background rejections, but with all front and back conversions in the tracker at all energies \citep{casandjian12}. The LIS measurement was based on the correlation between the \g rays and the \hi column densities derived from the Leiden/Argentine/Bonn (LAB) survey \citep{kalberla05}, for a spin temperature of 140\,K, in the local Galactic ring spanning  7 to 10\,kpc in Galactocentric distance. We have employed the LIS to apply the energy-dependent IRFs to model the gas emission. Large deviations from the LIS are unlikely in the nearby clouds of the Chamaeleon region  \citep{LAT12_Cham}, but small variations are possible in the complex as a whole or between gas phases. The absolute intensity of the LIS also changes for different choices of \hi spin temperature. This prompted us to leave the \g-ray emissivities of the different gaseous components free to renormalize in each energy band.

Sixty individual point sources have been detected inside the analysis region. Most of them are listed in the 2FGL catalogue \citep{LAT12_2FGL}. New ones have been added from the source list in preparation within the LAT collaboration for the next catalogue. A number of ``c'' sources have been flagged in the 2FGL catalogue for their likely confusion with ISM clumps or with temperature artefacts in the dust map of \citet{schlegel98} that was part of the interstellar background model used for source detection. The ``c'' sources have not been confirmed as significant point sources in the present analysis, and they have been removed from the fits. We have used the spectral characteristics given in the catalogues to compute the source flux spectra, $S_j(E)$. Their individual flux normalizations, $q_{S_j}$, have been left free in each energy band to compensate for the fact that their input spectral characteristics have been derived above a different interstellar background model. The contribution from sources lying within 4$\degr$ outside the analysis perimeter has been summed into a single map,  $S_{\rm{ext}}(l,b,E)$, for each energy band, and its global normalization, $q_{S\rm{ext}}$, has been left free. Similarly, because the IC model and isotropic intensity have been studied over the whole sky and with less data, we have left their normalization free in each band.

We have modelled the $I(l,b,E)$ intensity inside the analysis region and in a $4\degr$-wide peripheral band to account for its faint contribution inside the analysis perimeter through the wings of the LAT PSF. The modelled intensity $I(l,b,E)$ has been processed through the LAT IRFs to account for the position- and energy-dependent exposure on the sky and for the energy-dependent PSF. The resulting photon map, integrated over a specific energy band, can be directly compared to the observed data. We have used a binned maximum-likelihood with Poisson statistics to fit the model coefficients ($q$) to the LAT data in each of the four energy bands and in the overall one.

Figure~\ref{ISMCMP} shows the photon yields obtained for the various components of the model in the overall energy band, with the DNM template provided by the \avq extinction. The photon yields from the ISM dominate the total signal. The variety of spatial distributions and the relative strengths of the interstellar components allow their effective separation despite the limited resolving power of the LAT. 

\subsection{Dust models}
\label{sect_dustmodel}

We have considered three linear models for the dust analyses, using either the \avq$(l,b)$ extinction, the $\tau_{353}(l,b)$ optical depth, or the $R(l,b)$ radiance as a tracer of the total dust column density.
Mild variations in dust emissivity over spatial scales comparable to the cloud size would preserve a significant correlation between the structures observed with dust and the \nh distribution. We have therefore modelled the \avq and \taunu data in each direction as the linear combination of the different gaseous contributions, with free normalizations. The correlation visible in Fig.~\ref{dustgam_maps} between the dust radiance and either the interstellar \g rays or the other dust maps has prompted us to use the same linear model even though the radiance is more sensitive to small-scale variations in grain temperatures. We have added a free isotropic term to all models to account for the residual noise and the uncertainty in the zero level of the dust maps \citep{planck14_tau,planck14_pip103}
\begin{eqnarray}
      D(l,b) &=& \sum^3_1 y_{\ion{H}{i},i} \, N_{\ion{H}{i}\,i}(l,b) + y_{\rm{CO}} \, W_{\rm{CO}}(l,b) \nonumber \\
      &+& \, y_{\rm{DNM}} \, N_{\rm{H}\, \gamma}^{\rm{DNM}}(l,b) + y_{\rm iso},
   \label{equa_tau}
\end{eqnarray}
where $D(l,b)$  stands for \avq, \taunu, or $R$. The $y$ coefficients of the model are to be determined from fits to the data.
The $N_{\rm{H}\, \gamma}^{\rm{DNM}}(l,b)$ column-density map in the DNM phase has been constructed from the \g-ray data (see Sect.~\ref{sect_DNMiteration}).

The $y_{\ion{H}{i},i}$ coefficients in each analysis respectively give the average values of the \anh ratio, \opa opacity, and $R/N_H$ ratio (thus the specific power ratio \spw) in the different \hi maps. The \ydnm and \yco parameters can probe changes of these characteristics in the denser DNM and CO-bright phases. 


Toward dense regions, fitting a single modified blackbody spectrum to the combination of SEDs produced in various ISRF conditions along the sightlines yields an overestimate of the colour temperature, thus an underestimate of \taunu and of the opacity \citep{ysard12}. This bias is gradual, but significant only beyond the high \nh range of our sample. In any case, it would enhance rather than suppress any rising trend in opacity derived from the $y$ coefficients or in the curves and maps of Sect.~\ref{sect_dustevol}.


The dust models have been tested against the data using a least-squares (\chisq) minimization. We expect the uncertainties in the different models to exceed those of the observed dust maps because of our assumption of uniform grain distributions through the clouds and because of the limited capability of the \hi and CO data to trace the total gas (because of the data sampling, self-absorption, etc.). In the absence of a reliable estimate for the model uncertainties, we have set fractional error levels 
in order to obtain a reduced \chisq value of 1 in the dust fits. This has been achieved for fractions of 16\,\%, 18\,\%, and 13\,\%, respectively for the \avq, \taunu, and $R$ models. 

The results presented below, however, show curvature in the evolution of \anh, \opa, and \spw with increasing \nh (see Fig.~\ref{correl_gam_dust} of Sect.~\ref{sect_dustevol}). In this context, changing the statistical weight of the outlier data points can affect the values of the best-fit slopes of the linear model. We have therefore also performed the \chisq fits using the smaller uncertainties of the \taunu and $R$ maps. The results differ only slightly from those obtained with the model uncertainties set to achieve a unit reduced \chisq. We discuss this case in the rest of the paper as the results provide a better statistical description of the average slopes in the multivariate fits. None of our conclusions depends on this choice.


\subsection{DNM templates and analysis iterations}
\label{sect_DNMiteration}
\begin{figure}
\includegraphics[width=\hsize]{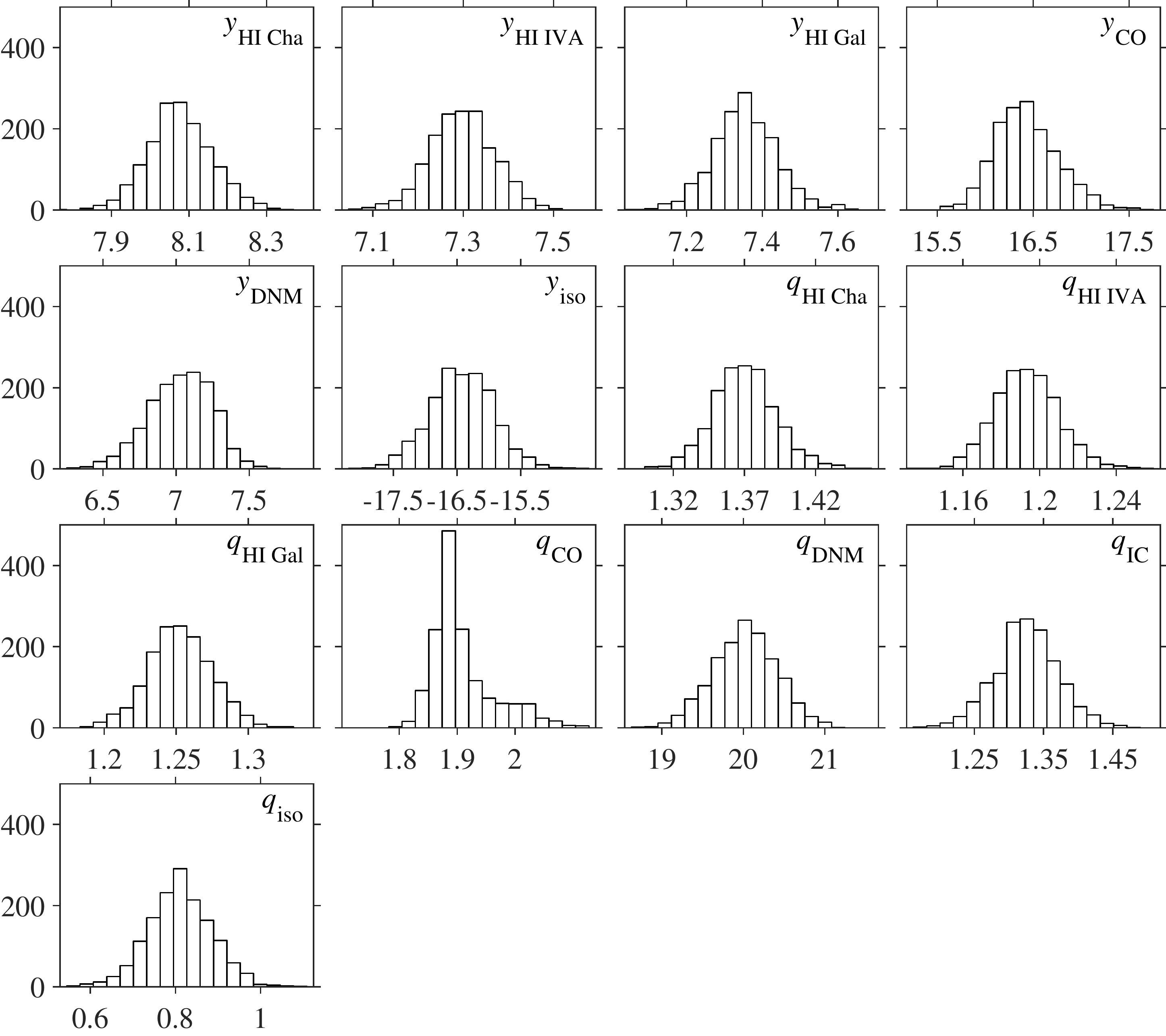}
\caption{Number distributions of the model coefficients obtained in the 1500 jackknife fits of the \anaQ analysis for the optically thin \hi case and overall energy band in \g rays. The $y_{\ion{H}{i},i}$ and \ydnm quantities are in units of \yhQunit, \yco in \ycoQunit, $y_{\rm{iso}}$ in \yisoQunit, \qco in \qcounit, and \qdnm in \qdnmQunit. The $q_{\ion{H}{i},i}$, $q_{\rm{IC}}$, and $q_{\rm{iso}}$ values are simple normalization factors.}
\label{qy_jack}
\end{figure}

Earlier \g-ray works cited in Sect.~\ref{intro} have shown that both the dust column density and the interstellar \g-ray intensity present significant and similarly structured residuals above the linear expectations from the \nhi column densities and \wco intensities. In the Chamaeleon region  analysis, we have \textit{independently} fitted the \g-ray intensity and the three dust maps according to Eqs.~(\ref{equa_gam}) and (\ref{equa_tau}) with only the \hi and CO maps as gaseous components. Figure~\ref{resi_noDNM}  shows extended regions where the data significantly exceed the best-fit models (positive residuals). These excesses have comparable spatial distributions in all data sets. They extend to several degrees (or parsecs) around the CO clouds. As these residuals delineate gas not accounted for by the \hi and CO line intensities, we can use their specific distribution, above the noise, to build a DNM template.

Since the work by \citet{grenier05}, dust data in optical depth or reddening have been used to construct DNM templates for \g-ray analyses to  complement  the \hi and CO data.
The present analysis allows a more reliable derivation of the DNM gas contribution in three ways. 
\begin{itemize}
\item First, by closing the loop between the \g-ray and dust fits. The DNM template estimated from the dust emission is provided to the \g-ray model;  conversely, the DNM map derived from the \g-ray intensity is provided to the dust model. The residuals are obtained in each case by subtracting from the observations the best-fit contributions from the \nhi, \wco, and ancillary (other than gas) components. Only positive residuals above the noise are kept (see below).
\item Second, by iterating between the dust and \g-ray fits in order to reach a solution where the $q$ and $y$ model coefficients, in particular those associated with the \hi and  CO maps, minimally compensate for the missing DNM gas structure (see Appendix~\ref{resi_noDNM}). They still do at some level because the DNM templates provided by the \g rays or dust emission are not perfect.
\item Third, by testing three different tracers of the total dust column density in parallel analyses.
\end{itemize} 

We have not smoothed the dust maps to the \g-ray resolution in the iteration. The dust maps have a finer angular resolution than the model templates and the fit results are not sensitive to structure on angular scales below the resolution of the template maps. It is therefore possible, and important, to keep the dust resolution to model the clumpy CO component. We also note that the diffuse DNM structures independently seen in the \g-ray and dust data (in Fig.~\ref{resi_noDNM}) extend over large angular scales, which can be resolved by the modest \g-ray or \hi resolutions. 

Special attention was paid to the construction of a DNM template from the positive residuals found in \g rays and in dust. A simple cut of the residuals at zero is not acceptable as it creates an offset bias by cutting out the negative noise, but not the positive noise. For both the dust and \g-ray emission, the residual histograms showed  Gaussian noise near or below zero, and a significant positive wing extending to large values. We have therefore denoised the residual maps using the multiresolution support method implemented in the \texttt{MR\,filter} software \citep{starck98}. We have used six scales in the B-spline-wavelet transform (\`a trous algorithm) and a hard $2\,\sigma$ threshold, using all scales for detection in dust and starting with the second scale in \g rays in order to limit the Poisson noise. We have also implemented a simple clipping method, first fitting a Gaussian to the noise-dominated part of the residual histogram, then setting the clipping threshold at the level where the histogram counts exceed the Gaussian. We have checked the consistency of the denoised and clipped maps in the regions rich in signal. We have adopted the former because the wavelet denoising is more efficient in the regions void or nearly void of signal.    

Figure~\ref{dustgam_maps} shows that the Poisson noise in the \g-ray map is still large after five years of data acquisition. To gather the largest photon statistics, we have used all four energy bands to construct the \g-ray DNM templates by summing the residuals obtained in each band before denoising. This was preferred over the direct use of the residual map obtained in the overall-band fit because the emissivity spectra of all components are better adjusted. 

\subsection{Jackknife tests}
\label{sect_jackknife}

\begin{figure*}
\includegraphics[width=\hsize]{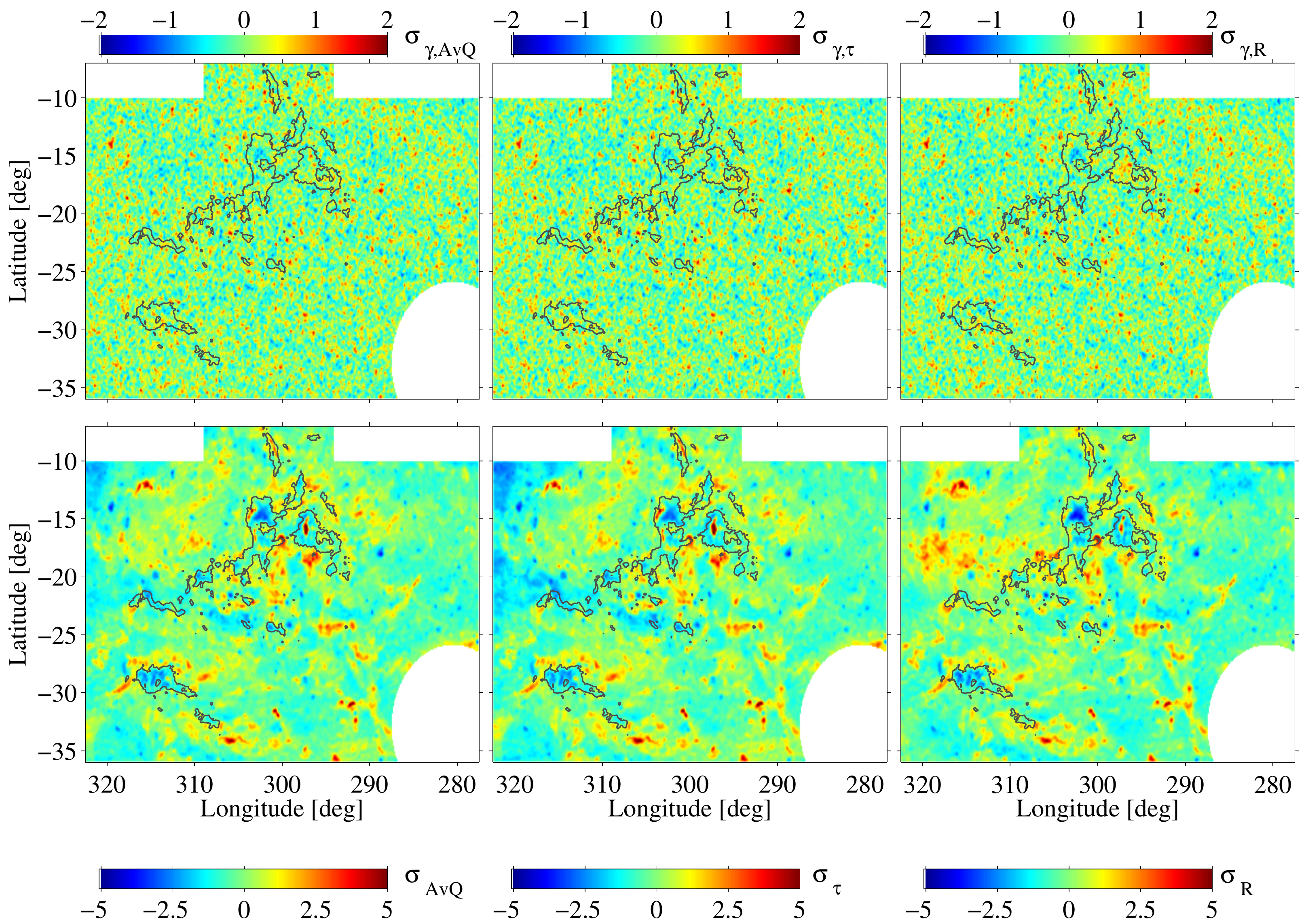}
\caption{\textit{Upper row}: photon count residuals between the data and best-fit model, in sigma units on a 0\fdg125-pixel grid, in the 0.4--100 GeV band, with the DNM template from the dust \avq (\textit{left}), \taunu (\textit{middle}), or radiance (\textit{right}). 
\textit{Lower row}: dust residuals between the data and best-fit model, in sigma units, for the fit in \avq (\textit{left}), \taunu (\textit{middle}), and radiance (\textit{right}). The grey contours outline the CO clouds at the 2.3\,\wcounit level.}
\label{GamRes}
\end{figure*}

The variation of the log-likelihood ratio and \chisq value around the best-fit parameters, namely the information matrix \citep[e.g.][]{strong85}, yields formal errors on each parameter. They include the effect of the correlation between parameters. Given the large number of pixels in the analysis, the small set of free parameters in each model, and the tight correlations present between the maps, the statistical errors on the best-fit coefficients are generally small (3--9\,\% for the gas \g-ray emissivities, 4--13\,\%  for $q_{\rm IC}$,  12--28\,\%  for $q'_{\rm IC}$, and 0.3--0.7\,\% for the dust parameters).

More systematic uncertainties may arise from spatial variations of the model coefficients across the field, from the presence of deviant sub-regions (e.g. near young stellar clusters), or from spatial variations in the mean level of \hi and CO self-absorption. To check the magnitude of these uncertainties, we have performed jackknife tests for the last analysis iteration. We have masked 20\,\% of the analysis region with a random set of 2\fdg625-wide squares and performed the \g-ray and dust fits on the unmasked zones. The process has been repeated 1500 times for each analysis. We have found robust distributions for the best-fit coefficients, as illustrated in Fig.~\ref{qy_jack} for the \anaQ model. All the parameters are well constrained in all analyses, with standard deviations of 2--6\,\% for the gas \g-ray emissivities and 1--3\,\% for the dust parameters. From a statistical point of view, the average coefficients that characterize our linear models apply to the whole region. They are not driven by a particular subset.
 
To construct the final statistical uncertainties on the $q$ and $y$ coefficients, we have added quadratically the standard deviations of the jackknife distributions and the $1\,\sigma$ errors inferred from the information matrices. 

\section{Results}
\label{sect_results}

The values of the best-fit $q$ and $y$ coefficients that have been obtained for the different \g-ray and dust fits are given in Table~\ref{table_qyfit_X2m}. In this section, we discuss the results on the relative quality of the fits obtained with the different dust tracers, with and without the DNM component, and for different optical depth corrections in the \hi. 

\subsection{Comparison of the dust tracers}

As a first test, we have replaced the combination of \hi, CO, and DNM templates in the \g-ray fits by a single dust map to trace the total gas. The quality of the fit greatly changes with the choice of dust tracer: the highest likelihood value is obtained with the $U_{\rm min}$-corrected \avq extinction, then with \taunu, and the poorest fit with the radiance. The values obtained for the log-likelihood ratios and the Neyman-Pearson lemma \citep{lemma33} indicate that \avq is a better representation of the \g-ray observations than the other two dust maps with rejection probabilities $<2\,{\times}\,10^{-11}$. 
This statement remains valid for all choices of \hi spin temperature. The \taunu and \avq quantities are both drawn from the dust emission SEDs, but the latter incorporates a $U_{\rm min}$-dependent correction to better match the dust reddening constraints. The present test against the \g rays, independent of dust, strongly confirms that the renormalization of \avq brings it in closer linear agreement with the total gas. It implies that the $U_{\rm min}$ parameter of the \citet{draineli07} model does not only  trace the ISRF, but also opacity variations. 

\subsection{Detection of the DNM component}
We have then checked that the \g-ray fits considerably improve when adding the dust-derived DNM template to the \hi and CO data. We obtain very large log-likelihood ratios between the best-fit models with and without a DNM component (respectively 1463, 1418, and 1354 for the  \anaQ, \anaT, and \anaR analyses), so the DNM structures are detected with a formal significance greater than $36\,\sigma$. Reciprocally, the \g-ray DNM template is detected at even larger confidence levels in the dust fits when we use a \chisq minimization with the observed uncertainties, when they are available (for \taunu and the radiance). We cannot obtain  a measure of the DNM detection when we set the dust-model uncertainties to achieve a reduced \chisq of 1. 

We then note that the combination of \hi, CO, and DNM data  represents the \g-ray emission better than a single dust map. The large confidence probabilities of the improvement (log-likelihood ratios of 68, 419, and 189, for \avq, \taunu, and $R$, respectively)  indicate the presence of significant differences in the average dust properties in each gas phase. 

\subsection{\hi optical depth correction}
The \g rays can help constrain the \textit{average} level of \hi optical-depth correction by comparing the $T_{\rm{S}}$-dependent contrast of the \nhi maps with the structure of the  \g-ray flux emerging from the \hi gas. It has been shown in the case of the Cepheus and Cygnus clouds that the mean spin temperature so inferred agrees with the more precise, but sparse, measurements obtained from paired absorption/emission \hi spectra \citep{LAT10_Cep,LAT12_Cyg}. We have found no such pairs toward the Chamaeleon region  in the literature \citep[e.g. ][]{heiles03,mohan04}, but the results of the three analyses indicate that the \g-ray fits significantly improve with decreasing  \hi opacity correction, for all energy bands. Figure~\ref{Tspin} indicates that a uniform temperature $T_{\rm{S}} > 340$\,K,  $> 300$\,K, and $> 640$\,K is preferred at the 95\,\% confidence level in the \anaQ, \anaT, and \anaR analyses, respectively. The results indicate that optically thin conditions largely prevail in the local and IVA \hi clouds, in agreement with the low mean brightness temperature of 4.1\,K in the sample and with the large fraction of lines that peak below 100\,K in brightness temperature (98.9\,\%). 

As the (uniform) spin temperature is decreased, the correction to \nhi increases, the \g-ray emissivity of the \hi clouds decreases and the opacity and specific power of their dust grains decreases.
The \hi-related coefficients of the \g-ray and dust models increase by 10--15\,\% as the spin temperature rises from 125\,K to optically thin conditions. \hi optical depth corrections have therefore a small effect on the derivation of \xco factors and dust properties per gas nucleon in this region.

In view of these results and with the added arguments that we detect no change of the CR spectrum in the present \hi structures (see Sect.~\ref{sect_spec}), nor in the larger, less transparent column densities probed in other clouds
\citep{LAT12_Cyg}, we consider the optically thin \hi case as that which best represents the Chamaeleon region  data. \textit{Unless otherwise mentioned, all plots and results hereafter have been generated for this case}.

Nonetheless, we find it useful to quote both types of uncertainties for our results: the statistical errors described above, and those related to the uncertain optical depth of the \hi lines. For the latter, we have taken the range of $q$ and $y$ coefficients obtained for the fits with $T_{\rm{S}}$ larger than the 95\,\% confidence limits quoted above. These ranges provide lower limits to the systematic uncertainties, since we can only explore models with uniform spin temperature, not properly representative of the variety of opacities present in the CNM \citep{heiles03}. We use the  notation, $x\,{=}\,x_0\,{\pm}\,\sigma_x$ $^{+\Delta x}_{-\Delta x'}$, to give both types of uncertainties.

\subsection{Residual maps}

The residual maps of Fig.~\ref{GamRes} indicate that the linear models provide excellent fits to the \g-ray data in the overall energy band. They do so in the four separate energy bands as well. The residuals are fully consistent with noise at all angular scales except, marginally, toward the brightest CO peaks of Cha I and Cha II when the DNM template comes from the dust radiance. 
Whereas the dust-derived DNM templates provide adequate structure and column density in addition to the \hi and CO contributions to fully account for the \g-ray observations, Fig.~\ref{GamRes} shows that significant residuals remain in all dust tracers. The positive residuals follow the bulk distribution of the DNM and they are partly due to the limitations in angular resolution and sensitivity of the \g-ray DNM template compared to its dust homologue. 
Clumps in the residual structure can also reflect localized variations in dust properties per gas nucleon that are not accounted for in the linear models. This is the case toward the dense CO clouds where the dust models often exceed the data because of a rapid increase in dust emissivity and decrease in specific power as \nh increases. These effects are discussed in Sect.~\ref{sect_dustevol}. The asymmetry of the residuals between the cores of Cham I and II in all dust and \g-ray fits requires further investigations with less optically-thick CO tracers such as $^{13}$CO.

\begin{figure}
\includegraphics[width=\hsize]{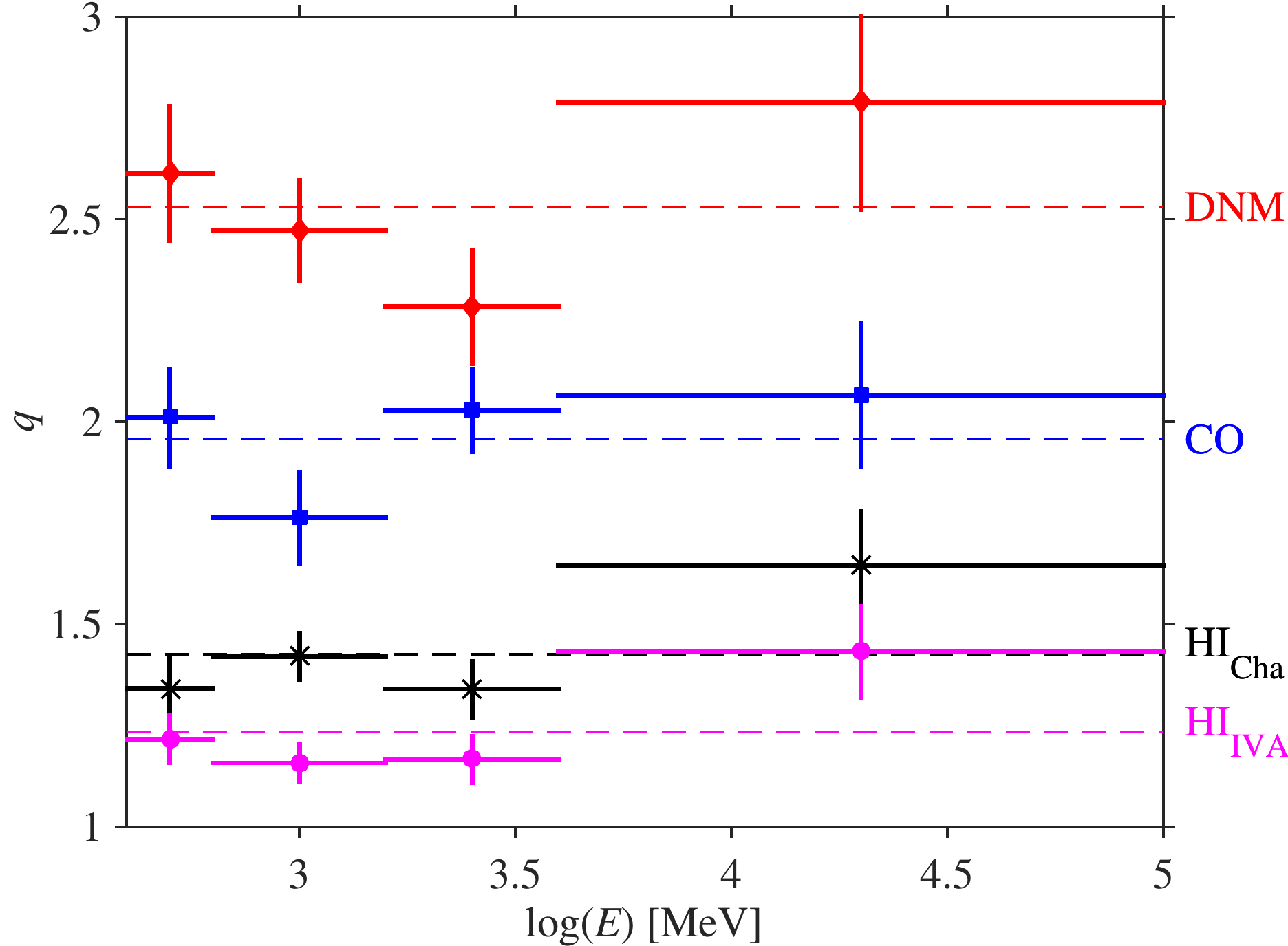}
\caption{Spectral variation, relative to the local interstellar spectrum $q_{\rm{LIS}}$, of the \g-ray emissivities obtained for the local gas components in the \anaQ analysis
. The \qco value is given in units of \qcounit, \qdnm in $8\,{\times}\,10^{20}\,\rm{cm}^{-2}\,\rm{mag}^{-1}$, and the \hi emissivities refer to the optically thin \hi case. The dotted lines mark the mean emissivities.}
\label{qem_LIS}
\end{figure}
\begin{figure}
\includegraphics[width=\hsize]{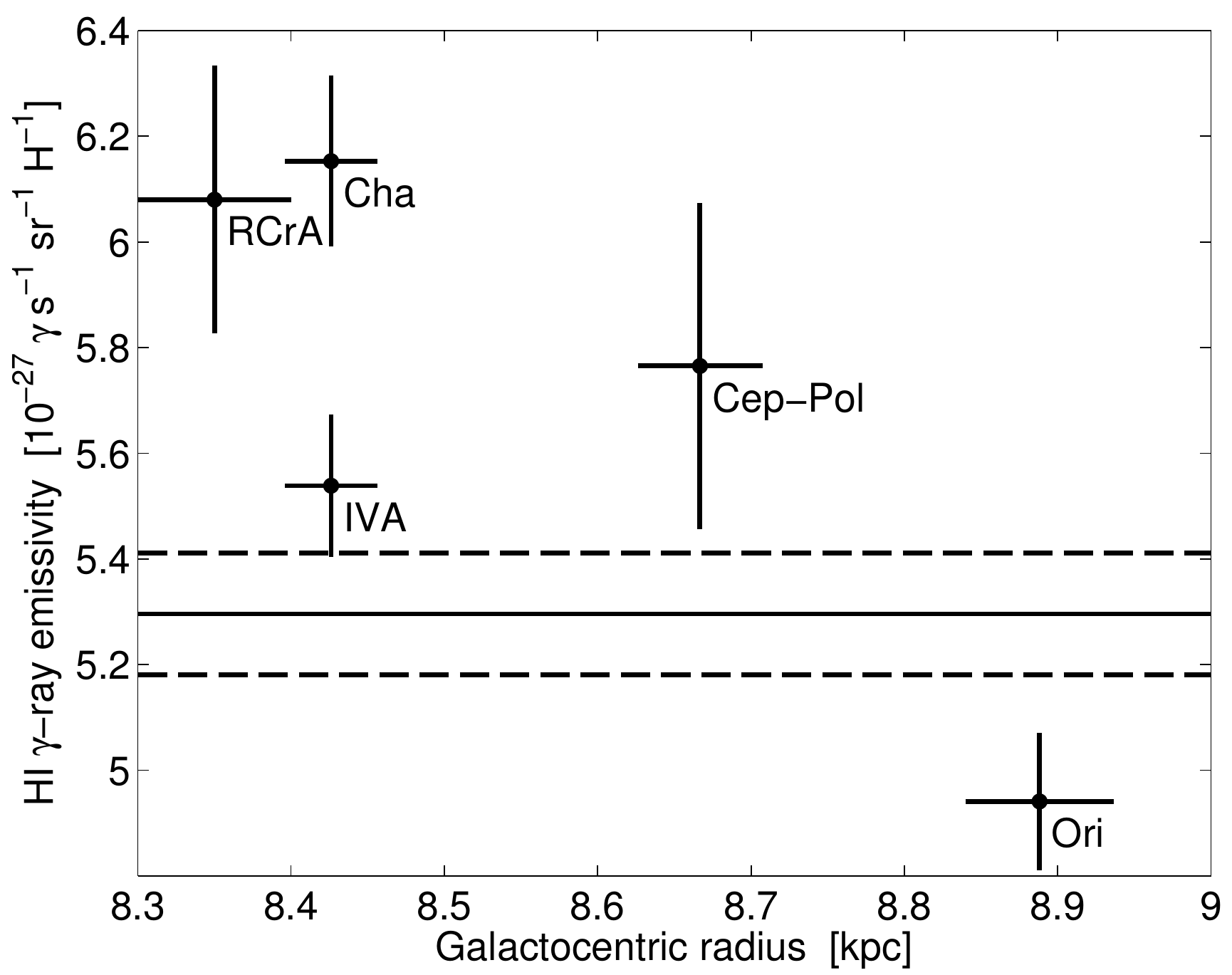}
\caption{Distribution, with Galactocentric radius, of the 0.4--10 GeV emissivities measured in the atomic gas of nearby clouds, for an \hi spin temperature of 125 K. The ${\pm}\,1\,\sigma$ error bars are purely statistical. The solid line marks the mean emissivity measured within 1.5 kpc about the solar circle, for an \hi spin temperature of 140 K. The dashed lines give the ${\pm}\,1\,\sigma$ error.}
\label{qHI_loc}
\end{figure}

\section{Cosmic-ray content of the clouds}
\label{sect_spec}

\begin{figure*}
\includegraphics[width=\hsize]{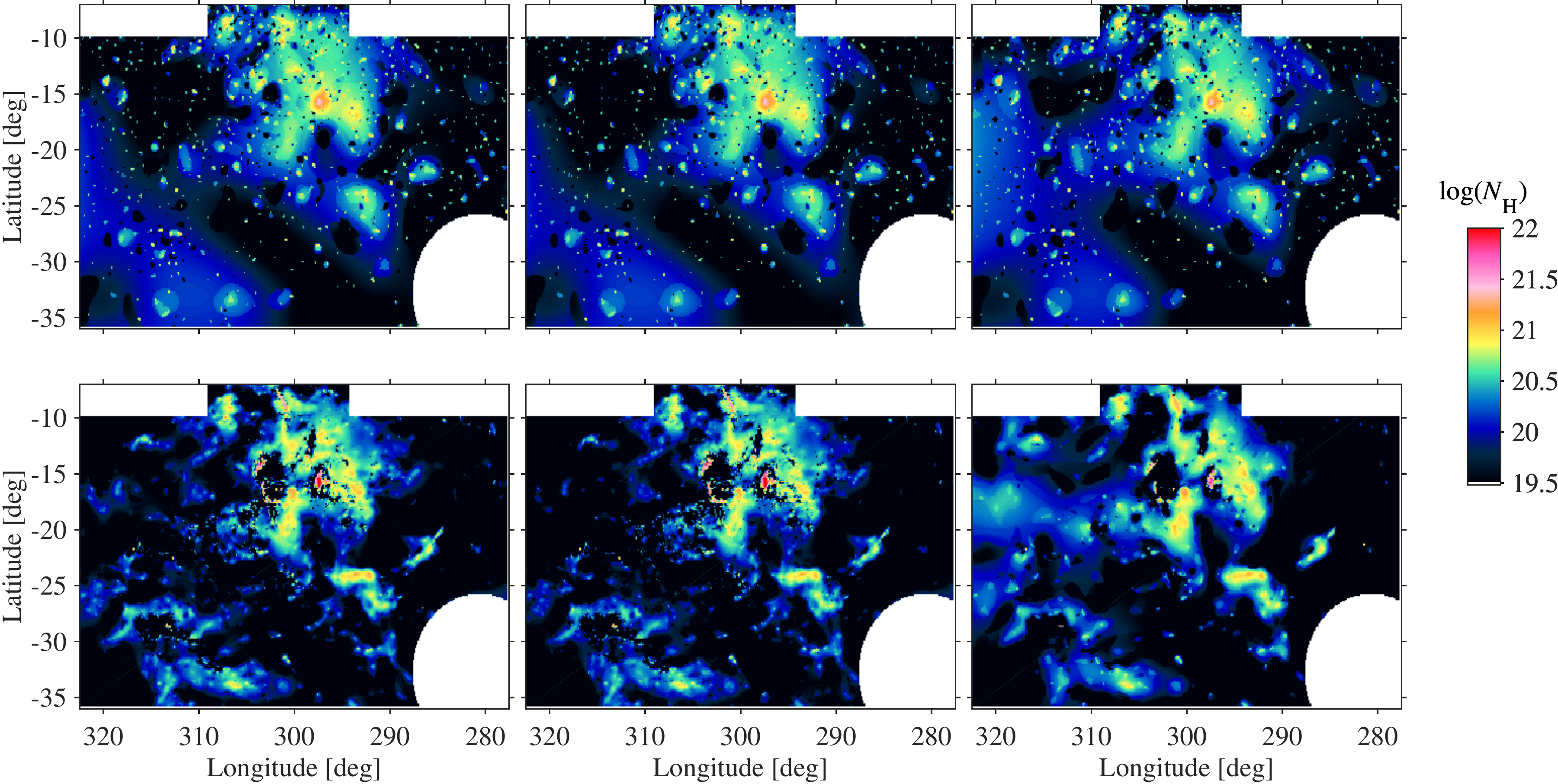}
\caption{Gas column-density maps in the DNM (in cm\msq) obtained from the \g-ray (\textit{upper row}) and the dust (\textit{lower row}) data after subtraction of the best-fit \nhi and \wco contributions and other components unrelated to gas. The dust tracers are respectively \avq, \taunu, and $R$ from left to right.}
\label{NH_DNM}
\end{figure*}

We have used wide \g-ray energy bands to increase the photon statistics to better separate the spatial patterns of the different emission components. Nevertheless, Fig.~\ref{qem_LIS} indicates that the $q$ normalizations relative to the local CR emissivity spectrum ($q_{\rm{LIS}}$) do not significantly change with energy. Figure~\ref{qem_LIS} illustrates this point for the \anaQ analysis, and we find the same trends for the \anaT and \anaR ones. The spectra of the emission originating from the different gas phases and in the different clouds are therefore all consistent with the shape of the input $q_{\rm{LIS}}$ spectrum. At the precision level of the current data, we find no spectral evidence for concentration or exclusion of CRs with increasing gas volume density, up to the $10^{3-4}$\,\percc densities sampled by CO observations. 

The input $q_{\rm{LIS}}$ emissivity spectrum is the average found over the large masses of atomic gas lying in the Galaxy within 1.5\,kpc about the solar circle. Its normalization in terms of emission rate per nucleon corresponds to a low \hi spin temperature of 140\,K \citep{casandjian12}. The Chamaeleon complex  results are given for the optically thin case preferred by the fits. We therefore expect the relative $q$  normalizations in Table~\ref{table_qyfit_X2m} to exceed unity for the same CR flux as in the LIS. To ease the comparison, we have calculated the present \g-ray emissivities for the same spin temperature of 140\,K. The results indicate that the CR flux in the atomic gas of the Chamaeleon complex  and of the IVA clouds is respectively ($22\,{\pm}\,5$)\,\% and ($8\,{\pm}\,4$)\,\% higher than the solar-circle average. The concordance is remarkable given the large differences in size, mass, and linear resolution between these small clouds and the broad Galactic ring. The emissivities in the Chamaeleon clouds and in the less massive and velocity-sheared IVA also compare well, within 20\,\% at all energies. It will be important to determine the distance of the IVA in order to investigate whether the 20\,\% difference is due to the unusual dynamical state of the cloud or to a larger altitude above the Galactic disc. Both cases would bring important constraints on CR diffusion.     

In order to compare with previous measurements in the solar neighbourhood \citep{LAT12_Cham_erratum,LAT12_Orion}, 
we have calculated the integral emissivity of the \hi gas 
between 400 MeV and 10 GeV, for a spin temperature of 125\,K. The values are shown in Fig.~\ref{qHI_loc} together with the average emissivity at the solar circle, integrated over the same energy band, but for a spin temperature of 140\,K. To calculate the Galactocentric positions of the clouds, we have taken distance ranges of 250--400\,pc for the Cepheus-Polaris complex \citep{schlafly14}, of 390--500\,pc for the Orion clouds \citep{LAT12_Orion,schlafly14}, and of 100--200\, pc for the R CrA, Chamaeleon, and IVA clouds \citep{mizuno01,corradi04,LAT12_Cham}. We separate the clouds in Galactocentric radius, but do not expect a CR gradient over such a small distance range. The data points indicate that an equivalent CR flux pervades the nearby Chamaeleon, R CrA, Cepheus, and Polaris clouds. We note that the low CR flux anomaly reported earlier for the Chamaeleon complex was due to an error in evaluating the exposure \citep{LAT12_Cham,LAT12_Cham_erratum}. 
The emissivity in Orion may be ${\sim}\,25\,$\% lower than in the other nearby complexes of the Chamaeleon, R CrA, Cepheus, and Polaris. This difference is small and commensurate with systematic uncertainties in component separations, so a careful re-analysis of the region with three times more \g-ray data (now available), higher \hi resolution, and iterative construction of the DNM map is required to assess this difference.

The uniformity of the CR spectrum across the different gas phases gives weight to the total-gas tracing capability of the \g-ray map. Various mechanisms can alter the CR flux inside dense clouds, but most are inefficient for particle energies in the GeV-TeV range corresponding to the LAT observations. The particles can  diffuse efficiently on magnetic irregularities with wavelengths commensurate with their gyroradii ($\ll$ 1 mpc), but the required power  to maintain the Alfv\'en waves against ion friction with the predominant gas neutrals inside dense clouds \citep{cesarsky78} would be too large. 
Particle depletion inside a dense core may also happen because of increased \g-ray losses in dense gas. It leads to a net CR streaming flux inward, which in turn generates Alfven waves on the outskirts of the core, on the flux tubes connected to the surrounding medium. These waves impede particle progression into the core \citep{skilling76}. The exclusion is strongly energy dependent and only efficient at particle energies below 0.1 GeV if one ignores  the magnetic field compression inside the dense cloud  or 1 GeV if one includes it
\citep{skilling76,cesarsky78}. In the Chamaeleon region, with a CR flux inferred to be near the local ISM average and with maximum \nhd column densities around $2\,{\times}\,10^{22}$\,\persqcm, exclusion is predicted to be negligible (< 2\,\%) for the particles that produce \g rays in the LAT band \citep{skilling76}. Random magnetic mirrors in the clouds have also been investigated, but they affect only CR particles at low energies, invisible to the LAT \citep{cesarsky78,padovani11}. Only CR trapping in the magnetic bottles created between dense cores might affect the emerging \g-ray intensity if the trapped particles die owing to radiative then ionization losses before escaping the bottles or before being replenished by residual diffusion. The prediction of a 3- to 5-fold increase in contrast in \g-ray intensity, compared to that in gas density \citep{cesarsky78}, would strongly bias the \xco factor upward. However, more detailed numerical simulations indicate that TeV particles effectively scatter off magnetic turbulence and smoothly diffuse throughout the complex uniform and turbulent field of a molecular cloud \citep{fatuzzo10}. All these concentration and exclusion processes would leave an energy-dependent signature that we do not detect.

\section{Gas column-densities in the dark neutral medium}
\label{sect_DNM}

The \qdnm coefficients of the \g-ray model (Eq.~\ref{equa_gam}) provide spectral information on the radiation produced in the DNM. The lack of energy dependence of these coefficients (see Fig.~\ref{qem_LIS}) indicates that the spectrum of the DNM-related \g-ray emission closely follows that produced by CR interactions with gas in the local ISM in general, and with the atomic and molecular gas of the Chamaeleon complex in particular. The fact that the \g-ray-derived and dust-derived DNM templates jointly yield reasonable values for the dust properties per gas nucleon in the DNM provide further evidence  that both the dust and \g rays reveal large quantities of gas missed with the \hi and CO data. 

We have converted the \g-ray and dust DNM templates into gas column densities, \nhdnm, under the assumption that the same CR flux permeates the diffuse \hi and DNM phases. The spectral uniformity of the \g rays borne in the two phases supports this assumption (see Fig.~\ref{qem_LIS}). To derive the \nhdnm maps, we have used the templates built of the \g-ray and dust fits in the three analyses. The conversion of the \g-ray templates uses the emission rate per nucleon measured in the atomic gas of the Chamaeleon complex , $q_{\rm{HI\,Cha}}(E)\,{\times}\,q_{\rm LIS}(E)$. The conversion of the dust templates uses the average \anhavgdnm, \opavgdnm, and \spwavgdnm ratios measured in \g rays as
$(\overline{D/N}_{\rm{H}})^{\rm{DNM}}\,{=}\,q_{\rm{HI\,Cha}} / q_{\rm{DNM}}$ (see Sect.~\ref{sect_anhopaspw}). We have applied the weighted means of the ratios obtained in the four \g-ray energy bands to produce the maps shown in Fig.~\ref{NH_DNM}.

The faintest signals shown in the plots are $2\,\sigma$ above the noise. We have noted in Sect.~\ref{sect_results} how strongly the quality of the different fits responds to the inclusion of the DNM maps in all the models. Figure~\ref{NH_DNM} also indicates that they share remarkably similar spatial distributions in the different analyses. The \nhdnm values obtained with the different tracers are remarkably close, considering the lower contrast due to the relatively broad LAT PSF. The layout of the DNM phase is intimately connected to the overall structure of the complex, and exhibits column densities similar to those in \hi. These moderate densities give further grounds to the assumption of a uniform CR flux through the \hi and DNM phases, since the diffusion and loss mechanisms discussed in the previous section set in at much larger densities in the compact molecular cores. The most sensitive CO map from \Planck (\texttt{TYPE\,3}, \citealt{planck14_co}) confirms that the extended DNM clouds cannot be explained by overlooked CO beyond the faint edges mapped with NANTEN down to about 1\,\wcounit. 

High \nhdnm peaks reaching $10^{22}$\,\persqcm are found that partially overlap bright CO cores. Their detection in \g rays and with three dust tracers confirms their reality, which plausibly relates to the saturation of the \wco intensities because of the large optical thickness of the dense molecular gas to $^{12}$CO ($J\,{=}\,1\rightarrow0$) line emission. We have checked that the factor of 2 difference between the peak values measured in \g rays and with the dust radiance is due to the reduced angular resolution of the LAT. The additional three-fold increase to the peak column densities measured with \taunu relates to the dust evolution discussed in Sect.~\ref{sect_dustevol}, so values beyond $5\,{\times}\,10^{21}$\,\persqcm should be considered with care. The high \nhdnm  peaks are surrounded by regions devoid of DNM, where the \wco structure traces the \g-ray and dust distributions fairly well. We need more \g-ray statistics at high energy to further test these peaks at the best LAT angular resolution against $^{12}$CO and $^{13}$CO line emissions. The \nhdnm values below about $2\,{\times}\,10^{20}$\,\persqcm near $l\,{=}\,319\degr$, $b\,{=}\,{-}18\degr$ and $l\,{=}\,318\degr$, $b\,{=}\,{-}11\fdg5$, should also be taken with care as they depend on the choice of dust tracer in a region of warm grains. 

\section{The \xco factor}
\label{sect_xcopaspw}

\begin{table}[htbp]
\begingroup
\newdimen\tblskip \tblskip=5pt
\caption{
\xco conversion factors obtained from the \g-ray and
dust fits in the separate analyses.}
\label{table_Xco}      
\vskip -3mm
\footnotesize
\setbox\tablebox=\vbox{
 \newdimen\digitwidth
 \setbox0=\hbox{\rm 0}
 \digitwidth=\wd0
 \catcode`*=\active
 \def*{\kern\digitwidth}
 \newdimen\signwidth
 \setbox0=\hbox{+}
 \signwidth=\wd0
 \catcode`!=\active
 \def!{\kern\signwidth}
 \halign{\tabskip=0pt\hbox to 1.5in{#\leaderfil}\tabskip=1em&
 \hfil#\hfil\tabskip=1em&
 \hfil#\hfil\tabskip=0pt\cr
\noalign{\doubleline}
\omit\hfil\xco$^{\rm{a}}$\hfil& \g-ray fits$^{\rm{b}}$ & Dust fits\cr
\noalign{\vskip 4pt\hrule\vskip 6pt}
\anaQ& $0.69\pm0.02^{+0.03}_{-0!**}$& $1.01\pm0.02^{+0.05}_{-0!**}$\cr \noalign{\vskip 4 pt}
\anaT& $0.65\pm0.02^{+0.04}_{-0!**}$& $1.27\pm0.03^{+0.07}_{-0!**}$\cr \noalign{\vskip 4 pt}
\anaR& $0.79\pm0.02^{+0.02}_{-0!**}$& $0.66\pm0.02^{+0.02}_{-0!**}$\cr
\noalign{\vskip 4pt\hrule\vskip 6pt}
}}
\endPlancktable
\tablenote {{\rm a}} In units of \xcounit \par
\tablenote {{\rm b}} The \g-ray values are the weighted averages of the results obtained in the four energy bands. \par
\endgroup
\end{table}
\begin{figure*}[!ht]
\includegraphics[width=\hsize]{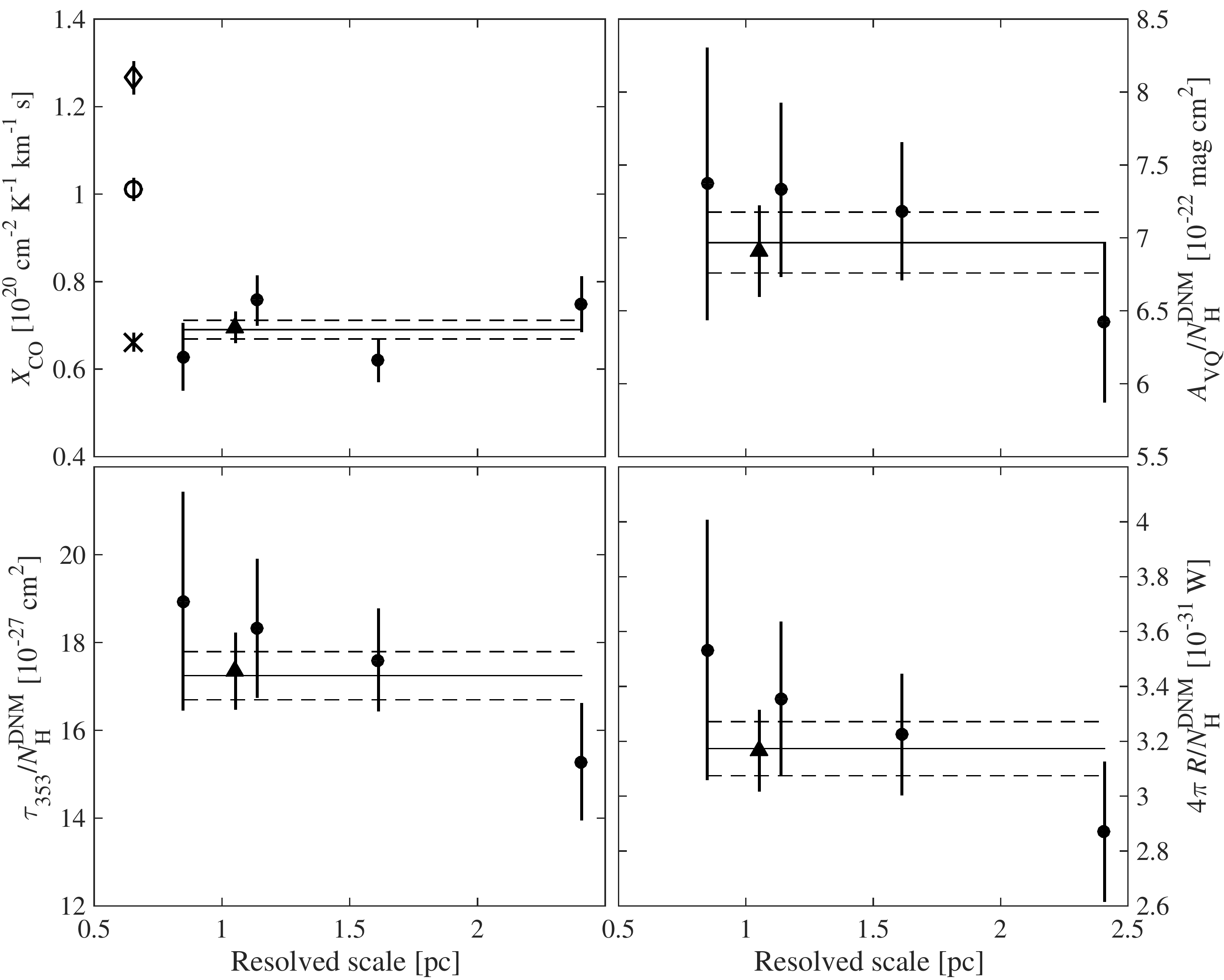}
\caption{Evolution of the \xco factors and of the average dust properties per gas nucleon in the DNM, as measured in \g rays for different linear resolutions in the clouds. The black triangles mark the \g-ray measurements in the overall energy band, in close agreement with the weighted average of the other four independent estimates (thin lines) and their ${\pm}1\,\sigma$ errors (dashed lines). The \xco factors obtained from the dust fits are shown as open symbols (circle for \avq, diamond for \taunu, and cross for $R$).}
\label{Xco_TauNH}
\end{figure*}

The fits in the separate \g-ray energy bands provide independent measures of the \xco conversion factor relating the \wco intensity and \nhd column density. Assuming the same CR flux in the \hi and CO-bright phases, the factor is given by \xcoG$\,{=}\,q_{\rm{CO}} /(2 q_{\rm{HI\,Cha}})$. We can take advantage of the energy-dependent variation of the LAT resolving power (FWHM of the PSF) to probe \xco at different linear scales in the clouds (here we assume a distance of 150\,pc). We have computed the effective PSF widths for the $q_{\rm LIS}$ spectrum and for the energy-dependent exposure of the LAT in this region. Figure~\ref{Xco_TauNH} shows no modification of the \xco factors at parsec scales in the clouds in the case of the \anaQ analysis. We find the same lack of any trend in the other two analyses. Table~\ref{table_Xco} lists the weighted averages of the values obtained in the four energy bands. They closely match the result obtained in the overall energy band, which combines photons obtained with different angular resolutions into a single map, but which has more robust photon statistics. The \xcoG results for the three analyses are consistent within the band-to-band dispersion. 

These findings agree with the theoretical prediction that CR exclusion be negligible in the less massive Chamaeleon clouds, with a loss of less than a few per cent from slower convection into the CO clouds \citep{skilling76}. Conversely, CR concentration inside the CO phase, or magnetic trapping between the dense cloudlets that populate the CO clouds, would bias \xcoG upward, but the effect is expected to be small because high-energy CRs effectively scatter off magnetic turbulence \citep{fatuzzo10}. We find no evidence for such trapping at the smallest linear scales probed by the LAT.

The \xcoG results compare well with other \g-ray measurements in nearby clouds, which range from $(0.63\,{\pm}\,{0.02}\, ^{+0.09}_{-0.07})\,{\times}\,$\xcounit in Cepheus-Polaris to $(0.99\,{\pm}\,{0.08}\, ^{+0.18}_{-0.10})\,{\times}\,$\xcounit in R CrA \citep{LAT12_Cham,LAT10_Cep}. A slightly higher factor, close to $(1.07\,{\pm}\,0.02)\,{\times}\,$\xcounit, has been measured to higher \wco intensities, with no departure from linearity, in the more massive Orion clouds \citep{LAT12_Orion}. 
The factors found in this work are moderately smaller than the previous estimate of $(0.96\,{\pm}\,{0.06}_{stat}\, ^{+0.15}_{-0.12^{sys}})\,{\times}\,$\xcounit obtained in the Chamaeleon clouds \citep{LAT12_Cham}. The difference stems from the improved component separation performed here, in particular to extract the local \qhi emissivity that enters the \xco calculation: higher angular resolution of the \hi data to reduce the cross talk with other components; separation of the local  \hi gas from the contributions in the IVA and Galactic disc; better separation of the diffuse \hi and IC components  across a wider region in latitude. The difference also stems from the use of optically thin \hi data to improve the \g-ray fit. For a spin temperature of 125\,K, as in previous works, we obtain a larger \xco factor of $(0.79\,{\pm}\,{0.03})\,{\times}\,$\xcounit. 

The \xco factors obtained so far in \g rays in nearby clouds are consistent with a value of 0.9$\,{\times}\,$\xcounit and rms dispersion of 0.3$\,{\times}\,$\xcounit. The latter is mostly driven by uncertainties in \hi spin temperature and in component separation between gas phases. These uncertainties prevent any claim of cloud-to-cloud variations in \xco until all clouds are modelled with the same set of approximations and same set of linear resolutions.

Similarly hypothesizing a uniform dust-to-gas mass ratio and uniform emission coefficient $\kappa_{353}$ for the grains, one can infer \xco from the dust fits as $X_{\rm{CO}}\,{=}\,y_{\rm{CO}} /(2 y_{\rm{HI\,Cha}})$. The \xcoQ and \xcoT values inferred from the dust extinction and optical depth are found to be at variance with those obtained with the dust radiance and with the \g-ray estimates (Fig.~\ref{Xco_TauNH}). The latter estimates, on the contrary, are consistent around $0.7\,{\times}\,$\xcounit. Variations in the dust-to-gas mass ratio, or the per cent level of contamination of the dust SED by CO line emission leaking into the \Planck filters.,would affect all dust estimates similarly. The 40\,\% (80\,\%) discrepancy between \xcoQ (\xcoT) and the other values thus has another cause. It can stem from dust evolution, where dust properties change with environment, and which has been invoked to explain a roughly 3-fold increase in dust emissivity in molecular regions \citep{stepnik03,martin12,roy13,planck13_pip82}. The evidence we present in Sect.~\ref{sect_dustevol} of a marked increase in \anh and \opa in the molecular environment indeed biases the dust derivation of \xco upward. 

Variations in dust emissivity or opacity in the CO phase can explain why the \xco factors derived in previous studies, from the intensity of the thermal dust emission or from its colour-corrected optical depth, were systematically higher than the \g-ray estimates, typically by a factor of more than 2 (e.g. \citealt{dame01,planck11_dark,grenier05}). Finding the cause of the discrepancy was hampered by the use of different \hi and CO calibrations, different correlation methods, and different angular resolutions. 
These limitations have been alleviated here. The results provide new insight into this recurrent problem by establishing, through the convergence between \xcoG and \xcoR, that the difference does not stem from a \g-ray versus dust-tracing problem, but rather from dust evolution that must be compensated for to trace the total dust column.

\cite{bolatto13} commented that the lower Xco values obtained from \g-ray analyses relative to dust ones is due to a difference in \xco definition, namely that the CO-faint \hd envelope of molecular clouds is included in the dust derivation and not in the \g-ray one because of the use of a DNM template. This is not the case here since both dust and \g-ray analyses include a DNM component, but this reason cannot be invoked in general because, by construction, the DNM component contains mass with column densities \textit{that do not correlate with the \wco intensity}. Its inclusion (or not) in the analysis does not remove (or add) gas that scales with \wco. Thus to first order, its inclusion does not have an impact on the linear scaling factor $\frac{N_{\rm{H}_2}}{W_{\rm{CO}}}$ that defines \xco. In the component separation of the total gas, the CO-related component gathers the whole gas column density that \textit{correlates} with \wco, independently of its chemical form and of its location inside the CO-bright clumps or in their peripheral CO-faint envelopes. \cite{grenier05} have verified the stability of \xco with respect to the addition of a DNM component, for three different dust DNM templates, and taking advantage of extensive CO maps across the sky to alleviate the impact of the residual cross-talk between the different gas components. They found that \xco increased (instead of decreased) by only 2--4\% when adding a DNM component.
With the higher resolution of the present study, we further show in Appendix \ref{sect_noDNM} that \xco as well as the dust and \g-ray emissivities in the \hi are biased upward when omitting the DNM structure from the model; they artificially increase to partially compensate for the missing gas, but the best-fit model then largely over-predicts the data toward the CO cores, thereby signalling that the \xco ratio is too large.

The convergence between \xcoG and \xcoR opens the way to additional studies in nearby clouds to quantify the amplitude of cloud-to-cloud variations in \xco and to investigate why the dust-derived factors in the Chamaeleon clouds are significantly lower than previous estimates based on the same type of data, for instance $(1.8\,{\pm}\,0.3)\,{\times}\,$\xcounit at $|b| > 5\degr$ from the 100\,$\mu$m intensity data \citep{dame01}, and $(2.54\,{\pm}\,0.13)\,{\times}\,$\xcounit at $|b| \ge 10\degr$ from the 0.1--3 mm optical depth \citep{planck11_dark}. We defer the derivation of \xco from dust reddening to subsequent work. We note, however, that \xcoR, which is less biased by dust evolution in the CO clouds, is less than half the values measured elsewhere with reddening measurements \citep[e.g. an average of $(1.67\,{\pm}\,0.08)\,{\times}\,$\xcounit at $|b| > 10\degr$ and near $2.1\,{\times}\,$\xcounit in the Taurus cloud, ][]{pineda10,paradis12} using the Two Micron All Sky Survey \citep[2MASS]{skrutskie06}. 
As discussed above, detailed comparisons based on the same methods, gas tracers, and resolutions are required to investigate the origin of these differences.

We also note a systematic difference, by a factor greater than two, between the \xco factors measured in \g rays at parsec scales in well resolved nearby clouds and the averages closer to $2\,{\times}\,$\xcounit obtained on a kiloparsec scale in spiral arms \citep{LAT10_Cep,LAT11_3rdQ,bolatto13}. \xco may vary with metallicity and UV-flux gradients across the Galaxy, but the discrepancy is already present within 1--1.5\,kpc in the local spiral arm \citep{LAT10_Cep,LAT11_3rdQ,LAT12_Cyg}, while \Herschel observations of nearby galaxies indicate rather uniform \xco values to large radii past the central kiloparsec \citep{sandstrom13}. Other observational and physical explanations for this discrepancy include the following. 
\begin{itemize}
\item Given the overwhelming mass locked up in the atomic phase, a small error in the \hi-CO phase separation can have a large impact on \xco. Wrongly attributing 10--20\,\% of the clumpy CNM to the CO structure, because of inadequate resolution and thus an increased level of cross-talk at large distances, would be sufficient. The study of nearby galaxies shows that \xco tends to increase when measured in confused environments. Systematically larger \xco values are found in highly inclined galaxies than in face-on galaxies with less pile-up along the lines of sight \citep{sandstrom13}. 
\item Separation of the DNM and CO phases when seen at large distance is even more problematic because of the DNM disposition around the CO and because of a more difficult DNM prediction from dust residuals along long sightlines. C$^+$ observations further suggest a systematic rise in the CO-dark to CO-bright \hd abundance with increasing radius in the Galaxy \citep{pineda13,langer14}. The larger DNM abundance and increasing difficulty in the DNM-CO separation conspire to bias the \xco factor upward in the outer Galaxy. 
\item \xco is expected to increase from the dense, cold molecular cores to the more diffuse, warmer molecular envelopes where CO is more exposed to photo-dissociation and the lines are weakly excited \citep{bolatto13}. Sampling a larger fraction of envelopes in well resolved clouds should therefore bias \xco upward, contrary to the observations. On the other hand, the CO abundance relative to \hd gradually changes from a square root dependence ($N_{\rm CO}/N_{\rm{H_2}}\propto N_{\rm{H_2}}^{0.5\,{\pm}\,0.2}$) to a quadratic one ($\propto N_{\rm{H_2}}^{2.1\,{\pm}\,0.7}$) with a transition around \nhd $\sim2.5\,10^{20}$\,\persqcm \citep{sheffer08}, so local changes in the balance between chemistry and photo-dissociation can cause large variations in \xco. It was suggested, however, that the variations average out in the mix of situations along the lines of sight \citep{liszt10,liszt12}. 
\end{itemize}
More tests are needed to disentangle the origin of the discrepancy. With larger photon statistics at high energy in \g rays, we will soon be able to investigate how the angular resolution and cross-talk between gas phases affect the \xco calibration beyond the solar neighbourhood.

\section{Mean dust properties in each phase}
\label{sect_anhopaspw}

\begin{table*}[htbp]
\begingroup
\newdimen\tblskip \tblskip=5pt
\caption{\anh ratios, opacities, and specific powers of the dust averaged over the different gas phases or clouds.}
\label{table_opaspw}      
\footnotesize
\setbox\tablebox=\vbox{
 \newdimen\digitwidth
 \setbox0=\hbox{\rm 0}
 \digitwidth=\wd0
 \catcode`*=\active
 \def*{\kern\digitwidth}
 \newdimen\signwidth
 \setbox0=\hbox{+}
 \signwidth=\wd0
 \catcode`!=\active
 \def!{\kern\signwidth}
 \newdimen\pointwidth
 \setbox0=\hbox{\rm .}
 \pointwidth=\wd0
 \catcode`?=\active
 \def?{\kern\pointwidth}
 \halign{\tabskip=0pt\hbox to 1.5in{#\leaderfil}\tabskip=2em&
 \hfil#\hfil\tabskip=2em&
 \hfil#\hfil\tabskip=2em&
 \hfil#\hfil\tabskip=0pt\cr
\noalign{\doubleline}
\omit\hfil Component\hfil& $\overline{A_{V\rm{Q}}/N}_{\rm{H}}$&
 $\overline{\tau_{353}/N}_{\rm{H}}$& $4\pi \overline{R/N}_{\rm{H}}$\cr
\noalign{\vskip 3pt}
\omit& [\anhunit$\!$]& [\opaunit$\!$]& [\spwunit$\!$]\cr
\noalign{\vskip 4pt\hrule\vskip 6pt}
\hi Galactic disc& $*7.41\pm0.09^{+0!**}_{-0.22}$& $12.4\pm0.2^{+0!**}_{-0.5}$&
 $6.08\pm0.06^{+0!**}_{-0.08}$\cr
\noalign{\vskip 4 pt}
\hi IVA& $*7.31\pm0.07^{+0!**}_{-0.26}$& $14.8\pm0.2^{+0!**}_{-0.6}$&
 $4.59\pm0.05^{+0!**}_{-0.09}$\cr
\noalign{\vskip 4 pt}
\hi Cha& $*8.11\pm0.09^{+0!**}_{-0.37}$& $16.3\pm0.2^{+0!**}_{-0.8}$&
 $4.85\pm0.06^{+0!**}_{-0.13}$\cr
\noalign{\vskip 4 pt}
    DNM& $*7.0*\pm0.2*^{+0!**}_{-0.3*}$& $17.2\pm0.5^{+0!**}_{-1.0}$&
 $3.17\pm0.10^{+0!**}_{-0.07}$\cr
\noalign{\vskip 4 pt}
     CO$^{{\rm a}}$& $11.9*\pm0.4*^{+0!**}_{-0.5*}$& $\,32?*\pm1?*^{+0!**}_{-2?*}$&
 $4.06\pm0.16^{+0!**}_{-0.09}$\cr
\noalign{\vskip 4pt\hrule\vskip 6pt}
}}
\endPlancktablewide
\tablenote {{\rm a}} The values for the CO phase assume the corresponding \xcoG factor derived in \g rays. \par
\endgroup
\end{table*}
\begin{table*}[htbp]
\begingroup
\newdimen\tblskip \tblskip=5pt
\caption{Average \nhebv ratios measured in the different gas phases or clouds.}
\label{table_NHEBV}      
\footnotesize
\setbox\tablebox=\vbox{
 \newdimen\digitwidth
 \setbox0=\hbox{\rm 0}
 \digitwidth=\wd0
 \catcode`*=\active
 \def*{\kern\digitwidth}
 \newdimen\signwidth
 \setbox0=\hbox{+}
 \signwidth=\wd0
 \catcode`!=\active
 \def!{\kern\signwidth}
 \newdimen\pointwidth
 \setbox0=\hbox{\rm .}
 \pointwidth=\wd0
 \catcode`?=\active
 \def?{\kern\pointwidth}
 \halign{\tabskip=0pt\hbox to 1.5in{#\leaderfil}\tabskip=2em&
 \hfil#\hfil\tabskip=2em&
 \hfil#\hfil\tabskip=2em&
 \hfil#\hfil\tabskip=0pt\cr
\noalign{\doubleline}
\omit\hfil Component\hfil&
 $\left(\overline{\frac{N_{\rm{H}}}{E(B-V)}}\right)_{A_{V\rm{Q}}}$&
 $\left(\overline{\frac{N_{\rm{H}}}{E(B-V)}}\right)_{\tau}$&
 $\left(\overline{\frac{N_{\rm{H}}}{E(B-V)}}\right)_{R}$\cr
\omit& [$10^{21}\,{\rm cm}^{-2}$]& [$10^{21}\,{\rm cm}^{-2}$]&
 [$10^{21}\,{\rm cm}^{-2}$]\cr
\noalign{\vskip 4pt\hrule\vskip 6pt}
\hi Galactic disc& $4.19\pm0.05^{+0.13}_{-0!**}$& $5.4*\pm0.1*^{+0.2*}_{-0!**}$&
 $3.83\pm0.07^{+0.05}_{-0!**}$\cr
 \noalign{\vskip 4 pt}
\hi IVA& $4.24\pm0.04^{+0.16}_{-0!**}$& $4.5*\pm0.1*^{+0.2*}_{-0!**}$&
 $5.07\pm0.10^{+0.10}_{-0!**}$\cr
\noalign{\vskip 4 pt}
\hi Cha& $3.82\pm0.04^{+0.19}_{-0!**}$& $4.11\pm0.10^{+0.23}_{-0!**}$&
 $4.80\pm0.10^{+0.13}_{-0!**}$\cr
 \noalign{\vskip 4 pt}
    DNM& $4.4*\pm0.1*^{+0.2*}_{-0!**}$& $3.9*\pm0.1*^{+0.2*}_{-0!**}$&
 $7.3*\pm0.3*^{+0.2*}_{-0!**}$\cr
 \noalign{\vskip 4 pt}
     CO& $2.6*\pm0.2*^{+0.1*}_{-0!**}$& $2.1*\pm0.2*^{+0.1*}_{-0!**}$&
 $5.7*\pm0.3*^{+0.1*}_{-0!**}$\cr
\noalign{\vskip 4pt\hrule\vskip 6pt}
}}
\endPlancktablewide
\endgroup
\end{table*}

Our analyses yield average dust properties per gas nucleon in each phase. The best-fit \yhi coefficients, respectively, give the average \anh ratio, opacity, and specific power in the three \hi structures; they are listed in Table~\ref{table_opaspw}. The dust properties compare reasonably well in the local Chamaeleon cloud and in the IVA, despite the broad wings of the \hi lines in the latter, which reflect an unusual dynamical state (perhaps shocked gas) or large internal shear. The relatively warm ($\gtrsim 20$\,K) dust in the Galactic disc, lying at large height above the plane, exhibits an equivalent \anh ratio, but a 30\,\% lower opacity and 30\,\% higher power than in the foreground clouds. All opacities and powers appear to be significantly larger than the values of $(7.1\,{\pm}\,0.6)\,{\times}\,$\opaunit and $3.6\,{\times}\,$\spwunit measured in high-latitude cirrus clouds exposed to the local ISRF \citep{planck13_pip82,planck14_tau}. We further discuss in Sect.~\ref{sect_dustevol} how the dust characteristics evolve with environment. 
  
The strong correlations found between the \g-ray intensity and dust emission in the DNM  yield a first measure of the average dust characteristics per gas nucleon in this phase. Their values are given by the $q_{\rm{HI\,Cha}} / q_{\rm{DNM}}$ ratios under the assumption of a uniform CR flux across the \hi and DNM phases. This hypothesis is corroborated by the same \g-ray emissivity spectra and moderate volume densities of the gas in both phases. From the fits in the four independent energy bands, we find the weighted averages listed in Table~\ref{table_opaspw}. They closely match those obtained with the high photon statistics of the overall energy band. Figure~\ref{Xco_TauNH} shows a marginally significant decreasing trend from 0.4 to 2.3\,pc in sampling scale. It relates to the dust evolution discussed in the next section.

We derive mean dust properties in the CO-bright phase from the values of \yco and the corresponding \g-ray \xcoG factors. The results are given in Table~\ref{table_opaspw}. They complement the measures in the \hi and DNM phases to reveal a pronounced rise in opacity, a milder one in \anh, and a 30\,\% decrease in specific power as the gas becomes denser across the phases. The power decrease may be due to the loss of optical/UV radiation from the diffuse envelopes to the dense CO clouds. This moderate evolution biases the \xcoR derivation only slightly downward. We note, however, that the average \spwavgco power in the CO-bright phase is larger than \spwavgdnm in the DNM, while we expect a fainter ISRF in the more shaded CO cores. This inversion is due to the presence of a few high-power spots in the CO clouds. Their high values maintain the average at a high level, despite the marked power decrease around them (see Figs.~\ref{pow_opa_maps} and \ref{pow_opa_relmaps} of section~\ref{sect_dustevol}).

All-sky averages of the $E(B-V)/R$ and $E(B-V)/$\taunu ratios have been measured from the correlation between the dust emission data and quasar colours \citep{planck14_tau}. We have used these ratios to convert the opacities and specific powers of Table~\ref{table_opaspw} into \nhebv ratios. We have also used $R_V\,{=}\,3.1$ to convert the \anh ratios. The results are presented in Table~\ref{table_NHEBV}. 
We note a significant dispersion according to the choice of dust tracer in all gas components.
The dispersion greatly exceeds the statistical errors. 
The values in the CO phase are at clear variance because of the dust evolution that is discussed in the next section. Departures from the canonical $N_{\rm{H}}\,{=}\,5.8\,{\times}\,10^{21}$\,\persqcm $E(B-V)$ relation \citep{bohlin78} have been recently noted as a function of Galactic latitude \citep{liszt14_H2}. The reported all-sky averages span values from around 4 to $9\,{\times}\,10^{21}$\,\persqcm. The values we find in the Chamaeleon complex are generally lower than the canonical one, but not unreasonably so for medium-latitude clouds \citep[compare with Fig.~5 of ][]{liszt14_H2}. 
Beyond these latitude variations, the results of our study call for caution in: (i) the choice of dust emission tracer to derive \nhebv estimates; and (ii) the application of an all-sky average to a particular environment as the dust properties evolve. 

\section{Dust evolution between gas phases}
\label{sect_dustevol}

\begin{figure*}
\sidecaption
\includegraphics[width=12cm]{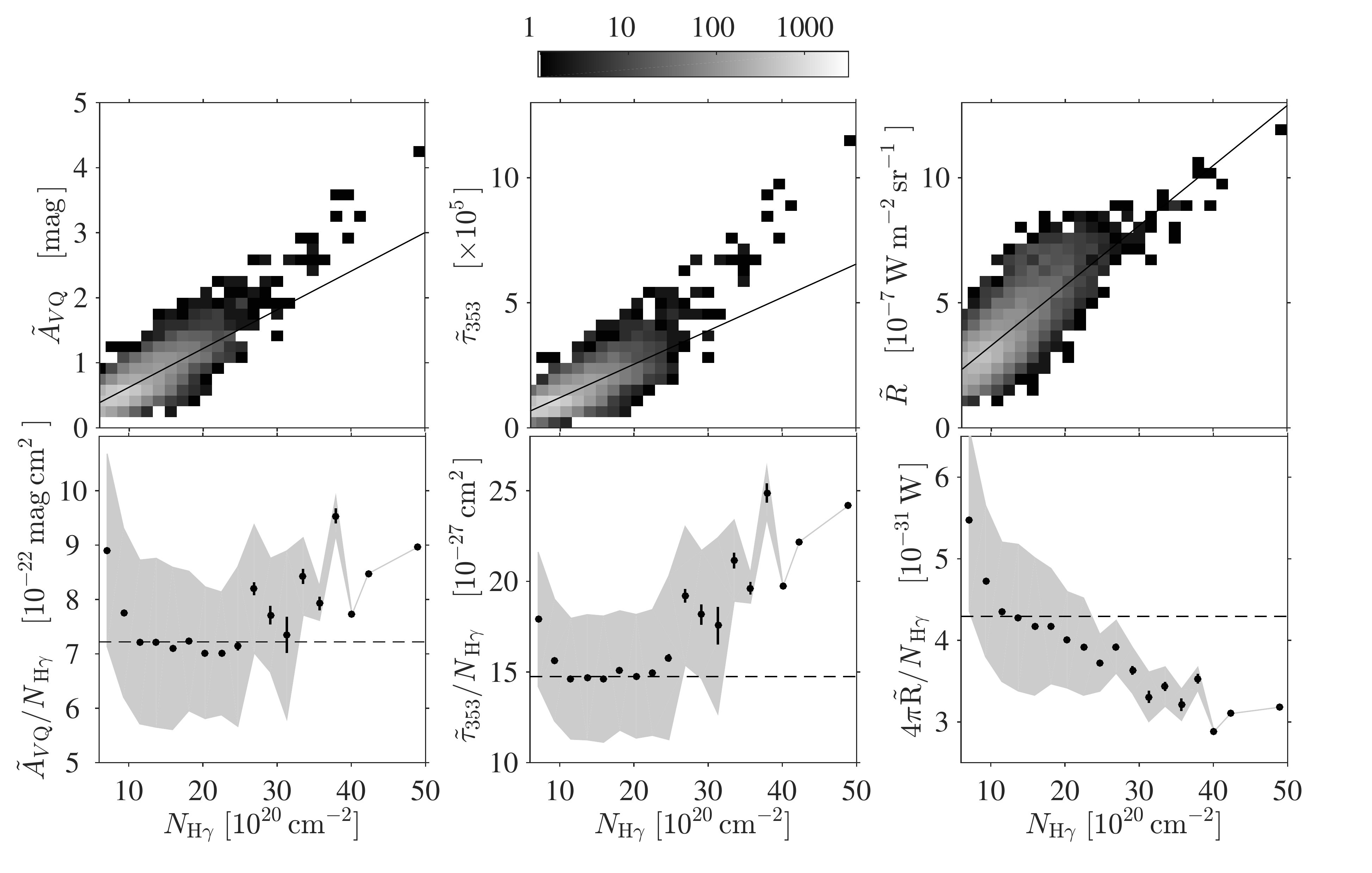}
\caption{\textit{Upper row:} 2D histograms of the correlations between the total gas column density, \nhgam, measured by the $10^{2.6-5}$ MeV interstellar \g rays, and the dust tracers convolved with the LAT response for an interstellar spectrum. The maps were sampled on a 0\fdg375 pixel grid. The solid lines mark the best linear regressions through the data points of the maps. \textit{Lower row:} evolution of the dust properties per gas nucleon in bins of  \nhgam. The error bars and shaded areas respectively give the standard error of the means and the standard deviations in each bin. The dashed lines give the mean ratios at low \nh, in the (1--2.2) $10^{21}$\,\persqcm interval.}
\label{correl_gam_dust}
\end{figure*}
\begin{figure*}
\sidecaption
\includegraphics[width=12cm]{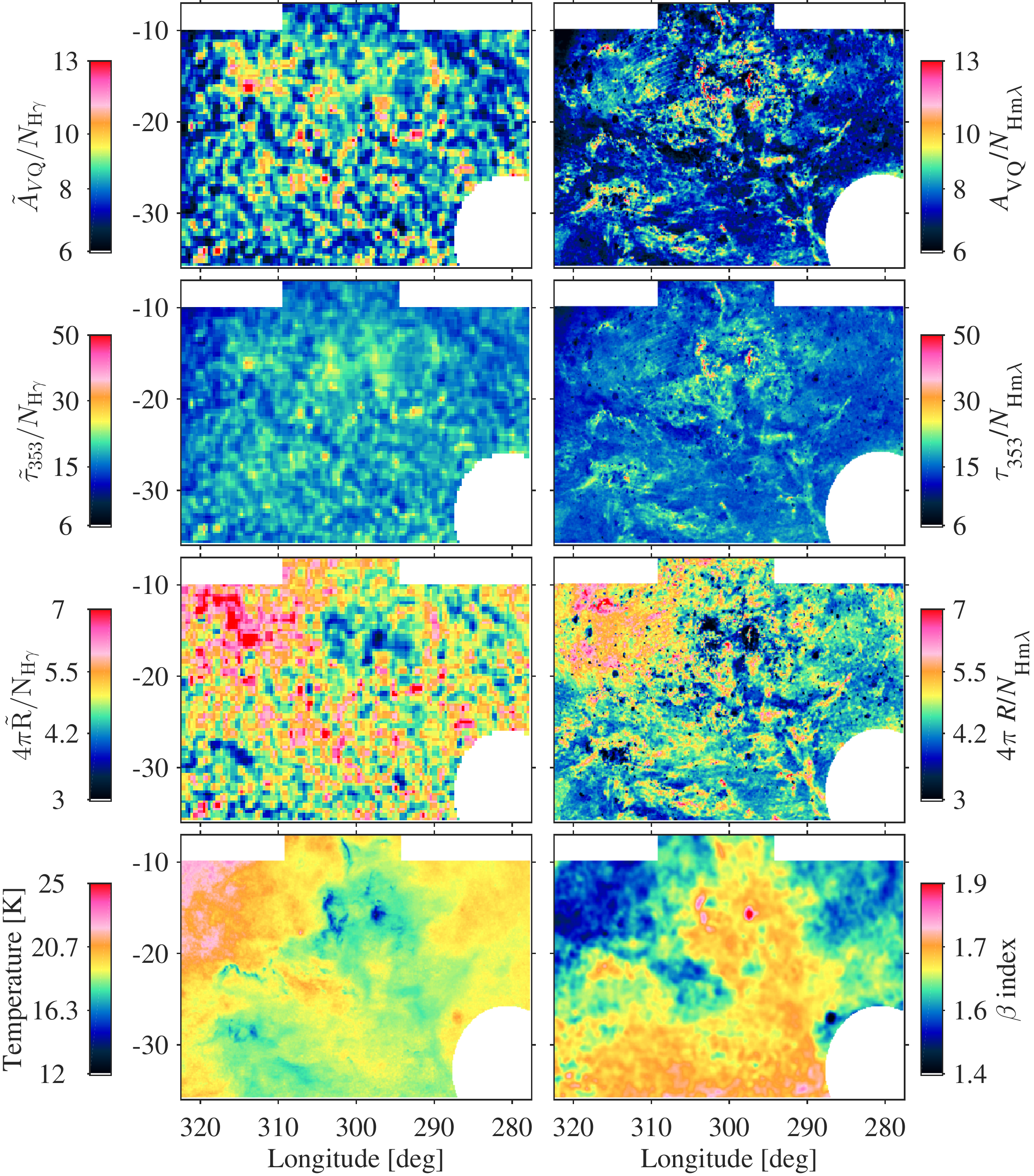}
\caption{\textit{Three upper rows}: spatial variations of the dust properties per gas nucleon with the total gas measured by \nhgam on a 0\fdg375 pixel grid (\textit{left}) and by \nhlam on a 0\fdg125 pixel grid (\textit{right}). \textit{Lower row}: colour temperature and $\beta$ index of the dust SED, measured at 5' resolution by \citet{planck14_tau} and displayed on a 0\fdg125 pixel grid. The \anh values are in units of \anhunit, \opa in \opaunit, and \spw in \spwunit.}
\label{pow_opa_maps}
\end{figure*}
\begin{figure*}
\sidecaption
\includegraphics[width=12cm]{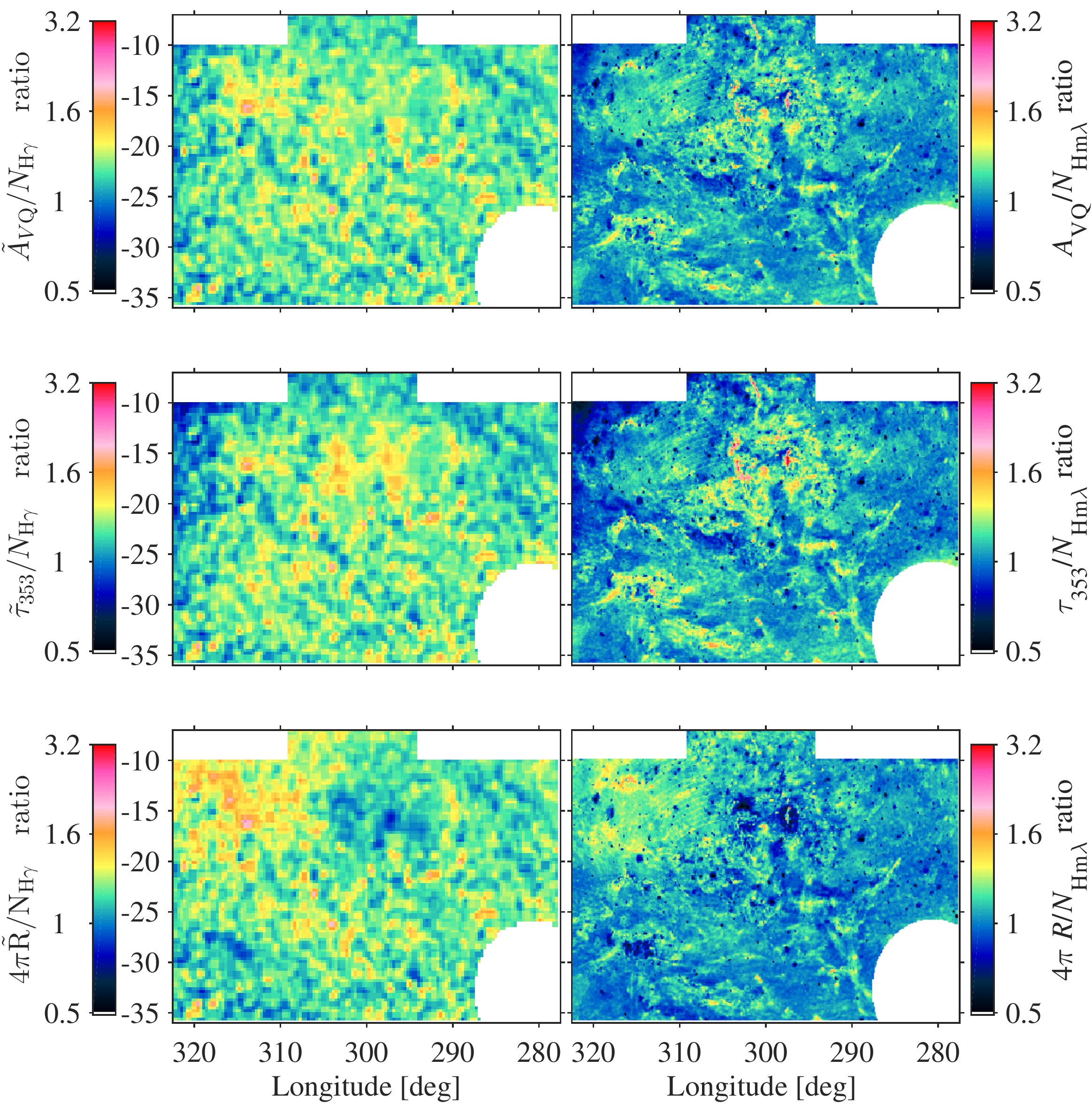}
\caption{Spatial variations of the dust properties per gas nucleon, relative to the average values obtained at low \nh, in the $(1-2.2)\,{\times}\,10^{21}$\,\persqcm interval for \nhgam (\textit{left column}), and in the  $(0.4-2)\,{\times}\,10^{21}$\,\persqcm interval for \nhlam (\textit{right column}). As in the previous figure, the total gas is measured by \nhgam on a 0\fdg375 pixel grid (\textit{left}) and by \nhlam on a 0\fdg125 pixel grid (\textit{right}). Colours saturate for ratios below 0.5 and above 3.2.}
\label{pow_opa_relmaps}
\end{figure*}
\begin{figure*}
\sidecaption
\includegraphics[width=12cm]{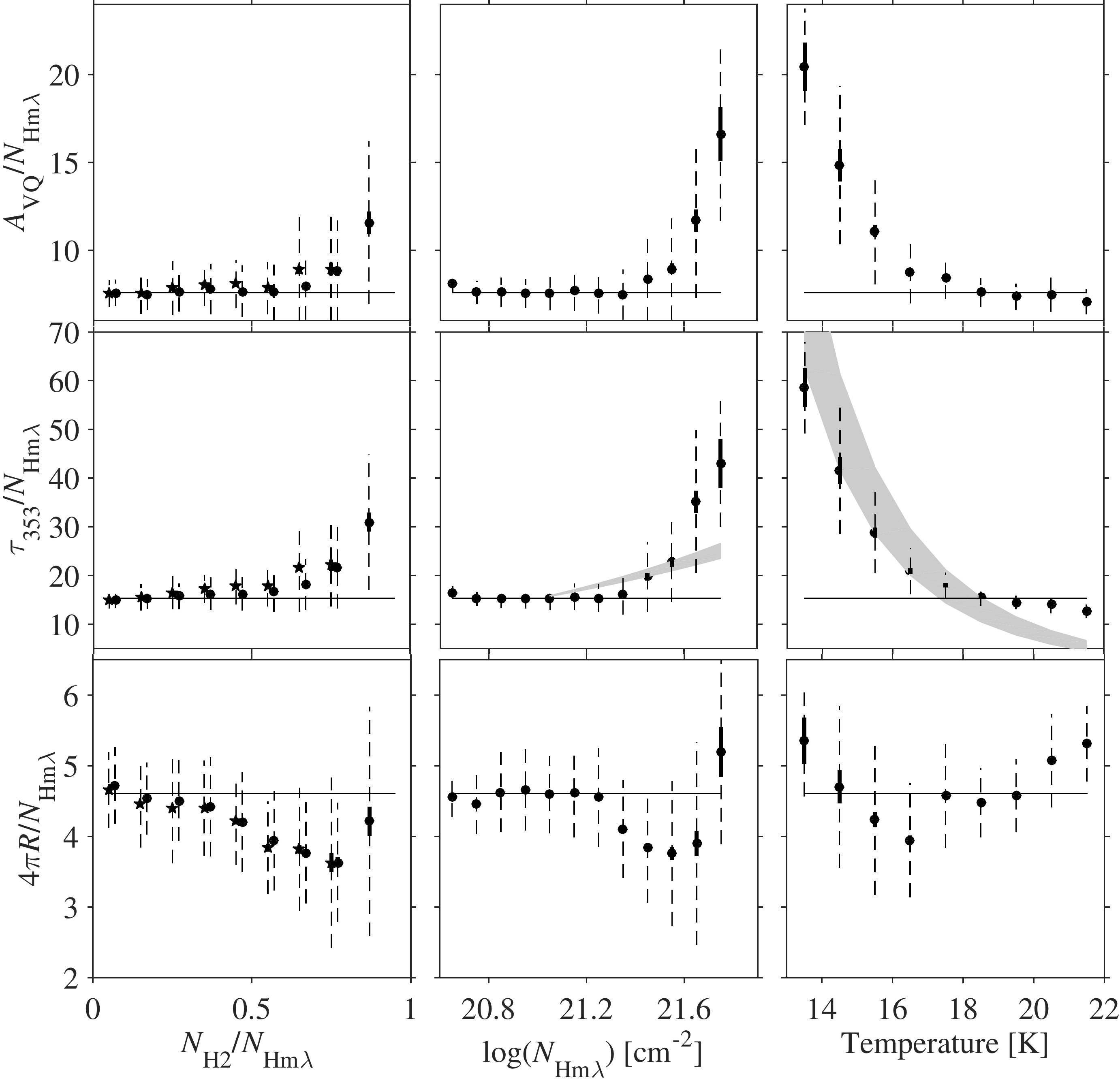}
\caption{Evolution of the dust properties per gas nucleon in intervals of molecular fraction in the gas column (\textit{left})), of total gas column density \nhlam (\textit{middle}), and of dust colour temperature (\textit{right}). The \anh ratios are in units of \anhunit, \opa in \opaunit, and \spw in \spwunit. To estimate the molecular fraction in \nh, we have assumed a DNM composition between half (stars) and fully (dots) molecular. The solid and dashed error bars respectively give the error on the means and the standard deviations in each sample. The thin lines mark the average values found in the $(0.4-2)\,{\times}\,10^{21}$\,\persqcm range. The shaded area in the temperature plot shows the expectation for uniform radiated powers ranging between 3.4 and $5\,{\times}\,10^{-31}$ W $H^{-1}$, with the mean index $\beta\,{=}\,1.65$ in the region. A power-law variation in opacity, $\sigma_{353} \propto$ \nh$^{0.28\,{\pm}\,0.03}$, has been reported above $10^{21}$\,\persqcm in Orion. The shaded curve in the central plot shows this trend scaled at the value of the Chamaeleon complex  opacity at \nhlam$\,{=}\,10^{21}$\,\persqcm.}
\label{dust_evol}
\end{figure*}
\begin{figure}
\centering
\includegraphics[width=0.8\hsize]{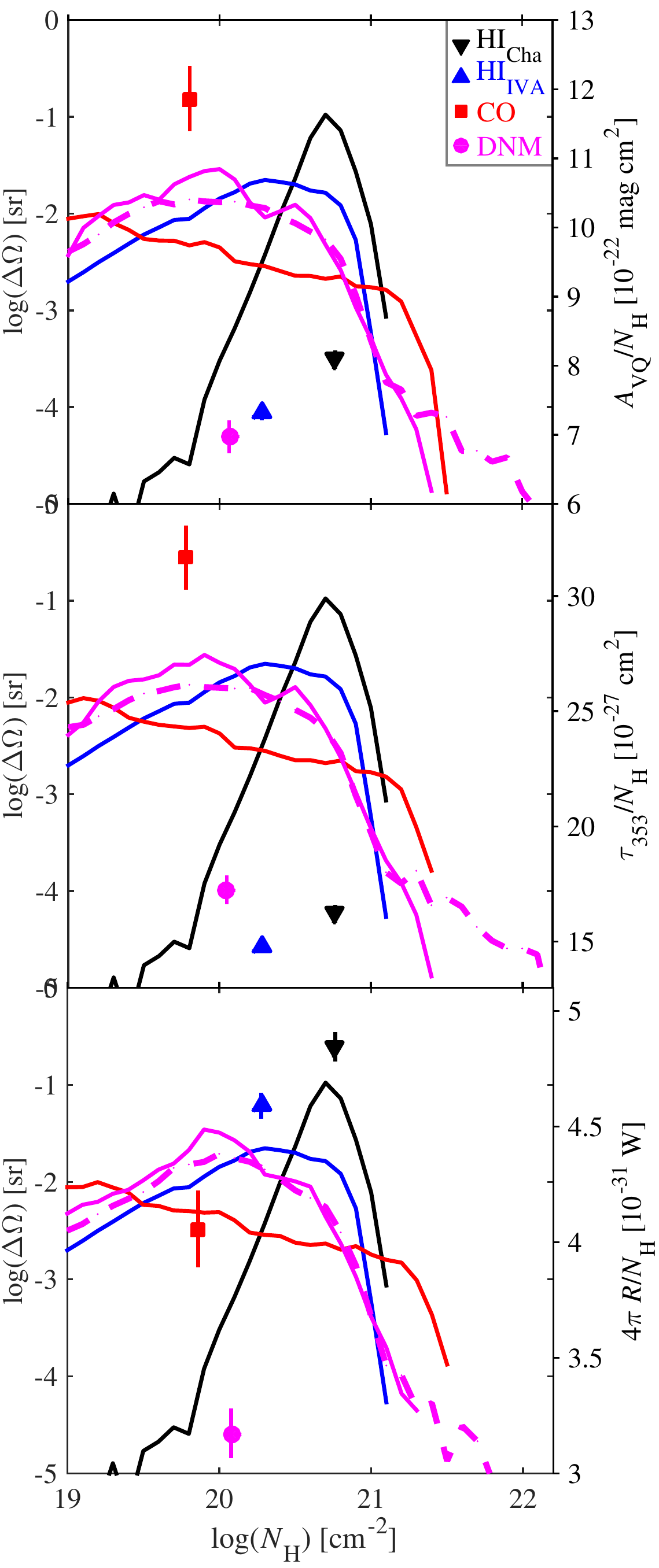}
\caption{Solid angle distributions of \nh in the different gas phases from the three analyses: in the optically thin \hi gas of the local cloud and intermediate-velocity arc; in the DNM gas as traced in \g rays (\textit{solid}) or in dust (\textit{dashed}); and in the CO phase with the \g-ray derived \xco factors. The data points give the mean dust properties (\anh, opacity, and specific power) in each phase. They are placed at the position of the average column density in each phase. The insert gives the colour code and symbol for each phase.}
\label{NH_histo}
\end{figure}

Dust opacities at submillimetre frequencies provide information on the dust-to-gas mass ratio and on the mass emission coefficient of the big grains. Their values often serve to estimate gas masses from dust emission at submillimetre and millimetre wavelengths within our Galaxy or in external galaxies. Establishing the evolution, or lack of evolution, of dust opacities across the various gas phases is therefore essential for numerous studies using dust to trace interstellar matter. 
Maps of \opa and \spw across the clouds can jointly serve to investigate whether the lower temperatures found in the dense regions result from an increased emission coefficient due to grain structural or chemical evolution, and/or from the ISRF attenuation in shaded areas. Under the assumptions that the large grains are well mixed with gas, that they have reached equilibrium temperatures between the heating rate and radiated power, and that the modified blackbody fits reliably characterize their SEDs, the specific-power map follows the first-order variations in heating rate. 

To map spatial variations of the dust properties per gas nucleon around the averages discussed in Sect.~\ref{sect_anhopaspw}, we have produced two different sets of total \nh column densities. They enable a check of the impact of the current limitations in tracers of the total gas.
\begin{itemize}
\item For the first set, which we denote \nhgam, we have converted the interstellar \g-ray intensity into gas column density using the emission rate found in the local atomic gas. The \g-ray intensity from the ISM is obtained from the LAT data in the overall energy band after subtraction of the \g-ray counts unrelated to gas (using the best fit parameters for the ancillary components of eq.~\ref{equa_gam}). We have checked that the other four independent energy bands yield similar maps. We have downgraded the binning to $0\fdg375$ to reduce the Poisson noise and photon discretization. On the one hand, \nhgam does not take into account the small ($<20$\,\%) \g-ray emissivity variations between the \hi structures; on the other hand no assumption is made, beyond the uniformity of the CR flux, on the irregular \hi optical depth, uneven CO-to-\hd conversion, and DNM structure. \nhgam therefore gives an estimate of the total \nh, independently of the chemical or thermodynamic state of the gas. 
\item For the second set, which we denote \nhlam, we have used the higher-resolution multiwavelength information from the \nhi, \wco, and \g-ray \nhdnm maps. We have included the gas from all phases and velocity components, with optically thin \hi and the \g-ray \xcoGQ factor deduced from the \anaQ analysis that best matches the LAT data. We warn the reader that the corresponding maps of the dust properties per gas nucleon are less reliable at the smallest angular scales because \nhlam uses approximations assessed over the whole complex (e.g. uniform \xcoG, optically thin \hi, etc.). We stress that the use of the higher values of \xcoQ and \xcoT for the construction of the \anh and \opa maps hardly lowers their contrast because of the large quantities of \hi and DNM gas that do not depend on \xco.
\end{itemize}

To investigate non-linear departures from the average properties discussed in Sect.~\ref{sect_anhopaspw}, we have produced four sets of plots in Figs.~\ref{correl_gam_dust} to \ref{dust_evol}. 
\begin{itemize}
\item Figure~\ref{correl_gam_dust} directly explores the correlations between the dust and \g-ray signals produced in the gas, with a common $0\fdg375$ binning. The slope follows the evolution of the dust properties per gas nucleon in intervals of increasing \nhgam.
\item Figures~\ref{pow_opa_maps} and \ref{pow_opa_relmaps} explore the spatial variations of the dust properties per gas nucleon as traced side by side with \nhgam and \nhlam, in absolute and in relative units. 
\item Figure~\ref{dust_evol} explores the evolution of the mean dust properties in intervals of increasing \nh, of increasing molecular fraction in \nh, and of increasing dust temperature. The values have been obtained with \nhlam binned at $0\fdg125$.
\end{itemize}

Together these investigations provide compelling evidence for dust evolution across the Chamaeleon complex. Figure~\ref{correl_gam_dust} illustrates the tight correlations that exist between the \g-ray flux emerging from the gas and the dust maps, \avqtilde, \tautilde and \Rtilde, convolved with the LAT response. We obtain Pearson's correlation coefficients of 0.8 for all cases. We detect curvature in all correlations, with a gradual downward curvature for \Rtilde, a marked upward curvature for \tautilde, and a milder one for \avqtilde. 

These curvatures, together with the significant trends shown in Fig.~\ref{dust_evol} and the spatial coherence of the variations seen in Fig.~\ref{pow_opa_maps}, provide evidence that the dust opacity at 353\,GHz varies by fa actor of 2 to 4.6 and that the radiative emissivity of the grains (via \anh) varies by a factor of 1.5 to 2.9, according to the environmental conditions in the gas density and chemistry. 

We can readily recognize the cloud structures in the \anh and \opa gradients across the field, so the maps depict a gradual evolution of the dust properties across the complex, with gradients of comparable magnitude in several clouds. The rise is already perceptible in the DNM phase, even though the column-density range over which it is extracted largely overlaps that of the \hi. The rise steepens near or in the brightest CO cores. The higher angular resolution of the \nhlam map captures the large dynamic range in the compact regions of highest density, with a 4-fold contrast in opacity and a 3-fold contrast in \anh ratio between those peaks and the diffuse \hi. The opacity peaks are confirmed in the \nhgam map, with a lower contrast due to the lower angular resolution. The gradual evolution of the dust properties is responsible for the change in mean \anh and \opa values across the gas phases in Table~\ref{table_opaspw}. The values in the DNM phase may be driven up by the large column densities in excess of $10^{21}$\,\persqcm observed around the brightest CO cores of Cham I, II, and III. However, Fig.~\ref{dust_evol} indicates that these ratios start to rise as soon as \hd molecules dominate over H atoms in the gas column. 

The opacities measured at 353\,GHz in the Chamaeleon complex  are interestingly larger than the average trends previously found over the whole sky. In pure \hi, in the $(0.4-2.0)\,{\times}\,10^{21}$\,\persqcm range, we measure a mean (standard deviation) opacity of 14.3 (2.3)$\,{\times}\,$\opaunit for a radiated power of 4.5 (0.4)$\,{\times}\,$\spwunit. This is a factor of 2 larger than the value estimated in the high-latitude \hi cirrus clouds where the radiated power is 25\,\% lower than in the Chamaeleon complex \citep{planck13_pip82}. So, we confirm large opacity variations within the atomic phase at column densities below $10^{21}$\,\persqcm, away from the conditions that cause dust evolution in the cold molecular environments. This large change within the diffuse atomic phase challenges the dust evolution models. Figure~\ref{dust_evol} further shows a constant opacity near 
$15\,{\times}\,$\opaunit in the $(0.4-2.0)\,{\times}\,10^{21}$\,\persqcm interval in total \nh, whereas the all-sky average rises from 7.3 to $12\,{\times}\,$\opaunit over the same \nh range \citep{planck14_tau}. Augmenting the \hi optical depth correction at the $2\,\sigma$ confidence limit in spin temperature does not reconcile these measurements. The discrepancy extends to the CO phase, with peak opacities in the Chamaeleon clouds that are 3.5 times larger than the all-sky average over the same \nh range (see Fig. 21 of \citealt{planck14_tau}). Figure~\ref{dust_evol} shows that the opacity rise in the Chamaeleon complex  is also steeper than the \opa $\propto N_{\rm{Htot}}^{\hbox{\hglue 0.7pt}0.28\,{\pm}\,0.01\,{\pm}\,0.03}$ dependence found in the Orion clouds, even though it was measured to much larger gas column densities (using near-infrared stellar reddening, \citealt{roy13}). The \xco factor in the Chamaeleon clouds is not abnormally low compared to that in other nearby clouds (see Sect.~\ref{sect_xcopaspw}). Its value cannot cause a large overestimation of the grain opacities in the Chamaeleon molecular cores.

The dust specific powers vary only moderately, by less than a factor of 2, inside the DNM and CO clouds in Figs.~\ref{pow_opa_maps} and \ref{pow_opa_relmaps}. 
The power variations hardly relate to the gas structure, except for two notable trends. The first is seen toward the denser DNM filaments, where both the radiated power and opacity exceed the surrounding values by 30--50\,\%. The second is a power decline by a factor of 2 toward the bright CO clouds of Cham I, II, III, and East II, in regions where the dust temperature drops below 18\,K and the column density exceeds about $2\,{\times}\,10^{21}$\,\persqcm. Figure~\ref{dust_evol} shows that the power drop relates to the presence of molecular gas. The low-power regions extend well beyond the densest filaments with the largest opacities (see Fig.~\ref{pow_opa_maps}); they are detected with both \nhgam and \nhlam. 

We have found no explanation for the 30\,\% to 60\,\% larger powers measured in the north-eastern corner of the field, where the opacities are close to the average, but the grain temperatures exceed 20\,K. 
The IC emission map of Fig.~\ref{ISMCMP} gives the integral along sightlines of the CR interactions with the global Galactic ISRF. Its asymmetry in longitude at $b > -20\degr$ suggests an enhanced ISRF, thus an enhanced heating rate, toward the warm grains. Yet, the warm region extends beyond our analysis perimeter and its global spatial distribution does not follow the smooth distribution of the Galactic ISRF, nor any gas structure (see Fig. 9 of \citealt{planck14_tau}). The abrupt change in dust SEDs, with $\beta < 1.6$ in this zone, also warns us that the temperature excess may not relate to large-scale stellar distributions. 

The clear anti-correlation we see in Fig.~\ref{dust_evol} between the dust opacity and temperature confirms the early results obtained in the diffuse ISM \citep{planck11_cirrus} and in high-latitude cirrus clouds \citep{planck13_pip82}, now with the independent gas-tracing capability of the \g rays. It reveals, however, a more complex situation than the anticipated temperature response of the grains to an opacity change while exposed to a uniform heating rate. On the one hand, since \avq has been corrected to first order for an ISRF-related bias, the residual change in \anh revealed in Figs.~\ref{correl_gam_dust} and \ref{dust_evol} suggests an emissivity rise due to a structural or chemical evolution of the grains. On the other hand, the grains are less heated in the regions of low \spw in the CO phase. The maps show a variety of situations inside the clouds, but two notable trends emerge. The lowest temperatures near 15\,K correspond to regions of median power and highest opacity, so they may be due to the enhanced radiative cooling of the grains. Conversely, the regions of lowest power exhibit median opacities, so the low temperatures near 17\,K in these environments may primarily result from a reduced heating rate. 

The chemical composition of the DNM is likely to encompass varying fractions of optically thick \hi and CO-dark \hd along different lines of sight, as they approach the CO edges. Optically thick \hi has been proposed to explain the excesses of dust emission over the thin \nhi and \wco expectations, which would make it the dominant form of DNM \citep{fukui14a,fukui14b}. Thick \hi would not explain the \opa variations by a factor above 3 seen in the thin \hi cirrus clouds at \nhi column densities predominantly below a few $10^{20}$\,\persqcm \citep{planck13_pip82}. Nor would it explain the \taunu$/$\nhgam variations seen in \g rays across the \hi and DNM phases, independently of \hi depth corrections and \xco conversions. It would be difficult to invoke thick \hi to explain large cloud-to-cloud variations in \opa over the same \nh range in the different clouds. Dust evolution thus currently prevents the use of dust \textit{emission} to study optically thick \hi. The magnitude of the observed rise in \opa or of the ISRF-corrected \anh with \nh is such that it would systematically lead to a substantial overestimation of \nhi. 

\begin{figure*}
\sidecaption
\includegraphics[width=12cm]{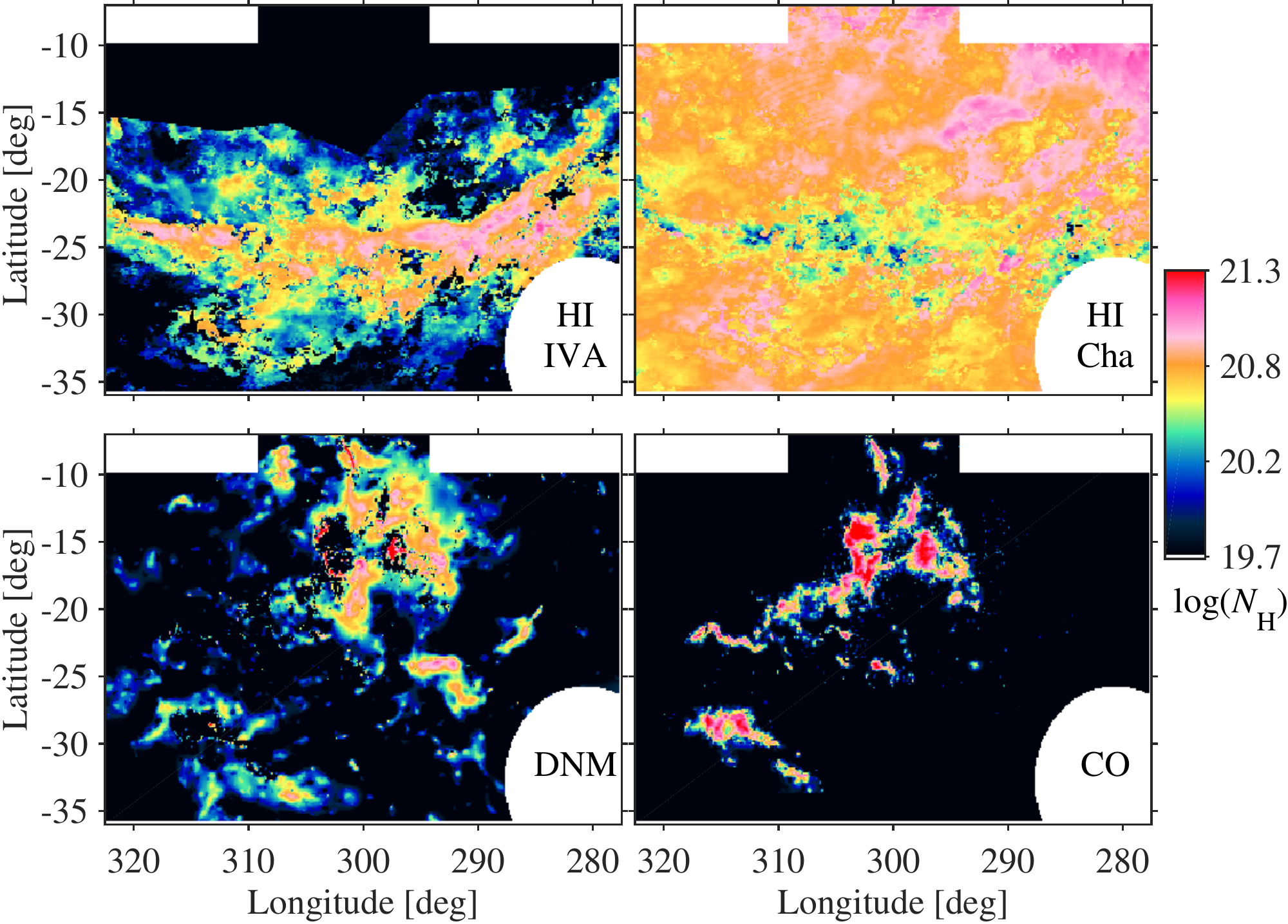}
\caption{\nh maps (in cm\msq) for the different gas phases in the analysis region: \hi in the intermediate velocity arc and, inside the Chamaeleon complex, gas in the \hi, DNM, and CO-bright phases. The maps are based on the \anaQ analysis, with the assumption of optically thin \hi, the value of \xcoG, and the dust DNM template calibrated in mass with the \g rays.}
\label{NH_all}
\end{figure*}

\section{Gas phase interfaces}
\label{sect_phase}

In this section we compare the gas column densities inferred in the different phases of the local clouds.

We note in the solid angle distributions of Fig.~\ref{NH_histo} that all three phases reach comparable peak column densities between 1 and 3 $\times10^{21}$\,\persqcm in these modest clouds. For the DNM, the histograms confirm the good correspondence in the three analyses between the column densities derived in \g rays and those derived from the dust and calibrated in mass with the \g rays.
The differences stem from the lower resolution in \g rays (see Fig.~\ref{NH_DNM}). Figure~\ref{NH_histo} stresses that the mean dust properties given for each phase in Table~\ref{table_opaspw} originate from broad and largely overlapping $N_{\rm{H}}$ ranges.

Figure~\ref{NH_all} compares the spatial distributions in each phase for the \anaQ analysis. The other analyses yield comparable maps. They illustrate how the DNM filamentary structure generally extends at the transition between the diffuse \hi and the compact CO cores of the Chamaeleon complex, but that the transition substantially varies inside the complex. For instance, the elongated, snaky CO filament that stretches along $b= -22\degr$ in Cha-East I is not surrounded by DNM gas, while the similar Musca CO filament, extending at low latitude along $l\,{=}\,301\degr$, is embedded in a rich DNM structure. The CO cloudlet at $l\,{=}\,301\degr$, $b\,{=}\,{-}24\degr$ is free of DNM while the DNM cloud near $l\,{=}\,307\degr$, $b\,{=}\,{-}10\degr$ is almost free of CO emission \citep[see the \Planck \texttt{TYPE\,3} CO map, ][]{planck14_co}. 

The bright DNM clouds at $l < 296\degr$ and $b < -20\degr$ are almost free of CO emission down to 1\,\wcounit. Whether they relate to the IVA would be worth investigating in the hope of finding a test case to explore the impact of unusual gas dynamics on the transition from atomic to  molecular gas.

Masses in the local Chamaeleon system have been derived for a common distance of 150\,pc and with the \g-ray DNM maps, despite their lower angular resolution, in order to be less sensitive to dust evolution. With the \anaQ analysis, we obtain masses of
44900 $^{+ 1600}_{-0}$ M$_{\sun}$ in the \hi, 
$9000\,{\pm}\,500 ^{+400}_{-0}$ M$_{\sun}$ in the DNM, and
$5000\,{\pm}\,200 ^{+  200}_{-0}$ M$_{\sun}$ in the CO-bright \hd. They
primarily reflect the compactness of the various phases. The DNM is the second most important contributor to the total mass because of its wide $N_{\rm{H}}$ range and large spatial extent. 
We obtain comparable masses within ${\pm}\,4$\,\% in the DNM and within $_{-5}^{+15}$\,\% in the CO phase with the other analyses. 

\begin{figure}
\includegraphics[width=\hsize]{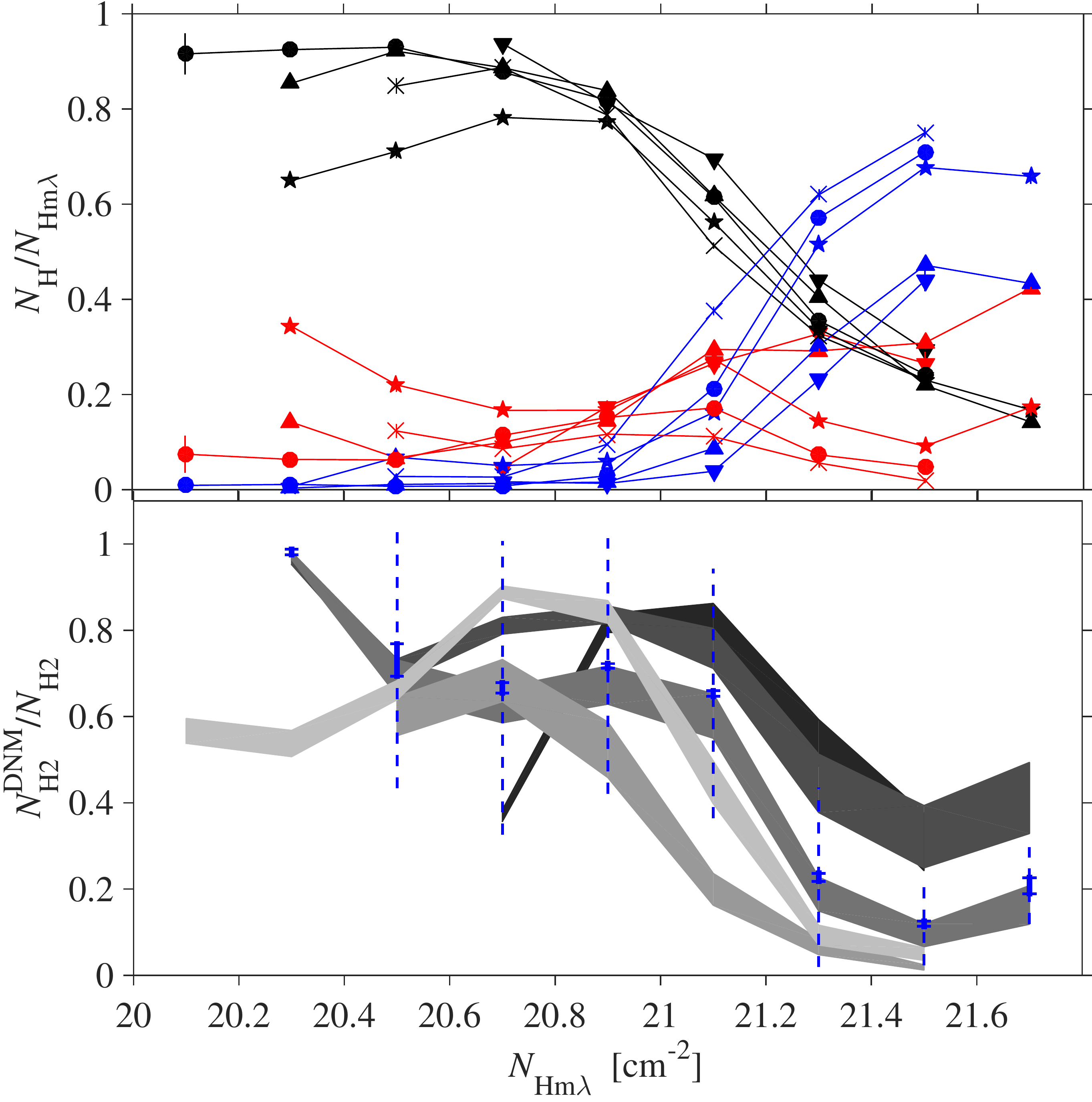}
\caption{Variations with \nhlam, in five separate clouds, of the fraction of the total column density in each phase (\textit{upper panel}, \hi in black, DNM in red, CO in blue) and of the fraction of the total \hd column density that is CO dark (\textit{lower panel}). Downward and upward triangles, stars, crosses, and circles refer to Musca, Cham I, Cham II+III, Cham East I, and Cham East II, respectively. The dark to light grey curves correspond to the clouds in the same order. The shaded areas cover the chosen uncertainty in DNM composition from 50\,\% to 100\,\% \hd. The dotted and solid error bars give the standard deviations and errors on the mean in each \nh bin; they are shown for only one cloud and the fully molecular DNM case.}
\label{NHfrac_H2dkfrac}
\end{figure}

We have studied the relative contributions and transitions between the different phases for the local Chamaeleon complex and for five of its sub-structures, 
as listed in Sect.~\ref{sec_clouds} and outlined in Fig.~\ref{NHI_WCO}. They have been chosen to lie outside the zone where \hi lines may overlap between the local and IVA components. To build the fractional curves of Fig.~\ref{NHfrac_H2dkfrac}, we have used the \anaQ analysis, the \xcoG factor, the \g-ray \nhdnm map, and a DNM composition between half and fully molecular. The other analyses yield similar curves. In the text below, we have translated \nh into visible extinction with the $N_{\rm H}/A_{V}$ ratio of $(2.15\,{\pm}\,0.14)\,{\times}\,10^{21}\,\rm{cm}^{-2}\,\rm{mag}^{-1}$ measured with the \textit{Far Ultraviolet Spectroscopic Explorer} \citep[\FUSE, ][]{rachford09}. To study the variations in DNM or dark-\hd fraction with $A_V$, we have directly used the $A_{V}$ map constructed from the 2MASS stellar data, with the median colour excess of the 49 stars nearest to each direction \citep{rowles09}. We have resampled the map into our analysis grid. The visual extinctions are quoted in magnitudes.

All the clouds exhibit similar fractional trends. The \hi fractions sharply decline beyond about $8\,{\times}\,10^{20}$\,\persqcm (or $A_V \sim 0.4$), in agreement with previous measurements, which indicate an \hi saturation at $(3-5)$, $(4-5)$, and $(8-14)\,{\times}\,10^{20}$\,\persqcm respectively with \FUSE observations of \hd \citep{gillmon06}, OH observations \citep{barriault10}, and dust observations in Perseus \citep{lee12}. Models indeed require \nhi near $10^{21}$\,\persqcm to shield \hd against UV dissociation for clouds of solar metallicity \citep{krumholz09,liszt14_H2}.  

The DNM is concentrated in the $10^{20}-10^{21}$\,\persqcm interval. It is systematically present before the onset of CO. The \hi-DNM transition occurs well into the translucent zone \citep{dishoeckblack88}, yet at different thresholds ranging from 2 to $8\,{\times}\,10^{20}$\,\persqcm (or 0.09 to 0.4 in $A_V$) for the different clouds. The transition to CO appears to be more stable, with all five clouds requiring about $1.5\,{\times}\,10^{21}$\,\persqcm (or $A_V \simeq 0.7$) to efficiently shield CO against photodissociation in the local ISRF. The transition cannot be attributed to the sensitivity threshold of the CO survey. It is consistent with the $A_V \simeq 0.5$ transition noted for the formation of the OH molecule that is one of the precursors of CO in the chemical evolution \citep{barriault10}.

We have not noted any obvious difference in \hi column densities or line widths that could relate to the variable DNM onset at low \nh. However, simultaneous \hi and \hd observations with \FUSE have suggested a rather complex and variable \hi--to--\hd interface, with $2N_{\rm{H}_2}/(N_{\ion{H}{i}}+2N_{\rm{H}_2})$ rapidly fluctuating between 0.1 and 0.7 up to $A_V \simeq 2.6$ \citep{rachford09}. Figure~\ref{NH_all} exhibits rather steep CO edges that occur in an \nh regime where an equilibrium set of reactions involving C$^+$ and OH leads to CO formation \citep{sheffer08}. So the DNM variations at lower \nh may conversely reflect the non-equilibrium chemistry that prevails in more diffuse gas. They may also reflect local variations in the \hi optical thickness that could not be taken into account in our analyses.

\begin{figure}
\includegraphics[width=\hsize]{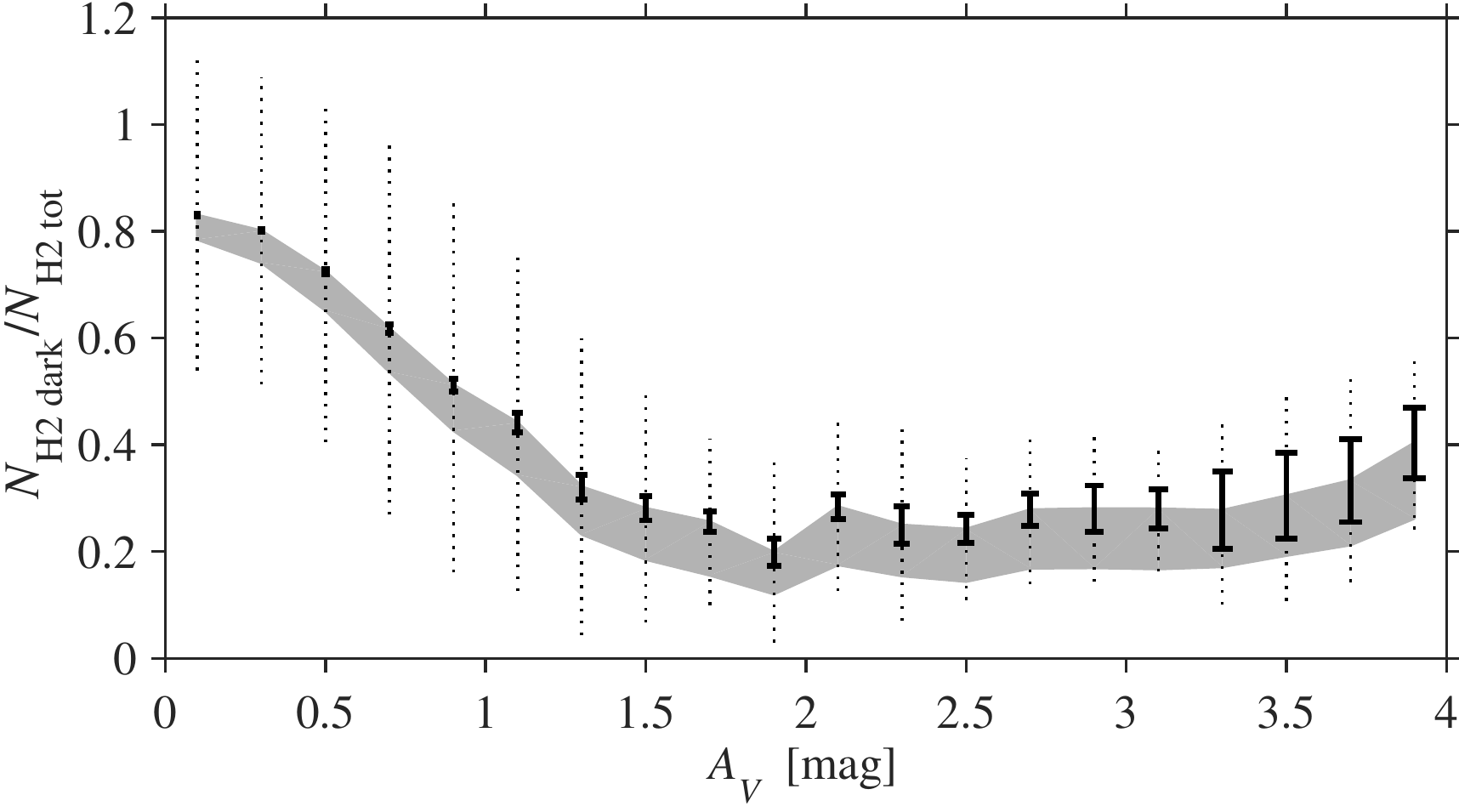}
\caption{Evolution of the CO-dark to total \hd column-density ratio with the 2MASS visual extinction. The molecular fraction in the DNM varies from 50\,\% (lower grey bound) to 100\,\% (upper bound). The gas column densities in the CO and DNM phases have been obtained with the \xcoGQ factor and the \g-ray \nhdnm map from the \anaQ analysis. The other analyses yield equivalent curves. The thick and dotted lines respectively give the standard error of the mean and the standard deviation of the ratios for each $A_V$ bin, in the case of a fully molecular DNM.}
\label{H2dkfrac_Av}
\end{figure}

We have followed the variation of the CO-dark to total \hd ratio in column density, assuming that the DNM consists of 50\,\% or 100\,\% molecular hydrogen. Figure~\ref{NHfrac_H2dkfrac} shows the trend with \nh in the individual clouds and Fig.~\ref{H2dkfrac_Av} shows the average evolution with $A_V$ over the whole complex of local and IVA clouds. To compute the latter, we have subtracted the extinction associated with the Galactic background \nhi using the $N_{\rm{H}}/A_V$ ratio of \FUSE. We obtain substantially the same profiles with the \anaT and \anaR analyses. 
As expected, the dark-\hd fraction steeply rises to 80\,\% in regions heavily exposed to the ISRF and the CO-bright phase rapidly takes over the molecular fraction once CO is fully shielded, at $A_V > 1.2$. This is deeper into the clouds than the theoretical prediction of $A_V \gtrsim 0.5-0.7$ for optically thick CO in a $10^5\,\rm{M}_{\sun}$ cloud with an incident UV flux extrapolated to the local ISRF (see Fig. 7 of \citealt{wolfire10}). The DNM contributes more than half of the molecular gas up to $A_V \simeq 0.9$. It retains 10--30\,\% of the molecular column densities to high $A_V$ as the lines of sight intersect envelopes of the CO-bright clouds. This is consistent with the theoretical finding that 20\,\% of \hd is not traced by CO even at the density peak \citep{levrier12}.

The fractions of CO-dark to total \hd in mass, $f_{\rm{dark\,H_2}}$, are listed for each cloud in Table~\ref{table_fDG} for the two choices of DNM composition. The fractions indicate there is often as much molecular mass in the inconspicuous DNM as in the CO-bright cores. They also often exceed the 32\,\% prediction for CO cloudlets exposed to the local ISRF \citep{levrier12}, or the 25\,\% prediction based on the PDR modelling of the outer layers of a $10^5\,\rm{M}_{\sun}$ spherical cloud if we extrapolate its illumination to the local ISRF \citep{wolfire10}. The latter model suggests that the extinction difference, $\Delta A_V$, between the \hi-\hd and the \hd-CO transitions is a weak function of the outside UV flux. However, as the mean extinction, $\overline{A_V}$, through the cloud measures the total molecular mass, the dark-\hd mass fraction should decline with increasing $\overline{A_V}$. It is difficult to transpose the mean extinction in the homogenous, spherical, and giant ($10^6\,\rm{M}_{\sun}$) cloud of the model with the average extinction in the observations. To help the comparison, we have taken averages, \avavgco, within the well-defined CO edges at \wco $>1$\,\wcounit. The model predicts $f_{\rm{dark\,H_2}} > 70$\,\% for the $0.4 \lesssim \overline{A_V} \lesssim 0.9$ range of the present clouds. We find lower fractions in the observations in Table~\ref{table_fDG}. The model also predicts a strong decline in $f_{\rm{dark\,H_2}}$ with increasing $\overline{A_V}$, whereas the data in Table~\ref{table_fDG} hint at an opposite trend, with a Pearson's correlation coefficient of 0.81 for a rise. 

\begin{table*}[htbp]
\begingroup
\newdimen\tblskip \tblskip=5pt
\caption{
For the whole local complex and for separate substructures: total gas
mass; fractions of the total mass in the three gas phases; mean and peak
extinctions within the CO boundaries; and mass fractions of the \hd that is
CO dark.}             
\label{table_fDG}      
\footnotesize
\setbox\tablebox=\vbox{
 \newdimen\digitwidth
 \setbox0=\hbox{\rm 0}
 \digitwidth=\wd0
 \catcode`*=\active
 \def*{\kern\digitwidth}
 \newdimen\signwidth
 \setbox0=\hbox{+}
 \signwidth=\wd0
 \catcode`!=\active
 \def!{\kern\signwidth}
 \halign{\tabskip=0pt\hbox to 1.5in{#\leaderfil}\tabskip=2em&
 \hfil#\hfil\tabskip=2em&
 \hfil#\hfil\tabskip=2em&
 \hfil#\hfil\tabskip=2em&
 \hfil#\hfil\tabskip=0pt\cr
\noalign{\doubleline}
\omit\hfil Structure\hfil& $M_{\rm{tot}}$& $f_{\ion{H}{i}}^{{\rm a}}$& ${f_{\rm{DNM}}}$&
 $f_{\rm{H_2}\,CO}$\cr
\omit& [M$_{\sun}$]& & &\cr
\noalign{\vskip 4pt\hrule\vskip 6pt}
Cham complex& $58900\pm500$& $0.76\pm0.01$& $0.15\pm0.01$& $0.08\pm0.01$\cr
       Musca& $*5100\pm100$& $0.70\pm0.01$& $0.24\pm0.01$& $0.06\pm0.01$\cr
       Cha I& $10200\pm100$& $0.64\pm0.01$& $0.23\pm0.01$& $0.12\pm0.01$\cr
  Cha II+III& $*7700\pm100$& $0.57\pm0.01$& $0.19\pm0.01$& $0.24\pm0.01$\cr
  Cha East I& $*3100\pm100$& $0.74\pm0.01$& $0.10\pm0.01$& $0.15\pm0.01$\cr
 Cha East II& $10100\pm100$& $0.78\pm0.01$& $0.13\pm0.01$& $0.08\pm0.01$\cr
\noalign{\vskip 4pt\hrule\vskip 6pt}
\omit& $\overline{A_V}$& $A_V^{\rm{max}}$& ${f_{\rm{dark\,H_2}}}^{\rm{b}}$&
 ${f_{\rm{dark\,H_2}}}^{\rm{c}}$\cr
\omit& [mag]& [mag]& &\cr
\noalign{\vskip 4pt\hrule\vskip 6pt}
Cham complex& 0.7& 7.1& $0.48\pm0.03$& $0.64\pm0.04$\cr
       Musca& 0.9& 4.3& $0.68\pm0.05$& $0.81\pm0.06$\cr
       Cha I& 0.8& 7.1& $0.48\pm0.03$& $0.65\pm0.04$\cr
  Cha II+III& 0.7& 6.5& $0.29\pm0.02$& $0.45\pm0.03$\cr
  Cha East I& 0.4& 2.2& $0.25\pm0.02$& $0.40\pm0.03$\cr
 Cha East II& 0.6& 2.9& $0.45\pm0.03$& $0.62\pm0.04$\cr
\noalign{\vskip 4pt\hrule\vskip 6pt}
}}
\endPlancktablewide
\tablenote {{\rm a}} For optically thin \hi. \par
\tablenote {{\rm b}} For a half molecular DNM.\par
\tablenote {{\rm c}} For a fully molecular DNM.\par
\endgroup
\end{table*}
\begin{figure}
\includegraphics[width=\hsize]{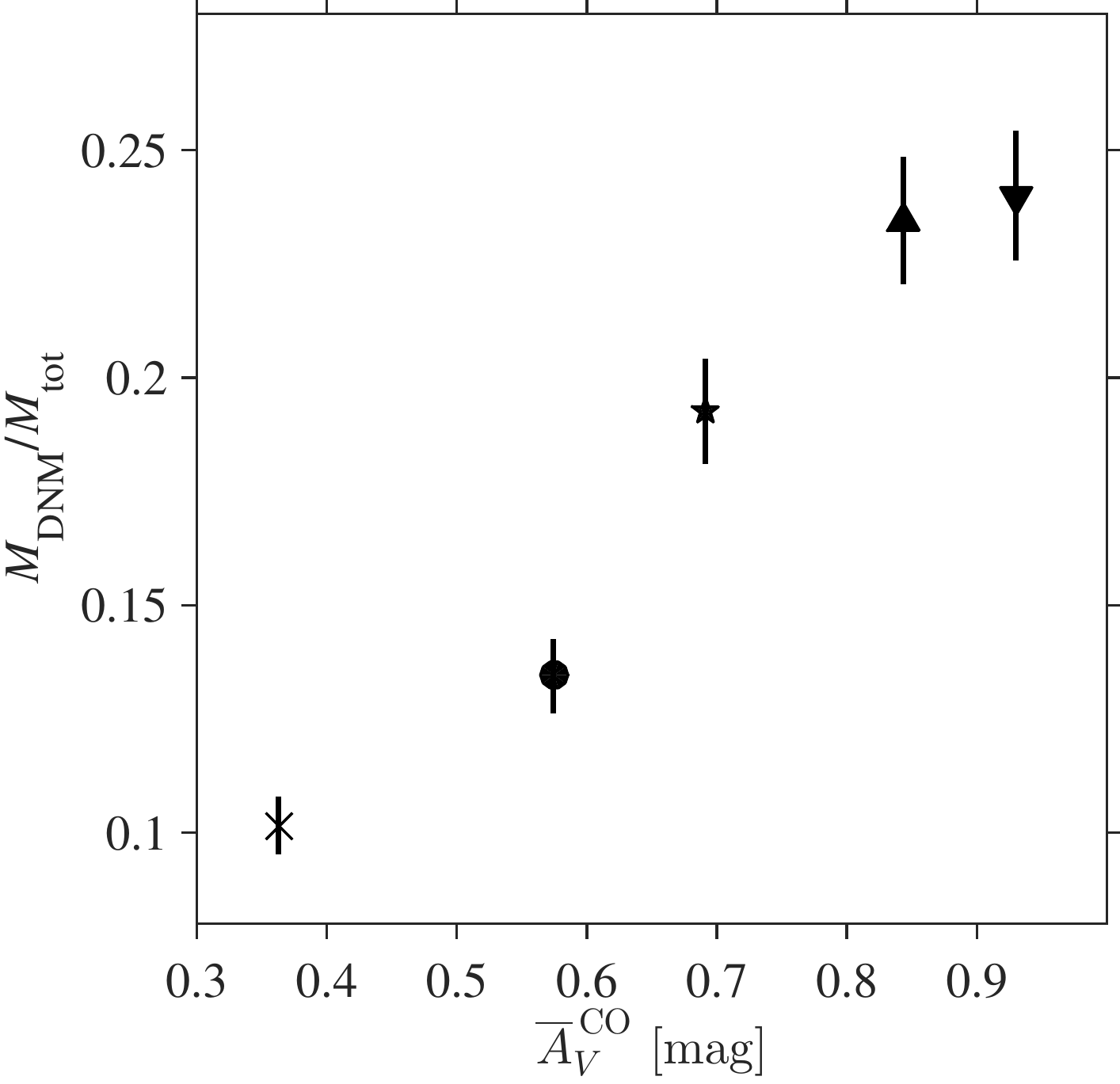}
\caption{Evolution of the DNM mass fraction with the average 2MASS extinction measured within the CO boundaries of each cloud. Downward and upward triangles, stars, crosses, and circles refer to Musca, Cham I, Cham II+III, Cham East I, and Cham East II, respectively.}
\label{fDNM}
\end{figure}

The DNM contains 10\,\% to 24\,\% of the total gas mass in individual clouds as well as in the whole complex (see $f_{\rm DNM}$ in Table~\ref{table_fDG}). These mass fractions do not depend on the DNM chemical composition. The fraction for the whole complex is consistent with the early estimate based on more limited, but independent data (EGRET in \g rays and DIRBE-\IRAS for the dust; \citealt{grenier05}). It is, however, lower than the mean ($43\,{\pm}\,18$)\,\% fraction derived from C$^+$ lines in CO clouds of the Galactic disc for a comparable sensitivity in $^{12}$CO \citep{langer14}. The early \g-ray analyses suggested that $f_{\rm DNM}$ decreases as the mass locked in the CO phase grows. The recent C$^+$ line results also indicate reduced fractions of 18\,\% toward dense $^{13}$CO clouds and 13\,\% toward the denser C$^{18}$O cores \citep{langer14}. In the Chamaeleon complex, we find no correlation between $f_{\rm DNM}$ and any of the cloud masses (correlation coefficients of 0.1, 0.2, and 0.3, respectively for the \hi, CO-bright, and total-cloud mass). Figure~\ref{fDNM}, however, exhibits a correlation, with a Pearson's coefficient of 0.98, between $f_{\rm DNM}$ and the \avavgco extinction averaged inside the CO contours. We stress that the mass fractions and extinctions are based on independent data. Whereas a decline of $f_{\rm DNM}$ or $f_{\rm{dark\,H_2}}$ with increasing $\overline{A_ V}$ can be explained by a shorter screening length to protect the CO phase against photo-dissociation, a rising trend with \avavgco is unexpected. It is confirmed for both $f_{\rm DNM}$ and $f_{\rm{dark\,H_2}}$ in the \anaT analysis. 
Five clouds constitute too scarce a sample to claim a definite increase in DNM abundance with \avavgco, but the correlation in Fig.~\ref{fDNM} calls for the observation of other test cases in more massive molecular complexes, and the discrepancy with theory calls for more detailed predictions for medium-size clouds and more realistic geometries. 

\section{Conclusions and perspectives}
We have explored the gas, dust, and cosmic-ray content of several clouds in the local Chamaeleon complex and in an intermediate-velocity \hi arc crossing the field. We have conducted three parallel analyses, coupling the \hi, CO, and \g-ray data with different dust tracers, namely the optical depth at 353\,GHz, the radiance, and the $U_{\rm min}$-corrected \avq extinction. Jackknife tests have verified that the uniform set of parameters of the \g-ray and dust models  apply statistically to the whole region under analysis. 
We find that the \avq map, which includes a linear correction for the ISRF strength in the Draine \& Li (2007) model, provides the best fit to the interstellar \g rays. Yet, the corrected extinction still rises significantly above the linear \g-ray expectation as the gas becomes denser. We find an even more pronounced upward curvature in dust optical depth with increasing interstellar \g-ray intensity, and conversely a saturation in dust radiance. 

We summarize the main results for each topic as follows.
\begin{itemize}

\item \textbf{On cosmic rays:} at the precision level of the current \g-ray data, the CR spectrum is shown to be uniform across the gas phases and to follow the energy distribution of the local-ISM average. The \g-ray emissivity per nucleon in the Chamaeleon complex  is equivalent to other measurements in the solar neighbourhood. It exceeds the average emissivity found along the solar circle in the Galaxy by only $(22\,{\pm}\,5)$\,\%.
We find no spectral signature of non-uniform CR penetration to the denser molecular cores traced in $^{12}$CO. We provide a first measurement of the \g-ray emissivity in an intermediate-velocity cloud. It is 20\,\% lower than in the Chamaeleon clouds at all energies.
We need further measurements in similarly sheared clouds and a distance estimate to the IVA to  assess whether the small change in CR flux is due to a larger altitude above the Galactic plane or to the unusual dynamical state of the clouds. 

\item \textbf{On the DNM:} the \g-ray flux and dust tracers reveal large amounts of DNM gas with comparable spatial distributions and mass columns at the interface between the \hi-bright and the CO-bright phases of the local complex. We have combined the dust and \g-ray analyses to build reliable DNM templates and to reduce the bias on the \hi- and CO-related parameters due to the DNM presence. With the equivalent of a fifth of the \hi mass and nearly twice the CO-bright mass, the inconspicuous DNM appears as a major constituent of the complex. Its spatial extent is intermediate between the diffuse \hi and compact CO phases. It dominates the molecular column densities up to $A_V \simeq 0.9$. 

\item \textbf{On dust evolution:} we provide average dust properties per gas nucleon (\anh ratio, \opa opacity, and \spw specific power) in the different gas phases and we follow their spatial variations across the clouds by means of two separate \nh maps. The lower resolution one, \nhgam, is inferred from the \g-ray data. It traces the total gas without any assumption on the in situ, non-uniform \hi opacities, \xco conversions, and DNM extraction. It relies only on the uniform CR flux. The higher resolution map, \nhlam, uses the radio data and \hi, DNM, and CO decomposition with the \anhavgdnm and \xcoG factors measured in \g rays. 
Both means provide evidence for a 2 to 4.6-fold rise in \opa and a more limited 1.5 to 2.9-fold rise in \anh over a single decade in \nh. The dust emissivity is seen to gradually evolve from the diffuse \hi to the modest CO cores of the Chamaeleon complex. This variation cannot be attributed to changes in the heating rate of the grains, since we find little variation in specific power in the \hi and DNM phases and a 2-fold decline in the dense CO clouds. These results confirm and extend into the DNM and CO-bright phases the earlier indications of opacity changes found in the atomic gas \citep{planck14_tau,planck13_pip82}. They confirm with independent radio and \g-ray data the variations suggested by the comparison of dust emission and reddening \citep{martin12,roy13}. These variations appear to be intimately linked to the gas structure as the density and molecular fractions grow. They presumably reflect a chemical or structural evolution of the grains.
Their magnitude severely limits the use of dust emission to trace the total gas to \nh $< 2\,{\times}\,10^{21}$\,\persqcm in the Chamaeleon clouds. We also find that the dust grains in the Chamaeleon complex  radiate 2 to 3 times more per unit mass, or are 2 to 3 times more numerous per gas nucleon, than on average over the whole sky. Within the \hi gas, the origin of the two-fold increase in opacity for a 25\,\% higher power compared to the high-latitude \hi cirrus clouds also requires elucidation. Cloud-to-cloud variations of this magnitude further limit the gas-tracing capability of the thermal dust emission until we understand their cause. The ISRF-related correction applied to \avq partially, but not completely, alleviates the upward curvature in dust emissivity. We may be witnessing structural/chemical evolution of the dust grains.

\item \textbf{On \xco:} we provide mean \xco conversion factors in the local clouds. The results elucidate a recurrent disparity between earlier \g-ray and dust calibrations of this factor. The disparity likely finds its origin in the pronounced \opa rise in the molecular clouds, a rise that induces a significant upward bias, by a factor of 1.9, on \xcoT compared to \xcoG or \xcoR. 
Further assessing the magnitude of the bias in more massive, but well resolved, molecular complexes has important implications for extragalactic studies that compare dust emission and CO observations to infer star formation efficiencies. 
The preferred \g-ray calibration of \xco $\simeq 0.7$ \xcounit in the Chamaeleon clouds agrees with other estimates in the solar neighbourhood. It stresses, however, an unexplained discrepancy, by a factor of 2, between the measurements in nearby clouds (at parsec scales and with limited spatial confusion between the gas phases) and the averages obtained at a kiloparsec scale in spiral arms (in particular in the Local Arm for the same metallicity and UV flux as in the local ISM). Larger photon statistics are required  above a GeV in order to investigate how the sampling resolution and cross-talk between gas phases affect the \xco calibration beyond the solar neighbourhood.
 
\item \textbf{On phase transitions:} we have explored how each phase contributes to the total gas column density and to the total mass in five separate clouds. They all show a marked decline in \hi fraction around $8\,{\times}\,10^{20}$\,\persqcm 
and an onset of CO near $1.5\,{\times}\,10^{21}$\,\persqcm. 
The DNM retains 10--30\,\% of the gas column densities to large extinctions because the lines of sight intercept DNM-rich envelopes around the CO-bright interiors. 
We find that the \hi-DNM transition varies from cloud to cloud across the $(2-8)\,{\times}\,10^{20}$\,\persqcm range, without an obvious explanation in \hi intensities or kinematics. The CO-dark to CO-bright \hd mass fraction often exceeds 50\,\% in the clouds of the complex, as predicted by theory for rather translucent clouds. The DNM contributions to the total cloud masses are low (10\,\%--24\,\%) and they surprisingly scale with the stellar extinction averaged within the boundaries of the CO phase. This trend needs confirmation in a larger sample.

\end{itemize}

Short-term plans include a thorough investigation of the non-linear rise of the dust emission coefficient per unit mass to higher \nh values, together with measurements of the ratio of the emission to absorption cross-sections of the grains as a function of \nh. The present analyses illustrate the potential of confronting \g-ray, dust, and radio tracers to gauge the amount of gas in the DNM and CO phases in order to follow the dust evolution. We have started to exploit the \Fermi LAT, \Planck, and stellar data toward these goals in the case of more massive, well resolved clouds.

Future prospects also include a study of the variable DNM abundance and its relation to the total cloud mass and \avavgco. The pronounced dust evolution prevents the use of the thermal emission of the grains to gauge the amount of optically thick \hi gas in the complex \hi--to--\hd transition \citep{fukui14a,fukui14b}. In the absence of extensive \hi absorption measurements, OH and CH surveys of the DNM interface can help uncover the cause of the variable DNM abundance. 
UV spectroscopy shows that the CH-to-\hd abundance is constant over two decades in \nhd ($10^{19.5-21.5}$\,\persqcm) in the translucent regime. 
OH is widespread at $A_V > 0.5$ and it rarely correlates with the surveyed CO \citep{barriault10,allen12}. CH and OH surveys 
can therefore provide key information on the DNM composition and why its relative mass varies from cloud to cloud. Constraining the latter is essential to extrapolate the local DNM abundances to Galaxy-wide values. Early \g-ray estimates suggested a Galactic DNM mass as large as in the CO-bright phase \citep{grenier05}. In dust emission, the DNM contribution to \nh ranges from 10\,\% in the outer Galaxy to 60\,\% in the inner regions \citep{planck11_3D}. In C$^+$ line emission, the DNM amounts to 30\,\% of the Galactic molecular mass and it is as massive as the cold \hi (CNM) and CO-bright phases in the outer Galaxy \citep{pineda13}. However, the limited angular resolution of the \g-ray data, the changes in the dust radiative properties across gas phases, and the difficult separation of C$^+$ from the CNM, DNM, and ionized regions, all hamper our ability to reliably measure the DNM mass to large distances. Adding the kinematical information of the CH and OH lines opens promising avenues.

\begin{acknowledgements}
The development of \Planck\ has been supported by: ESA; CNES and
CNRS/INSU-IN2P3-INP (France); ASI, CNR, and INAF (Italy); NASA and DoE
(USA); STFC and UKSA (UK); CSIC, MICINN, JA, and RES (Spain); Tekes,
AoF, and CSC (Finland); DLR and MPG (Germany); CSA (Canada); DTU Space
(Denmark); SER/SSO (Switzerland); RCN (Norway); SFI (Ireland);
FCT/MCTES (Portugal); and PRACE (EU). A description of the Planck
Collaboration and a list of its members, including the technical or
scientific activities in which they have been involved, can be found
at
\url{http://www.sciops.esa.int/index.php?project=planck&page=Planck_Collaboration}.
     
The \Fermi LAT Collaboration acknowledges generous ongoing support
from a number of agencies and institutes that have supported both the
development and the operation of the LAT as well as scientific data analysis.
These include the National Aeronautics and Space Administration and the
Department of Energy in the United States, the Commissariat \`a l'Energie Atomique
and the Centre National de la Recherche Scientifique / Institut National de Physique
Nucl\'eaire et de Physique des Particules in France, the Agenzia Spaziale Italiana
and the Istituto Nazionale di Fisica Nucleare in Italy, the Ministry of Education,
Culture, Sports, Science and Technology (MEXT), High Energy Accelerator Research
Organization (KEK) and Japan Aerospace Exploration Agency (JAXA) in Japan, and
the K.~A.~Wallenberg Foundation, the Swedish Research Council and the
Swedish National Space Board in Sweden. Additional support for science analysis during the operations phase is gratefully acknowledged from the Istituto Nazionale di Astrofisica in Italy and the Centre National d'\'Etudes Spatiales in France.

Support from the Institut Universitaire de France is acknowledged.
\end{acknowledgements}

\begin{appendix}

\section{\hi component separation}
\label{sect_HIcomp}
\begin{figure*}
\includegraphics[width=\hsize]{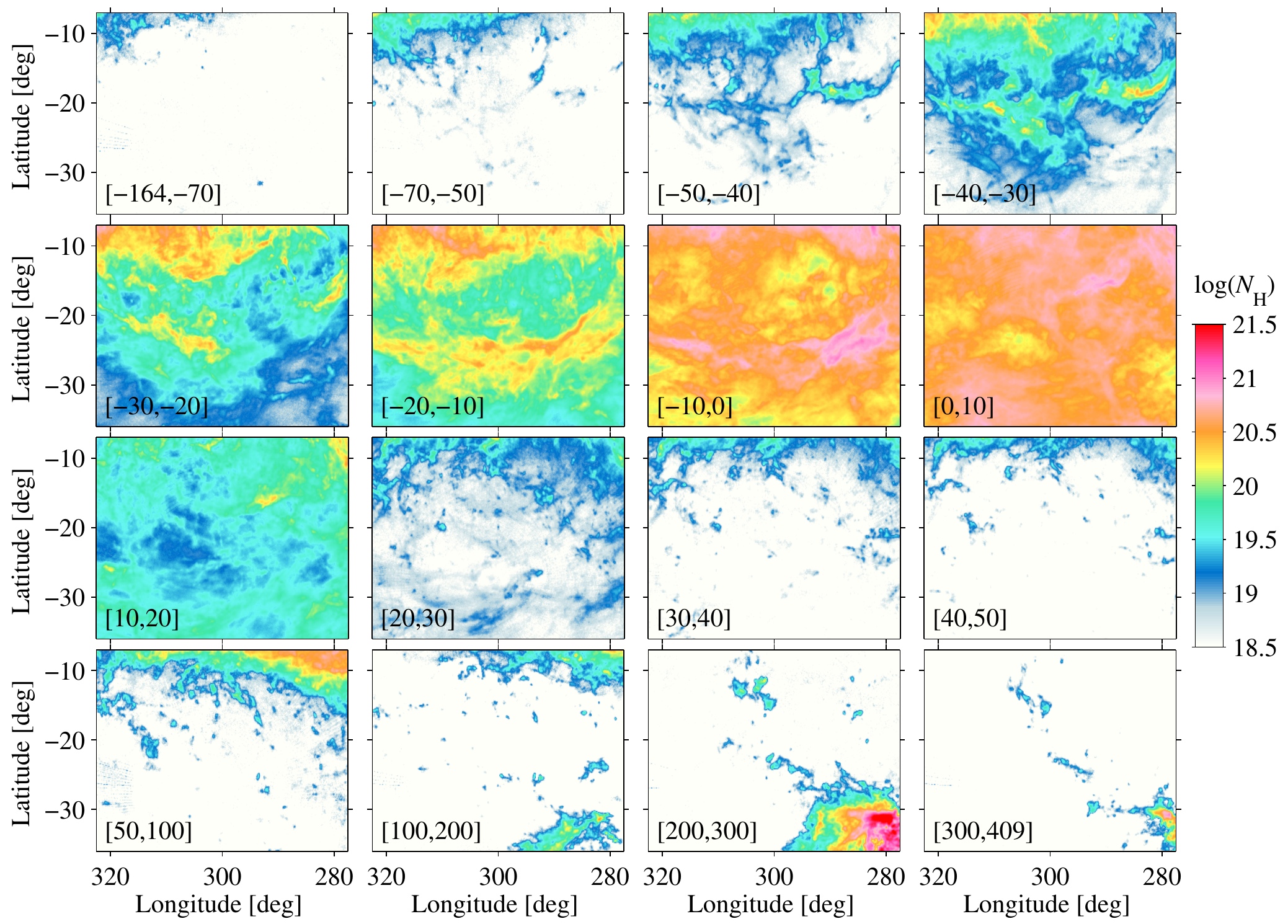}
\caption{Maps 
of the \nhi column densities (in cm$^{-2}$) obtained at 14\farcm5, resolution from the GASS survey for optically thin emission and integrated over contiguous velocity intervals between the values given in \kmpers on the lower left corner of each map.}
\label{NHI_slices}
\end{figure*}

We have developed a careful kinematical separation of the four \hi structures present within the analysis region, namely: 
   \begin{itemize}
      \item the \hi gas associated with the star-forming CO clouds of the Chamaeleon complex  at velocities between about $-4$ and $+15$\,\kmpers;
      \item the gas in an intermediate velocity arc (IVA), crossing the whole region around $-25\degr$ in latitude, at negative velocities down to  $-40$\,\kmpers; 
      \item the background \hi lying at high altitude above the Galactic disc, at velocities below about $+170$\,\kmpers;
      \item gas in the LMC and its tidal tails at the highest velocities.
   \end{itemize}
These components are visible in Fig.~\ref{NHI_slices}.

Because of the broad widths of \hi lines, the procedure aims to correct the velocity spillover from one component to the next in the column-density derivation. To do so for each direction in the sky, we have: detected all significant lines in the measured \hi spectrum; fitted them with pseudo-Voigt profiles; used the line centroids in velocity to distribute the lines between the components; and integrated each line profile to add its contribution to the column-density map of the relevant component.

The core of the method is based on fitting every spectrum in brightness temperature, $T_{\rm B}(v)$, with a sum of lines across the whole $-164.1 \leq v \leq 409$\,\kmpers velocity interval exhibiting significant line emission. The first step is to find the number and approximate velocity of the lines to be fitted. For each $(l,b)$ pixel direction, we have smoothed the spectrum in velocity with a Gaussian kernel of 1.48\,\kmpers. We have measured the rms dispersion in temperature, $T_{\rm rms}$, outside the bands with significant emission. We have clipped the data to zero outside regions with $T(v) > T_{\rm rms}$ in 3 adjacent channels. The clipping limits the number of fake line detections triggered by strong noise fluctuations in bands devoid of emission. We have then computed the curvature $d^2T_{\rm B}/dv^2$ in each channel by using the 5-point-Lagrangian differentiation twice. We have detected the line peaks and shoulders by finding the negative minima in $d^2T_{\rm B}/dv^2$. We have eliminated peak detections caused by the edges of the clipped bands. An average line FWHM of 8.2\,\kmpers was found in the line fits. We have merged potential lines when separated by less than one half width at half maximum (HWHM) in velocity. Weak lines ($T_{\rm peak} <$ 3\,K) were also merged when closer than one FWHM in order to limit the number of fake detections on noise fluctuations in faint line wings. When merging lines, the new velocity centroid was set to the average between the parent velocities. The final number of detected peaks and shoulders and their velocities have been used as input parameters for fitting multiple lines across the spectrum. 

We have fitted each spectrum with a sum of pseudo-Voigt line profiles, one for each detected peak or shoulder. Such a profile combines a Gaussian and a Lorentzian with the same velocity centroid, width, and height, and a relative weight that spans the interval from 0 (pure Gaussian) to 1 (pure Lorentzian). The velocity centroid was allowed to move within ${\pm}\,3.3$\,\kmpers (${\pm}\,4$ channels) around the original peak velocity. We have manually checked the precision of many fits across the region. All fits were checked to yield a total line intensity to better than 80--90\,\% of the data integral over the whole spectrum. 

In order to preserve the total \hi intensity observed in each direction, the small residuals (positive and negative) between the observed and modelled spectra have been distributed between the lines in proportion to the height of each line at each velocity. The line profiles have been corrected accordingly and integrated for a given choice of spin temperature to correct the resulting column-density for the \hi optical depth.

\begin{figure}[h!]
\includegraphics[width=\hsize]{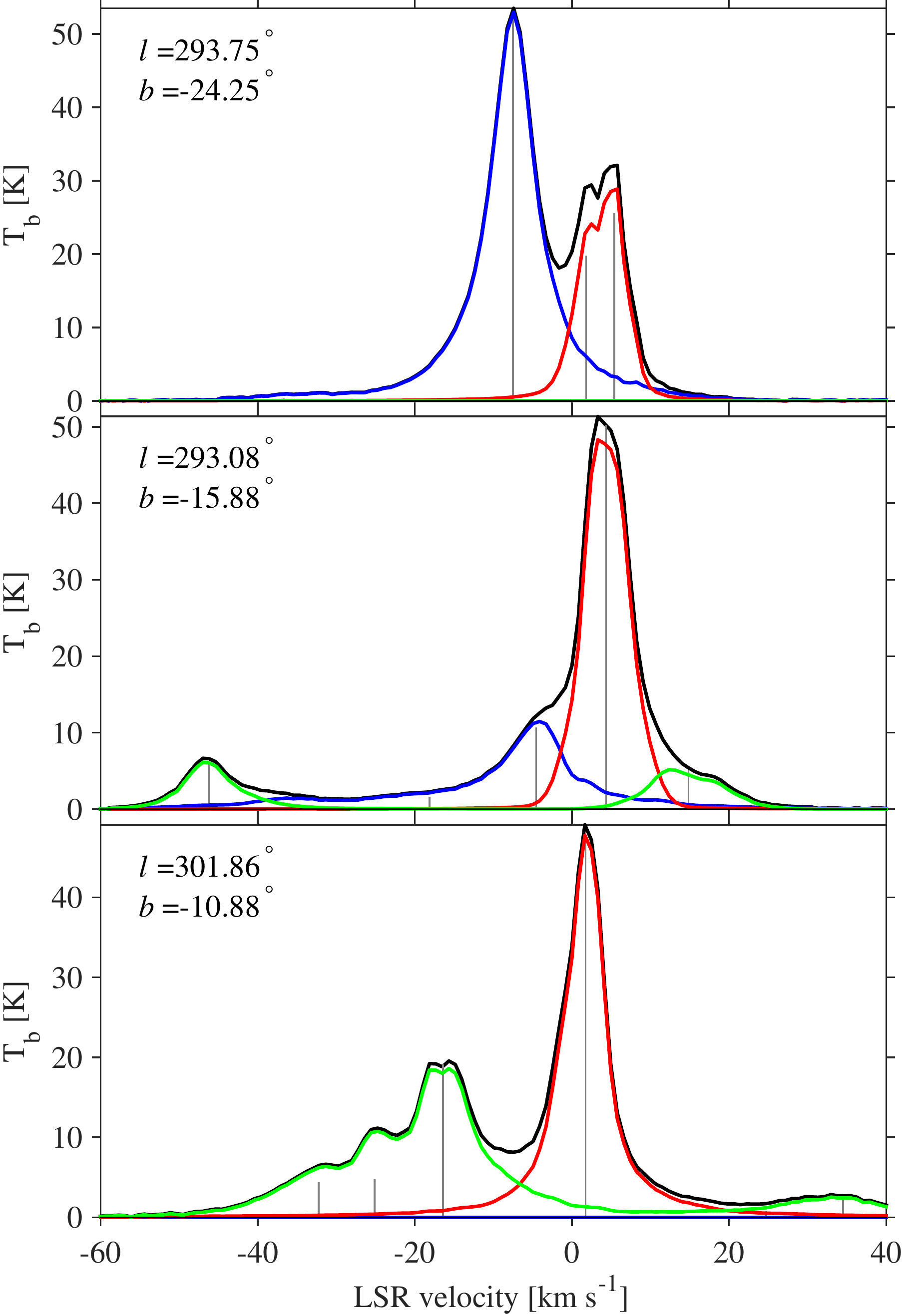}
\caption{\hi GASS spectra, in brightness temperature, measured toward the three directions indicated on the plots. The best line fits attributed to the local (red), IVA (blue), and Galactic (green) components sum up as the black curve, which closely follows the original data. The grey lines indicate the central velocity and temperature of the individual lines that contribute to a component.}
\label{HI_spec}
\end{figure}

Each line was attributed to one of the four components according to its velocity centroid in the $(l,b)$ direction. The spatial separation between the IVA and Galactic disc components runs along a broken line of minimum intensity starting at $b\,{=}\,{-}12\degr$ at low longitude, reaching $b\,{=}\,{-}18\fdg5$ at $l\,{=}\,300\degr$, and moving up to $b \simeq -16\degr$ at the highest longitude; the frontier can be seen in Fig.~\ref{NHI_WCO}. Lines pertaining to the local Chamaeleon clouds were selected at $-4$\,\kmpers $\leq v < 14.8$\,\kmpers for latitudes below this curve, and at $-7.4$\,\kmpers $\leq v < 14.8$\,\kmpers above the border. The lines of the IVA cloud were selected in the $-40$\,\kmpers $\leq v < -4$\,\kmpers interval at latitudes below the curve. The latitude cut between the Galactic disc and LMC components runs linearly from $(l,b)=(273\fdg4,-13\fdg75)$ to $(326\fdg6,-8\fdg5)$. The LMC lines were selected at $v \geq 131.9$\,\kmpers below the latitude cut and at $v \geq 174$\,\kmpers for all latitudes. The rest of the data was attributed to the Galactic disc component. A selection of spectra is given in Fig.~\ref{HI_spec} to illustrate the separation of the local, IVA, and Galactic components.

We have summed the column densities derived for each line associated with a component to produce the maps of Fig.~\ref{NHI_WCO}.  We have tested different velocity cuts around $-4$\,\kmpers between the local and IVA clouds. A change of 1 or 2\,\kmpers changes the related maps by 3 to 6\,\% in total mass, mostly in the overlap region near $l\,{=}\,283\degr$ and $b\,{=}\,{-}25\degr$. The overall structure and contrast in each map, which drive the correlation studies, remain largely unchanged.  

\section{\hi spin temperature}
\label{sect_spin}

\begin{figure}[h]
\includegraphics[width=\hsize]{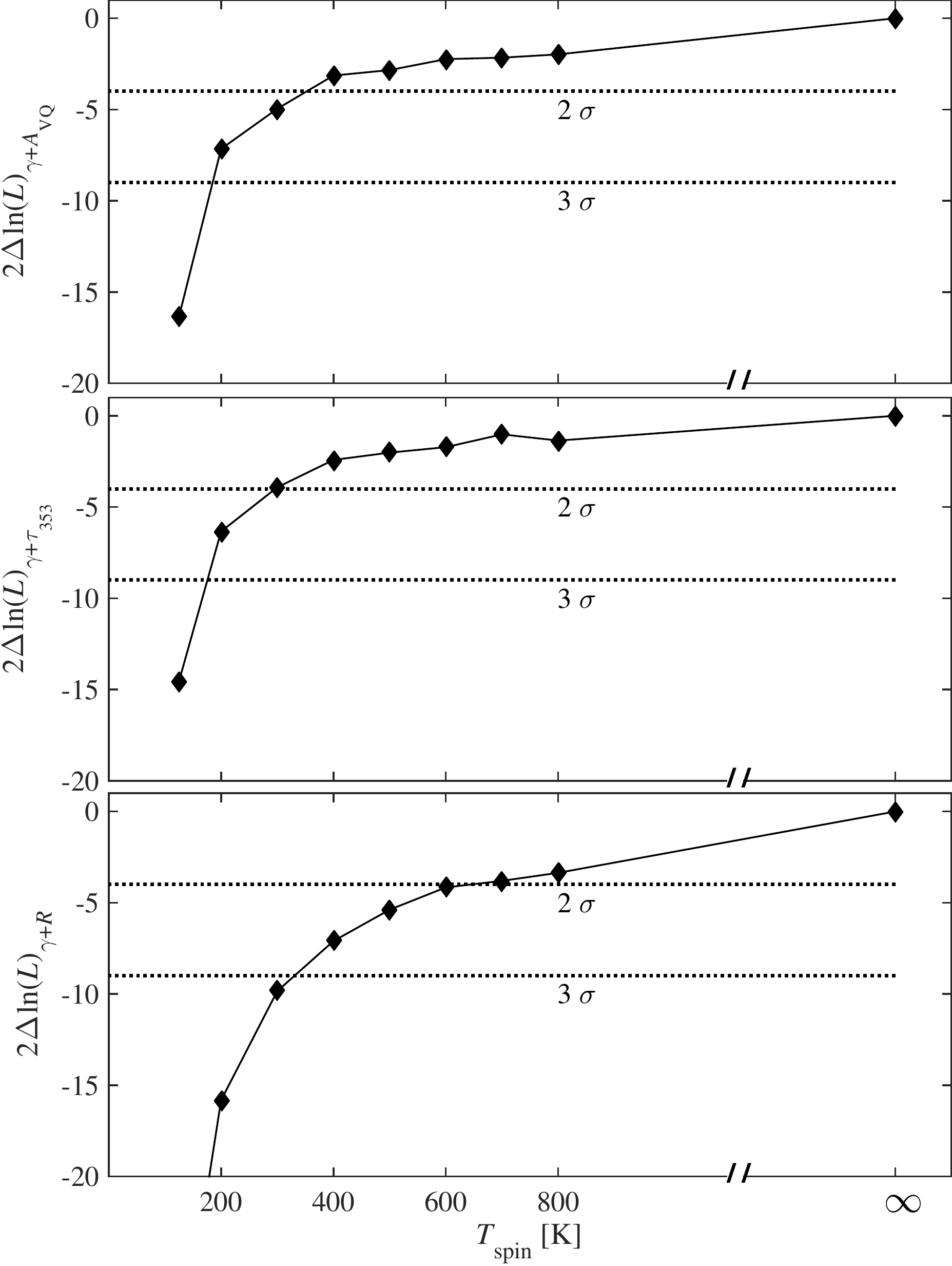}
\caption{Evolution of the log-likelihood ratios of the \g-ray fits with the \hi spin temperature, using the four independent energy bands in the \anaQ (top), \anaT (middle), and \anaR (bottom) analyses.}
\label{Tspin}
\end{figure}

For all analyses and for each energy band, we have found that the maximum likelihood value of the \g-ray fit increases with decreasing \hi opacity, thus with increasing spin temperature. The data in each band being independent, one can sum the log-likelihoods of each fit to constrain the spin temperature. Figure~\ref{Tspin} indicates that uniform temperatures of $T_{\rm{S}} > 340$\,K,  $> 300$\,K, and $> 640$\,K are preferred at the $2\,\sigma$ confidence levels for the \anaQ, \anaT, and \anaR analyses, respectively. The maximum in brightness temperature in the whole data cube is 152\,K. This indicates that optically thin conditions largely prevail across the whole velocity range.

In view of these results and with the added arguments that 
the CR spectrum inside the three \hi structures is close to the local one and that the \g-ray intensities have been shown to scale linearly with \nhi to higher, less transparent, column densities in more massive clouds \citep{LAT12_Cyg}, so that \g rays apparently trace all the \hi gas, we follow the \g-ray results and consider the optically thin \hi case as that which best represents the data.    

\section{CO calibration checks}
\label{sect_COcalib}

\begin{table*}[htbp]
\begingroup
\newdimen\tblskip \tblskip=5pt
\caption{Linear regression parameters and correlation coefficients between the \wco intensities measured with NANTEN, CfA, and \Planck.}
\label{table_COcal}
\vskip -3mm
\footnotesize
\setbox\tablebox=\vbox{
 \newdimen\digitwidth
 \setbox0=\hbox{\rm 0}
 \digitwidth=\wd0
 \catcode`*=\active
 \def*{\kern\digitwidth}
 \newdimen\signwidth
 \setbox0=\hbox{+}
 \signwidth=\wd0
 \catcode`!=\active
 \def!{\kern\signwidth}
 \halign{\tabskip=0pt\hfil#\hfil\tabskip=2em&
 \hfil#\hfil\tabskip=2em&
 \hfil#\hfil\tabskip=2em&
 \hfil#\hfil\tabskip=2em&
 \hfil#\hfil\tabskip=0pt\cr
\noalign{\doubleline}
$y$-axis& $x$-axis& Slope& Intercept& Corr.\ coeff.\cr
\noalign{\vskip 4pt\hrule\vskip 6pt}
                NANTEN&                    CfA& $1.015\pm0.008$&
 $0.92\pm0.06$& 0.95\cr
                NANTEN& \Planck\ CO ``Type~3''& $0.951\pm0.002$&
 $0.12\pm0.01$& 0.97\cr
\Planck\ CO ``Type~1''&                    CfA& $0.96*\pm0.02*$&
 $2.54\pm0.18$& 0.74\cr
CfA                   & \Planck\ CO ``Type~3''& $0.747\pm0.006$&
 $0.96\pm0.06$& 0.94\cr
\noalign{\vskip 4pt\hrule\vskip 6pt}
}}
\endPlancktablewide
\endgroup
\end{table*}
\begin{figure}[h]
\includegraphics[width=\hsize]{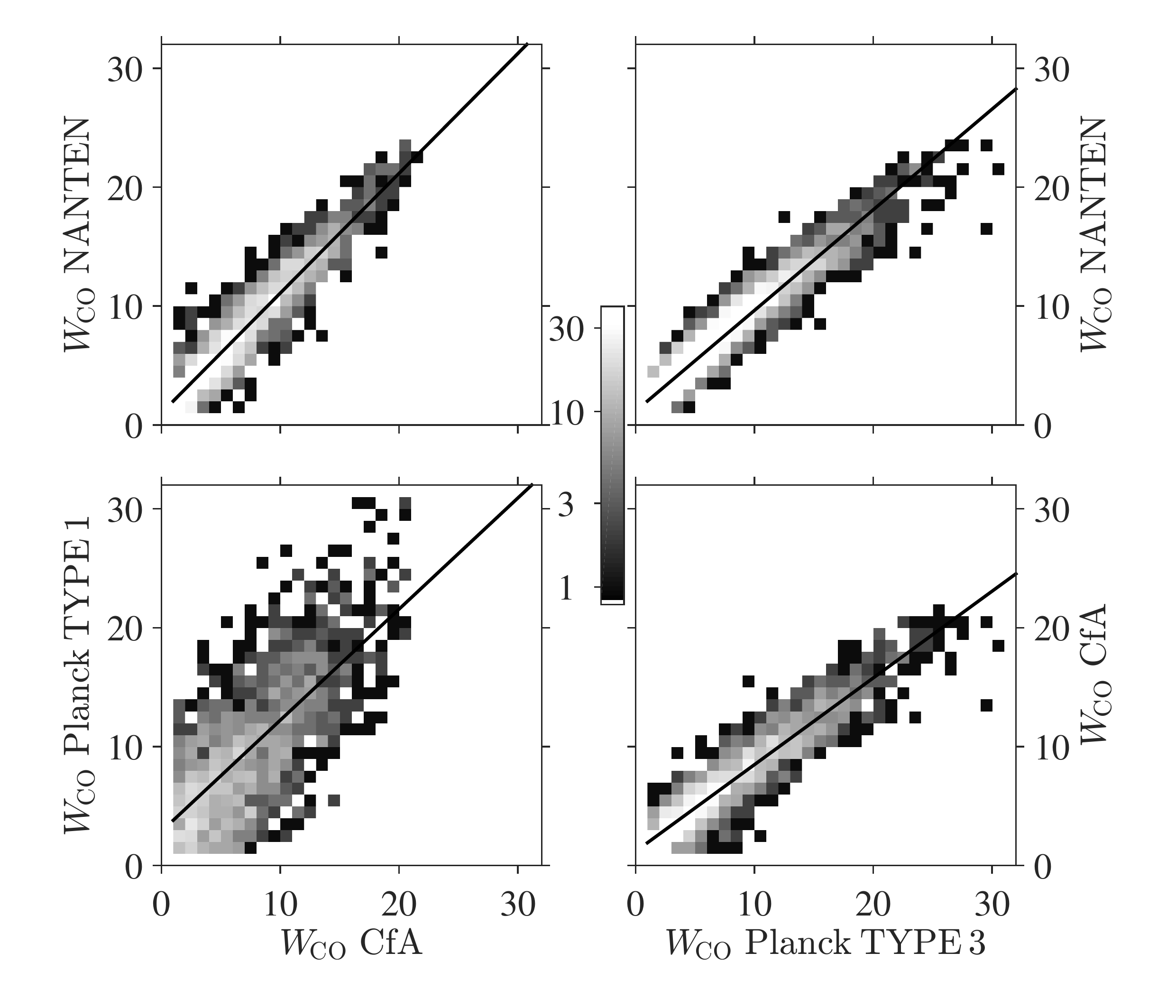}
\caption{Correlations and linear regressions between the NANTEN, CfA, and \Planck \texttt{TYPE\,3} and \texttt{TYPE\,1} \wco intensities in the analysis region. The data are sampled at 0\fdg125 and the intensities are given in \wcounit. The best-fit slopes are given in Table~\ref{table_COcal}}
\label{WCO_NANTEN_CfA}
\end{figure}

Checks on the NANTEN data cube have revealed several artefacts that we have corrected before integrating the spectra to obtain the \wco intensities given in Fig.~\ref{NHI_WCO}. We have removed significant negative lines, probably caused by the presence of a line in the off band in frequency-switching. High-order polynomial residuals were also present in the baseline profiles outside the main CO lines. They did not average out to zero in the \wco map, so we have filtered them from the regions void of significant CO intensity. Moment-masking is commonly used to clean \hi and \wco maps \citep{dame11}. It was not applicable here because the residuals, unlike noise, extended over several contiguous channels. We have filtered the original \wco map using the multiresolution support method implemented in the \texttt{MR \, filter} software \citep{starck98}, with seven scales in the bspline-wavelet transform (\`a trous algorithm). For the Gaussian noise of \wco, denoising with a hard $4\,\sigma$ threshold led to robust results. The final map shown in Fig.~\ref{NHI_WCO} is composed of the original, unfiltered, \wco intensity where the filtered one exceeded 1\,\wcounit, and of the filtered intensity outside these faint edges. Particular attention was paid to preserve the faint cloud edges which hold a fair fraction of the cloud mass because of their large volume.
 
Figure~\ref{WCO_NANTEN_CfA} illustrates the correlations found between the \wco intensities measured over the region with NANTEN (after the corrections described above), with CfA from the moment-masked, fully sampled observations of the Chamaeleon complex \citep{boulanger98}, and with \Planck \citep{planck14_co}. It shows that the data from  the \Planck \texttt{TYPE\,1} map is too noisy for our purpose. The other data sets show tight correlations. The correlation coefficients and the slope and intercept of the regression lines are given in Table~\ref{table_COcal}. They indicate that the \Planck \texttt{TYPE\,3} method systematically overpredicts the intensities in this region compared to the consistent yield of both radio telescopes. An earlier comparison of the NANTEN and CfA surveys across the sky had found 24\,\% larger intensities on average in the NANTEN maps \citep{planck11_dark}. Figure~\ref{WCO_NANTEN_CfA} shows that the NANTEN and CfA photometries fully agree in the Chamaeleon region. The data are consistent with a unit slope over the whole intensity range. The intercept is compatible with the sensitivity of the NANTEN observations, which varies between 1.5 and 2\,\wcounit across the region. 

\section{independent \g-ray and dust fits without DNM templates}
\label{sect_noDNM}

\begin{figure}
\includegraphics[width=\hsize]{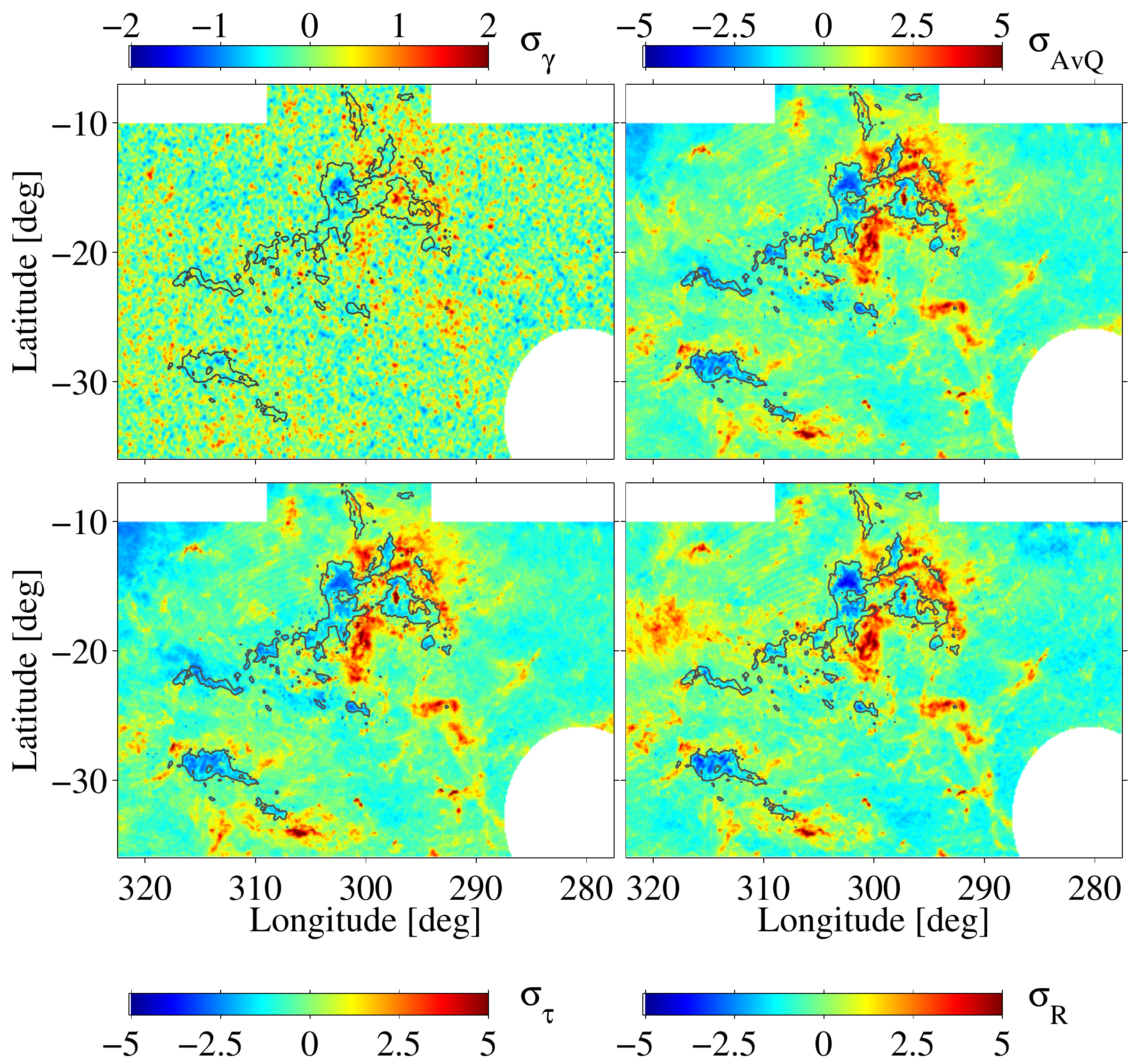}
\caption{Maps of residuals (data minus best-fit model, in sigma units) obtained when fitting the \g-ray, \avq, R, or \taunu data (clockwise) with optically thin \hi and \wco, but no DNM. In all plots, the grey contours outline the shape of the CO clouds at the 2.3\,\wcounit level. }
\label{resi_noDNM}
\end{figure}

In order to check for the presence of substantial amounts of gas not traced by the \hi and CO line intensities in the independent \g-ray and dust data sets, we have fitted the \g-ray and dust data with models that do not include a DNM template. All the other components described in Sects.~\ref{sect_gamodel} and \ref{sect_dustmodel} have been kept free. The resulting best fits are of significantly lower quality than those obtained with models including the DNM (see Sect.~\ref{sect_results}). The residual maps, however, are interesting in that they exhibit comparable regions of positive residuals in the independent dust and \g-ray data, as shown in Fig.~\ref{resi_noDNM}. 

We also note that, in the absence of a DNM template, the best fits yield systematically larger contributions from the \hi and CO components than in models that include the DNM. These components, in particular the CO one, are amplified to partially compensate for the missing gas structure. We find 4--13\,\% larger \g-ray emissivities for the local \hi and IVA components, and 22\,\% to 57\,\% larger CO \g-ray emissivities. The dust fits respond the same way, with 7--12\,\% larger \anh, 9--15\,\% larger opacities and 4--7\,\% larger specific powers for the local \hi and IVA components, and 25\,\% to 40\,\% larger CO contributions. As a consequence, the best-fit models often over-predict the data toward the CO clouds (see the negative residuals toward Cha II+III and Cha East I and II in Fig.~\ref{resi_noDNM}). These results prompted us to iterate the construction of DNM templates between the \g-ray and dust analyses in order to reduce the DNM bias on the determination of the \hi and CO parameters.

\section{Best-fit interstellar coefficients}
\label{sect_coeff}
\newpage 

\begin{sidewaystable*}
\begingroup
\newdimen\tblskip \tblskip=5pt
\caption{
Best-fit coefficients of the \g-ray models ($q$ for each energy band) and of
the dust models ($y$) for the \anaQ (top), \anaT (middle), and \anaR (bottom)
analyses.}
\label{table_qyfit_X2m}      
\vskip -3mm
\footnotesize
\setbox\tablebox=\vbox{
 \newdimen\digitwidth
 \setbox0=\hbox{\rm 0}
 \digitwidth=\wd0
 \catcode`*=\active
 \def*{\kern\digitwidth}
 \newdimen\signwidth
 \setbox0=\hbox{+}
 \signwidth=\wd0
 \catcode`!=\active
 \def!{\kern\signwidth}
 \halign{\tabskip=0pt\hbox to 1.5in{#\leaderfil}\tabskip=1em&
 \hfil#\hfil\tabskip=1em&
 \hfil#\hfil\tabskip=1em&
 \hfil#\hfil\tabskip=1em&
 \hfil#\hfil\tabskip=1em&
 \hfil#\hfil\tabskip=1em&
 \hfil#\hfil\tabskip=1em&
 \hfil#\hfil\tabskip=0pt\cr
\noalign{\doubleline}
\noalign{\vskip -3pt}
\omit\hfil Energy band\hfil& $q_{\rm{HI\,Cha}}$& $q_{\rm{HI\,IVA}}$&
 $q_{\rm{HI\,Gal}}$& ${q_{\rm{CO}}}^{\rm{a}}$& ${q_{\rm{DNM}}}^{\rm{b1}}$&
 $q_{\rm{IC}}$& $q_{\rm{iso}}$\cr
\noalign{\vskip 2pt\hrule\vskip 4pt}
$10^{2.6}$--$10^{2.8}$\,MeV&$1.34\pm0.07^{+0!**}_{-0.09}$&
$1.22\pm0.06^{+0!**}_{-0.04}$&$1.17\pm0.08^{+0!**}_{-0.03}$&
$2.01*\pm0.12*^{+0!**}_{-0.03*}$&$20.9*\pm1.3*^{+0.5*}_{-0!**}$&
$1.32*\pm0.13*^{+0.01*}_{-0!**}$&$0.8*\pm0.2*^{+0.1*}_{-0!**}$\cr \noalign{\vskip 4 pt}
 
$10^{2.8}$--$10^{3.2}$\,MeV& $1.42\pm0.06^{+0!**}_{-0.06}$&
$1.16\pm0.05^{+0!**}_{-0.04}$& $1.20\pm0.07^{+0!**}_{-0.03}$&
$1.762\pm0.112^{+0.006}_{-0!**}$& $19.8*\pm1.0*^{+0!**}_{-0.2*}$&
$1.44*\pm0.13*^{+0!**}_{-0.01*}$& $0.6*\pm0.3*^{+0.1*}_{-0!**}$\cr \noalign{\vskip 4 pt}
 
$10^{3.2}$--$10^{3.6}$\,MeV& $1.34\pm0.07^{+0!**}_{-0.07}$&
 $1.17\pm0.06^{+0!**}_{-0.04}$& $1.38\pm0.08^{+0!**}_{-0.04}$&
 $2.027\pm0.102^{+0!**}_{-0.009}$& $18.27\pm1.12^{+0.08}_{-0!**}$&
 $1.285\pm0.151^{+0.006}_{-0!**}$& $1.13\pm0.32^{+0.10}_{-0!**}$\cr \noalign{\vskip 4 pt}
 
$10^{3.6}$--$10^{5.0}$\,MeV& $1.64\pm0.13^{+0!**}_{-0.07}$&
 $1.43\pm0.11^{+0!**}_{-0.05}$& $1.29\pm0.17^{+0!**}_{-0.03}$&
 $2.065\pm0.176^{+0!**}_{-0.001}$& $22.31\pm2.12^{+0.07}_{-0!**}$&
 $1.25*\pm0.20*^{+0!**}_{-0.03*}$& $0.53\pm0.28^{+0.06}_{-0!**}$\cr \noalign{\vskip 4 pt}
 
$10^{2.6}$--$10^{5.0}$\,MeV& $1.38\pm0.04^{+0!**}_{-0.05}$&
 $1.20\pm0.03^{+0!**}_{-0.04}$& $1.25\pm0.04^{+0!**}_{-0.04}$&
 $1.921\pm0.072^{+0!**}_{-0.010}$& $19.99\pm0.67^{+0.02}_{-0!**}$&
 $1.33*\pm0.08*^{+0.01*}_{-0!**}$& $0.83\pm0.15^{+0.05}_{-0!**}$\cr 
\noalign{\vskip 2pt\hrule\vskip 4pt}
\omit& ${y_{\rm{HI\,Cha}}}^{\rm{c1}}$& ${y_{\rm{HI\,IVA}}}^{\rm{c1}}$&
 ${y_{\rm{HI\,Gal}}}^{\rm{c1}}$& ${y_{\rm{CO}}}^{\rm{d1}}$&
 ${y_{\rm{DNM}}}^{\rm{c1}}$&& ${y_{\rm{iso}}}^{\rm{e1}}$\cr 
\noalign{\vskip 2pt\hrule\vskip 4pt}
\omit& $8.11\pm0.09^{+0!**}_{-0.37}$& $7.31\pm0.07^{+0!**}_{-0.26}$&
 $7.41\pm0.09^{+0!**}_{-0.22}$& $16.38\pm0.34^{+0!**}_{-0.04}$&
 $7.1\pm0.2^{+0!**}_{-0.4}$& & $-16.4\pm0.5^{+1.2}_{-0!**}$\cr 
\noalign{\vskip 2pt\hrule\vskip 4pt}
\omit\hfil Energy band\hfil& $q_{\rm{HI\,Cha}}$& $q_{\rm{HI\,IVA}}$&
 $q_{\rm{HI\,Gal}}$& ${q_{\rm{CO}}}^{\rm{a}}$& ${q_{\rm{DNM}}}^{\rm{b2}}$&
 $q_{\rm{IC}}$& $q_{\rm{iso}}$\cr 
\noalign{\vskip 2pt\hrule\vskip 4pt}
$10^{2.6}$--$10^{2.8}$\,MeV& $1.36\pm0.07^{+0!**}_{-0.06}$&
 $1.22\pm0.06^{+0!**}_{-0.04}$& $1.05\pm0.08^{+0!**}_{-0.04}$&
 $1.91*\pm0.13*^{+0!**}_{-0.02*}$& $8.88\pm0.59^{+0.06}_{-0!**}$&
 $1.54*\pm0.13*^{+0.03*}_{-0!**}$& $0.63\pm0.23^{+0.05}_{-0!**}$\cr \noalign{\vskip 4 pt}
 
$10^{2.8}$--$10^{3.2}$\,MeV& $1.45\pm0.06^{+0!**}_{-0.04}$&
 $1.17\pm0.05^{+0!**}_{-0.03}$& $1.09\pm0.07^{+0!**}_{-0.03}$&
 $1.695\pm0.121^{+0!**}_{-0.007}$& $8.23\pm0.43^{+0!**}_{-0.03}$&
 $1.69*\pm0.13*^{+0!**}_{-0.01*}$& $0.28\pm0.27^{+0.07}_{-0!**}$\cr \noalign{\vskip 4 pt}
 
$10^{3.2}$--$10^{3.6}$\,MeV& $1.37\pm0.07^{+0!**}_{-0.07}$&
 $1.19\pm0.06^{+0!**}_{-0.05}$& $1.27\pm0.09^{+0!**}_{-0.02}$&
 $1.977\pm0.106^{+0!**}_{-0.004}$& $7.50\pm0.51^{+0!**}_{-0.02}$&
 $1.498\pm0.150^{+0!**}_{-0.008}$& $0.8*\pm0.3*^{+0.2*}_{-0!**}$\cr \noalign{\vskip 4 pt}
 
$10^{3.6}$--$10^{5.0}$\,MeV& $1.70\pm0.13^{+0!**}_{-0.09}$&
 $1.45\pm0.11^{+0!**}_{-0.04}$& $1.16\pm0.17^{+0!**}_{-0.03}$&
 $2.021\pm0.180^{+0!**}_{-0.004}$& $8.97\pm0.93^{+0.02}_{-0!**}$&
 $1.388\pm0.200^{+0.010}_{-0!**}$& $0.37\pm0.28^{+0.07}_{-0!**}$\cr \noalign{\vskip 4 pt}
 
$10^{2.6}$--$10^{5.0}$\,MeV& $1.42\pm0.04^{+0!**}_{-0.06}$&
 $1.21\pm0.03^{+0!**}_{-0.04}$& $1.13\pm0.05^{+0!**}_{-0.04}$&
 $1.87*\pm0.08*^{+0!**}_{-0.02*}$& $8.20\pm0.33^{+0.09}_{-0!**}$&
 $1.556\pm0.080^{+0.006}_{-0!**}$& $0.52\pm0.15^{+0.09}_{-0!**}$\cr 
\noalign{\vskip 2pt\hrule\vskip 4pt}
\omit& ${y_{\rm{HI\,Cha}}}^{\rm{c2}}$& ${y_{\rm{HI\,IVA}}}^{\rm{c2}}$&
 ${y_{\rm{HI\,Gal}}}^{\rm{c2}}$& ${y_{\rm{CO}}}^{\rm{d2}}$&
 ${y_{\rm{DNM}}}^{\rm{c2}}$&& ${y_{\rm{iso}}}^{\rm{e2}}$\cr 
\noalign{\vskip 2pt\hrule\vskip 4pt}
\omit & $1.63\pm0.02^{+0!**}_{-0.06}$& $1.48\pm0.02^{+0!**}_{-0.04}$&
 $1.24\pm0.02^{+0!**}_{-0.04}$& $4.130\pm0.102^{+0.002}_{-0!**}$&
 $1.73\pm0.05^{+0!**}_{-0.09}$& & $-4.0\pm0.1^{+0.2}_{-0!**}$\cr 
\noalign{\vskip 2pt\hrule\vskip 4pt}
\omit\hfil Energy band\hfil& $q_{\rm{HI\,Cha}}$& $q_{\rm{HI\,IVA}}$&
 $q_{\rm{HI\,Gal}}$& ${q_{\rm{CO}}}^{\rm{a}}$& ${q_{\rm{DNM}}}^{\rm{b3}}$&
 $q_{\rm{IC}}$& $q_{\rm{iso}}$\cr
\noalign{\vskip 2pt\hrule\vskip 4pt}
$10^{2.6}$--$10^{2.8}$\,MeV& $1.31\pm0.08^{+0!**}_{-0.01}$&
 $1.20*\pm0.06*^{+0!**}_{-0.06*}$& $1.55*\pm0.08*^{+0!**}_{-0.03*}$&
 $2.33*\pm0.12*^{+0!**}_{-0.01*}$& $5.73\pm0.38^{+0!**}_{-0.06}$&
 $0.59\pm0.14^{+0.09}_{-0!**}$& $1.45\pm0.24^{+0!**}_{-0.05}$\cr \noalign{\vskip 4 pt}

$10^{2.8}$--$10^{3.2}$\,MeV& $1.40\pm0.06^{+0!**}_{-0.17}$&
 $1.13*\pm0.05*^{+0!**}_{-0.06*}$& $1.59*\pm0.07*^{+0.07*}_{-0!**}$&
 $2.045\pm0.108^{+0.003}_{-0!**}$& $5.5*\pm0.3*^{+0.3*}_{-0!**}$&
 $0.6*\pm0.1*^{+0!**}_{-0.2*}$& $1.6*\pm0.3*^{+0.8*}_{-0!**}$\cr \noalign{\vskip 4 pt}

$10^{3.2}$--$10^{3.6}$\,MeV& $1.35\pm0.07^{+0!**}_{-0.05}$&
 $1.15*\pm0.06*^{+0!**}_{-0.07*}$& $1.73*\pm0.09*^{+0!**}_{-0.12*}$&
 $2.25*\pm0.10*^{+0!**}_{-0.02*}$& $5.0*\pm0.3*^{+0!**}_{-0.1*}$&
 $0.6*\pm0.2*^{+0.1*}_{-0!**}$& $1.81\pm0.32^{+0.08}_{-0!**}$\cr \noalign{\vskip 4 pt}
 
$10^{3.6}$--$10^{5.0}$\,MeV& $1.7*\pm0.1*^{+0!**}_{-0.2*}$&
 $1.42*\pm0.11*^{+0!**}_{-0.06*}$& $1.75*\pm0.18*^{+0!**}_{-0.01*}$&
 $2.32*\pm0.17*^{+0!**}_{-0.02*}$& $5.9*\pm0.6*^{+0!**}_{-0.2*}$&
 $0.68\pm0.22^{+0!**}_{-0.06}$& $0.9*\pm0.3*^{+0.3*}_{-0!**}$\cr \noalign{\vskip 4 pt}

$10^{2.6}$--$10^{5.0}$\,MeV& $1.38\pm0.04^{+0!**}_{-0.10}$&
 $1.173\pm0.031^{+0!**}_{-0.001}$& $1.626\pm0.045^{+0.008}_{-0!**}$&
 $2.201\pm0.078^{+0.005}_{-0!**}$& $5.49\pm0.19^{+0.01}_{-0!**}$&
 $0.57\pm0.08^{+0.02}_{-0!**}$& $1.5*\pm0.1*^{+0.2*}_{-0!**}$\cr
\noalign{\vskip 2pt\hrule\vskip 4pt}
\omit& ${y_{\rm{HI\,Cha}}}^{\rm{c3}}$& ${y_{\rm{HI\,IVA}}}^{\rm{c3}}$&
 ${y_{\rm{HI\,Gal}}}^{\rm{c3}}$& ${y_{\rm{CO}}}^{\rm{d3}}$&
 ${y_{\rm{DNM}}}^{\rm{c3}}$& & ${y_{\rm{iso}}}^{\rm{e3}}$\cr
\noalign{\vskip 2pt\hrule\vskip 4pt}
\omit &$3.86\pm0.04^{+0!**}_{-0.05}$& $3.65\pm0.04^{+0!**}_{-0.04}$&
  $4.84\pm0.05^{+0!**}_{-0.07}$& $5.10\pm0.13^{+0!**}_{-0.02}$&
  $2.72\pm0.09^{+0!**}_{-0.01}$& & $-4.7\pm0.3^{+0.2}_{-0!**}$\cr
\noalign{\vskip 2pt\hrule\vskip 4pt}
}}
\endPlancktablewide
\tablefoot{The first uncertainties are statistical, the second result from
changes in \hi spin temperature over the 2$\,\sigma$ confidence interval from
the optically thin case. The latter do not include the 8\,\% systematic
uncertainty in the LAT sensitive area for the $q$ parameters. Model
uncertainties have been used to optimize the dust fits. \par 
$^{{\rm a}}$ In \qcounit,   
\quad $^{{\rm b1}}$ In \qdnmQunit,
\quad $^{{\rm b2}}$ In \qdnmTunit,
\quad $^{{\rm b3}}$ In \qdnmRunit,
\quad $^{{\rm c1}}$ In \yhQunit,
\quad $^{{\rm c2}}$ In \yhTunit,
\quad $^{{\rm c3}}$ In \yhRunit,
\quad $^{{\rm d1}}$ In \ycoQunit, \par
$^{{\rm d2}}$ In \ycoTunit,
\quad $^{{\rm d3}}$ In \ycoRunit,
\quad $^{{\rm e1}}$ In \yisoQunit,
\quad $^{{\rm e2}}$ In \yisoTunit,
\quad $^{{\rm e3}}$ In \yisoRunit.}
\endgroup
\end{sidewaystable*}

\end{appendix}

\end{document}